\documentclass[12pt,preprint]{aastex}

\shorttitle{The molecular gas in LIRGs: mass estimates}
\shortauthors{Papadopoulos, van der Werf, Isaak, Xilouris,  M\"uhle, \& Gao}
\begin{document}

\title{The molecular gas in Luminous Infrared Galaxies II: extreme physical
conditions, and their effects on the X$_{\rm \bf co}$ factor}

\author{Padelis \ P.\ Papadopoulos}
\affil{Max Planck Institute for Radioastronomy,  Auf dem H\"ugel 69,  D-53121 Bonn,
 Germany}
\email{padelis@mpifr-bonn.mpg.de}

\author{Paul van der Werf}
\affil{Leiden Observatory,  Leiden University, P.O.~Box~9513, NL-2300 RA Leiden, 
The Netherlands}
\email{pvdwerf@strw.leidenuniv.nl}

\author{E. Xilouris}
\affil{Institute of Astronomy and Astrophysics, National Observatory of Athens,
I.Metaxa \& Vas.Pavlou str., GR-15236, Athens, Greece}
\email{xilouris@astro.noa.gr}

\author{Kate G. Isaak}
\affil{Research and Scientific Support Department, European Space Agency,
ESTEC, Keplerlaan~1, NL-2201, The Netherlands}
\email{kisaak@rssd.esa.int}

\and

\author{Yu Gao}
\affil{Purple Mountain Observatory, Chinese Academy of Sciences, Nanjing, Jiangsu 210008, China}
\email{pmogao@gmail.com}

\begin{abstract}

In  this work  we  conclude the  analysis  of our  CO  line survey  of
Luminous     Infrared     Galaxies     (LIRGs:    $\rm     L_{IR}$$\ga
$10$^{11}$\,L$_{\odot  }$)  in   the  local  Universe  (Paper\,I),  by
focusing  on the  influence of  their average  ISM properties  on the
total molecular gas mass estimates via the so-called $\rm X_{co}$=$\rm
M(H_2)/L_{co,1-0}$ factor. One-phase  radiative transfer models of the
global  CO Spectral Line  Energy Distributions  (SLEDs) yield  an $\rm
X_{co}$ distribution with: $\rm \langle X_{co}\rangle $$\sim
$$(0.6\pm   0.2)$\,$\rm   M_{\odot}$\,(K\,km\,s$^{-1}$\,pc$^2$)$^{-1}$
over a  significant range of  average gas densities,  temperatures and
dynamical states.  The latter  emerges as the most important parameter
in determining  $\rm X_{co}$, with unbound states  yielding low values
and self-gravitating  states the  highest ones.  Nevertheless  in many
(U)LIRGs  where  available higher-J  CO  lines  (J=3--2, 4--3,  and/or
J=6--5)  or  HCN  line  data  from the  literature  allow  a  separate
assesment     of     the     gas     mass    at     high     densities
($\geq$10$^{4}$\,cm$^{-3}$) rather than a simple one-phase analysis we
find     that    {\it    near-Galactic     $X_{co}$$\sim    $(3-6)\,$
  M_{\odot}$\,(K\,km\,s$^{-1}$\,pc$^2$)$^{-1}$      values      become
  possible.}  We  further show that in the  highly turbulent molecular
gas  in ULIRGs  a high-density  component will  be common  and  can be
massive enough for its high $\rm X_{co}$ to dominate the average value
for the entire galaxy.  Using  solely low-J CO lines to constrain $\rm
X_{co}$ in such  environments (as it has been the  practice up to now)
may have thus resulted  to {\it systematic underestimates of molecular
  gas mass in  ULIRGs} as such lines are dominated  by a warm, diffuse
and  unbound gas phase  with low  $\rm X_{co}$  but very  little mass.
Only well-sampled  high-J CO SLEDs (J=3--2 and  higher) and/or multi-J
observations of heavy rotor molecules (e.g. HCN) can circumvent such a
bias, and the  latter type of observations may  have actually provided
early evidence  of it in  local ULIRGs. The  only way that  the global
$\rm  X_{co}$  of  such  systems  could be  significantly  lower  than
Galactic is if the average dynamic  state of the dense gas is strongly
gravitationally unbound.  This is an unlikely possibility that must be
nevertheless examined,  with lines of  rare isotopologues of  high gas
density tracers (e.g.  H$^{13}$CN,  high-J $^{13}$CO lines) being very
valuable in yielding (along with  the lines of the main isotopes) such
constraints.  For  less IR  luminous, disk-dominated systems,  we find
that the  galaxy-averaged $\rm X_{co}$ deduced by  one-phase models of
global SLEDs can also underestimate  the total molecular gas mass when
much of  it lies  in a SF-quiescent  phase extending beyond  a central
star-forming region. This is because  such a phase (and its large $\rm
X_{co}$)  remain inconspicious  in global  CO SLEDs.   Finally detailed
studies of  a subsample  of galaxies finds  ULIRGs with  large amounts
($\sim $10$^{9}$\,M$_{\odot}$) of very  warm ($\geq $100\,K) {\it and}
dense  gas ($\ga  $10$^5$\,cm$^{-3}$) that  could represent  a serious
challenge to  photon-dominated regions as  the main energy  portals in
the molecular ISM of such systems.

\end{abstract}


\section{Introduction}

Soon after  the discovery of  the luminous infrared  galaxies (LIRGs),
whose bolometric  luminosities are dominated  by the infrared  part of
their   Spectral  Energy   Distributions  (SEDs)   ($\rm  L_{IR}$$\geq
$10$^{11}$\,L$_{\odot}$) (e.g.  Soifer et  al.  1987), single dish and
interferometric  CO  J=1--0,  2--1  line  observations  were  used  to
determine their total molecular gas mass and its distribution (Sanders
et al. 1988a;  Tinney et al.  1990; Wang et al.   1991; Sanders et al.
1991; Solomon et al.~1997; Downes  \& Solomon 1998; Bryant \& Scoville
1996, 1999).  These efforts  were paralleled by several investigations
of the  so-called $\rm X_{co}$=M(H$_2$)/$\rm  L_{CO}$(1--0) factor and
its dependance on the average ISM conditions both theoretical (Dickman
et al.   1988; Maloney \& Black  1988; Wolfire et  al.  1993; Sakamoto
1996;  Bryant   \&  Scoville   1996;  Wall  2007)   and  observational
(e.g. Israel 1988, 1993, 1997; Solomon et al.  1997; Downes \& Solomon
1998; Yao et al. 2003).  The average molecular gas conditions in LIRGs
used   in  such   studies  have   been  typically   constrained  using
CO(2--1)/(1--0)  and  CO/$^{13}$CO  J=1--0,  2--1  line  ratios  (e.g.
Braine  \& Combes  1992;  Horellou et  al  1995; Aalto  et al.   1995;
Papadopoulos  \&  Seaquist 1998).   Higher-J  transitions (J=3--2  and
higher)  were  used  only  sporadically and  mostly  for  star-forming
galactic  nuclei (e.g.   Devereux et  al.  1994;  White et  al.  1994;
G\"usten et al.  1996; Nieten et al.  1999; Mauersberger et al.  1999;
Dumke et al.  2001; Yao et al.  2003). This was a result of the larger
difficulties  such  observations  pose  in terms  of  available  submm
receivers, their  sensitivity, and  the dry weather  conditions needed
(especially   for  $\nu   $$\ga  $460\,GHz,   CO   J=4--3).   Receiver
sensitivity limitations  also hindered  large multi-J line  surveys of
the much  fainter lines  from heavy rotor  molecules such as  HCN that
probe  higher density  gas ($>$10$^{4}$\,cm$^{-3}$)  except  in nearby
galactic nuclei  (Jackson et al.  1995;  Paglione et al.   1997) and a
few luminous ULIRGs (e.g. Gracia-Carpio et al.~2008).

Such limitations will soon be overcome after the ongoing commissioning
of  the Atacama  Large  Millimeter Array  (ALMA)  is completed.   Then
routine  multi-J observations  of CO  and heavy  rotor  molecules will
yield unhindered view over the  entire range of physical conditions in
molecular  clouds, from  their quiescent  and low-density  phase ($\rm
n(H_2)$$\sim     $(10$^2$-10$^3$)\,cm$^{-3}$,    $\rm    T_{kin}$$\sim
$(10-15)\,K) to the dense and warm gas intimately associated with star
formation  ($\rm  n(H_2)$$\sim  $(10$^{4}$-10$^{7}$)\,cm$^{-3}$,  $\rm
T_{kin}$$\sim  $(30-150)\,K).  The  power  of interferometric  multi-J
line imaging in  revealing the mass distribution of  dense warm SF gas
in LIRGs has already  been demonstrated by pioneering SMA observations
(Sakamoto et al.  2008; Wilson et al.  2009; Iono et al.  2007, 2009),
while  in  the  grand spiral  M\,51  CO  line  ratio imaging  at  high
resolution  revealed  AGN-excited gas  in  its  nucleus  (Iono et  al.
2004).  The  influence of the  high-excitation conditions found  in SF
regions gas  on the  $\rm X_{co}$ in  galaxies may not  necessarily be
strong since dense and warm SF gas amounts to only $\sim $(0.5-3)\% of
typical Giant Molecular Clouds (GMCs) mass.  Even smaller fractions of
the  total molecular  gas in  spirals disks  resides in  their centers
($\sim $(0.1-1)\%)  where strong tidal fields, high  cosmic ray energy
densities  and/or AGN  can  drive a  high  molecular line  excitation.
Nevertheless  this  may  no  longer  be  true  for  the  merger-driven
starbursts in ULIRGs  where a dense SF gas phase  can contain the bulk
of their total  molecular gas mass (e.g. Solomon et  al.  1992; Gao \&
Solomon 2004).  Moreover, cases of AGN-driven mechanical and radiative
feedback affecting  the bulk of the  molecular gas of  the host galaxy
and the corresponding CO  SLEDs have now been identified (Papadopoulos
et al.   2008; van der Werf  et al.  2010).  These  systems along with
ULIRGs,  yield  a nearby  glimpse  of  ISM  conditions that  could  be
prevelailing in the distant~Universe.

In the present work we  examine the influence of the average molecular
gas  conditions found  in LIRGs  (Papadopoulos et  al  2011, hereafter
Paper\,I) on the  $\rm X_{co}$ factor.  We do so  by using the largest
combined database of  LIRGs/CO transitions for which such  a study has
been conducted,  while discussing  also the limitations  and potential
biases of past theoretical and observational studies.  We then outline
methods that  could be employed in  the upcoming era of  ALMA, and the
special role  the Herschel Space  Observatory (HSO) can  play, towards
improved  total molecular  gas mass  estimates, especially  for ULIRGs
($\rm  L_{IR}$$>$$10^{12}$\,$\rm L_{\odot}$).   Several  such galaxies
whose CO  line ratios indicate  extreme ISM conditions  (see Paper\,I)
are now studied  individualy, their impact on the  $\rm X_{co}$ values
examined in  detail.  Throughout this  paper we adopt a  flat $\Lambda
$-dominated cosmology  with $\rm H_0$=71\,km\,s$^{-1}$\,Mpc$^{-1}$ and
$\Omega_{\rm m}$=0.27.

\section{Molecular gas physical conditions and  mass estimates in LIRGs}

 The  formal dependance  of the  $\rm  X_{co}$ factor  on the  average
 density, temperature,  and kinematic  state of large  molecular cloud
 ensembles  (where  the statistical  notion  of  $\rm X_{co}$  remains
 applicable)  is explored  in  several papers  (e.g.   Dickman et  al.
 1986; Young \& Scoville 1991; Bryant \& Scoville 1996; Solomon et al.
 1997; Papadopoulos \&  Seaquist 1999; Downes \& Solomon  1998; Yao et
 al. 2003).  CO and $^{13}$CO lines can yield constraints on these ISM
 properties, and thus on the corresponding $\rm X_{co}$, via radiative
 transfer models (e.g. Mao et al.  2000; Weiss et al.  2001).  In this
 regard  low-J  CO  SLEDs  (up  to J=3--2)  with  $\rm  n_{crit}$$\sim
 $(400\,cm$^{-3}$--10$^4$)\,cm$^{-3}$     and    $\rm    E_J/K_B$$\sim
 $(5.5--33)\,K are  adequate for determining the average  state of the
 molecular gas  and thus the  appropriate $\rm X_{co}$,  provided that
 most of its mass is distributed  in ordinary GMCs.  The low-J CO SLED
 segment  and  its modeling  can  then  in  principle yield  the  mass
 normalization for the  entire CO or of any  other molecular line SLED
 (e.g.  of HCN) typically  emanating from much smaller mass fractions.
 For several ULIRGs  in our sample whose CO SLEDs  will be extended up
 to J=13--12 using the  HSO this normalization is especially important
 as it allows determining the mass of the highly-excited molecular gas
 emitting the very high-J  lines, and setting important constraints on
 the energy  source responsible  for its excitation  (van der  Werf et
 al.~2010; Rangwala et al.~2011).

\subsection{Prior work and some  methodological limitations}

 There are  currently only three  major observational studies  of $\rm
 X_{co}$ using CO  lines for substantial LIRG samples  (Solomon et al.
 1997;  Downes \& Solomon  1998; Yao  et al.   2003).  For  the highly
 turbulent  molecular  gas  in  (U)LIRGs  these  typically  find  $\rm
 X_{co}$=(1/10--1/3)$\rm X_{co,Gal}$,  with the low  values attributed
 mostly  to non  self-gravitating  gas distributions.   For SF  spiral
 disks   a  $\rm   X_{co,Gal}$$\sim   $(4-5)\,$\rm  X_l$\footnote{$\rm
 X_l$=M$_{\odot}$ (K\,km\,s$^{-1}$\,pc$^2$)$^{-1}$} remains applicable
 as  most  of their  molecular  gas  is  found in  cool,  low-density,
 self-gravitating GMCs  as those  in the Milky  Way, pockmarked  by SF
 ``spots''  of warm  and  dense  gas with  also  nearly Galactic  $\rm
 X_{co}$ (e.g.  Young \& Scoville 1991).  In isolated spirals low $\rm
 X_{co}$$\sim $(1/20--1/5)$\rm X_{co,Gal}$ values can be found only in
 their nuclei  (2r$\la $100--200\,pc)  (Regan 2000), but  involve only
 small fractions of the total  molecular gas reservoirs.  Thus for the
 metal-rich environments of LIRGs current observational work points to
 a bimodal  $\rm X_{co}$  distribution, with near-Galactic  values for
 all isolated spiral disks, and $\rm \langle X_{co}\rangle
 $$\sim  $(1/4-1/5)$\times  $$\rm  X_{co,Gal}$  in  the  merger-driven
 starbursts of  ULIRGs.  This view  has even been widely  adopted even
 for high-z star-forming LIRGs where  only sparse sampling of their CO
 SLEDs  exists  (e.g  Greve  et  al.   2005;  Tacconi  et  al.   2006;
 Dannerbauer  et al.   2009).  It  is  worth noting  that the  claimed
 bimodality of the Schmidt-Kennicutt star formation relations (linking
 SFR and molecular gas supply) between disks and mergers (Daddi et al.
 2010a;  Genzel et  al.  2010)  is nearly  equivalent to  such  an $\rm
 X_{co}$ bimodality (Narayanan et al.  2010 and references therein).

 Nevertheless much  of the aforementioned  observational work contains
 some serious  methodological limitations  borne out from  the limited
 molecular  line data  available per  object and  specific assumptions
 made about the radiative transfer models. These are:

\begin{itemize}

\item Setting  $\rm T_{kin}$=$\rm T_{dust}$  in the CO  line radiative
     transfer  models  used  to  constrain $\rm  X_{co}$  even  though
     photoelectric, turbulent, or cosmic-ray heating can easily induce
     $\rm T_{kin}$$\gg$$\rm T_{dust}$.

\item Using  a constant dV/dR  (usually 1\,km\,s$^{-1}$\,pc$^{-1}$) in
      the Large Velocity Gradient (LVG) radiative transfer models that
      interpret CO  line ratios even  if: a) virial gas  motions alone
      correspond to $\sim $8-10 times  higher values for the dense gas
      ($\rm n$$\ga $10$^5$\,cm$^{-3}$) that may be dominant in ULIRGs,
      and  b)  stellar  mass  concomitant with  molecular  gas  and/or
      tidally-induced  velocity fields  can easily  yield  much larger
      dV/dR values.

\item The  use of empirical relations  and object-specific assumptions
      for obtaining $\rm X_{co}$  expressions that can account for the
      potentially  significant  stellar   mass  concomitant  with  the
      molecular gas, an issue that arises especially in ULIRGs.

\item The  molecular lines used in  such studies (low-J  CO lines) are
         insensitive to the properties  of the gas with densities $\rm
         n(H_2)$$>$10$^3$\,cm$^{-3}$.

\end{itemize}
 
\noindent
The last  two set the  most serious methodological limitations  on the
very influential study of $\rm  X_{co}$ in ULIRGs by Downes \& Solomon
1998  (herafter DS98), whose  results are  widely used  for presumably
similar systems  at high  redshifts.  In that  study models of  the CO
J=1--0, 2--1  interferometric images of  5 ULIRGs were used  to deduce
$\rm  X_{co}$$\sim $(1/5-1/6)$\rm  X_{co,Gal}$, attributing  these low
values to  a continous  molecular gas distribution  encompassing large
fractions of stellar  mass and thus velocity fields  determined by the
total (gas)+(stellar) mass  (see also Downes et al.   1993; Solomon et
al.  1997).   The use of empirical relations  (e.g.  $\rm L_{IR}/L^{'}
_{co}$=200\,$\rm   L_{\odot}$\,(K\,km\,s$^{-1}$\,pc$^2$)$^{-1}$),  and
setting  a $\rm  r_{m,*}$=$\rm  M_{new,*}/M_{bulge,*}$$\sim $0.5  mass
ratio  of newly  formed stars  ($\rm M_{new,*}$)  to those  in  an old
stellar bulge  ($\rm M_{bulge,*}$) in the pre-merger  spirals make the
computed $\rm  X_{co}$ {\it  specific to the  few ULIRGs  studied} and
certainly not automatically applicable to other ULIRGs in the local or
the    distant    Universe.    Moreover    ``freezing''    dV/dR    to
1\,km\,s$^{-1}$\,pc$^{-1}$   renders    such   models   incapable   of
constraining the average dynamical state of the gas and thus exploring
its effect on $\rm X_{co}$ in a straightforward~way (see section 2.2).

 Furthermore, while the DS98  formalism can be generalized to arbitary
 $\rm r_{m,*}$  values, it remains impractical for  the dusty gas-rich
 systems  at  high redshifts  where  stellar  populations are  heavily
 dust-enshrouded.  Finally an $\rm  X_{co}$ factor deduced solely from
 CO J=1--0, 2--1  line emission models may be  inapplicable for a much
 denser gas phase  (n(H$_2$)$\ga $10$^4$\,cm$^{-3}$).  This will yield
 a small error when such a  phase represents only a few\% of the total
 mass  per typical GMC  (as is  the case  in spiral  disks), but  to a
 potentially  much  larger  error  if  the  dense  gas  dominates  the
 molecular  gas mass  budget.   These difficulties  are compounded  by
 setting  dV/dR=1\,km\,s$^{-1}$\,pc$^{-1}$, which  while  justified by
 the sole CO(2-1)/(1-0) line ratio available in the DS98 study, may be
 partly  responsible for  the contradictory  results obtained  for the
 average gas  density and  CO line excitation.   Indeed while  HCN and
 high-J CO observations indicate much of  the gas in some ULIRGs to be
 very  dense  ($\rm  n(H_2)$$>$10$^4$\,cm$^{-3}$), radiative  transfer
 models of CO J=1--0 and  J=2--1 interferometric images of some of the
 same ULIRGs  (e.g.  Arp\,220)  become incompatible with  these images
 already for n$\ga $10$^{3}$\,cm$^{-3}$ (DS98).  Velocity gradients of
 $\rm  dV/dR$$\gg$1\,km\,s$^{-1}$\,pc$^{-1}$ (expected for  the highly
 turbulent gas  disks of ULIRGs)  and high temperatures for  the dense
 gas  can much reduce  line optical  depths and  the emergent  CO line
 emission possibly aleviating these~discrepancies.

 In  summary all  current  observational studies  of  $\rm X_{co}$  in
 LIRGs,  leave  its  range  and  dependance on  the  average  density,
 temperature and  kinematic state of  the molecular gas  still largely
 unexplored.  The  small LIRG/CO line  datasets used in  such studies,
 and the  aforementioned limitations  of the analysis  methods, limits
 their applicability to other such systems in the local or the distant
 Universe.

\subsection{The average  state of the molecular gas and $\rm X_{co}$}

 Our dataset of  total CO lines luminosities for  a substantial sample
 of LIRGs provides  a good opportunity for exploring  the $\rm X_{co}$
 dependance  on  the  average  ISM state  in  vigorously  star-forming
 systems.  The average dV/dR for the gas is now a free parameter to be
 constrained by  LVG radiative transfer models along  with the density
 and    temperature.    This    provides   an    extinction-free   and
 straightforward method  for determining  the effects of  stellar mass
 concomitant with  molecular gas and/or tidal  disruption of molecular
 clouds on $\rm X_{co}$.

To  place our  discussion regarding  molecular gas  mass  estimates in
LIRGs in  an autonomous context  we reproduce the $\rm  X_{co}$ factor
(see Appendix) as

\begin{equation}
\rm X_{CO}=\frac{M(H_2)}{L^{'} _{CO(1-0)}}=\frac{3.25}{\sqrt{\alpha}} 
\frac{\sqrt{n(H_2)}}{T_{b,1-0}} K_{vir}^{-1}\, \left(\frac{M_{\odot}}{K\,km\,s^{-1}\,pc^2}\right),
\end{equation}

\noindent
where  $\alpha$=0.55--2.4  depending  on  the  assumed  cloud  density
profile  (see Bryant  \& Scoville  1996, Eqs  A\,11, A\,17),  and $\rm
n(H_2)$, $\rm  T_{b,1-0}$ are  the average gas  density and  CO J=1--0
brightness temperature of the molecular cloud ensemble.  The parameter
$\rm K_{vir}$ is given by

\begin{equation}
\rm K_{vir}=\frac{\left(dV/dR\right)}{\left(dV/dR\right)_{virial}}\sim 
1.54\frac{[CO/H_2]}{\sqrt{\alpha}\Lambda _{co}}\left(\frac{n(H_2)}{10^3\,
 cm^{-3}}\right)^{-1/2},
\end{equation}

\noindent
(with  $\rm \Lambda_{co}=[CO/H_2]/(dV/dR$  being one  of the  outputs  of a
typical LVG  model), and  determines the average  dynamic state  of the
molecular  gas,  with  $\rm   K_{vir}$$\sim  $1--3  for  the  (mostly)
self-gravitating GMCs  in the  Galactic disk (and  $\rm K_{vir}$$\ll$1
corresponding to dynamically unattainable gas motions).  For a typical
value of $\alpha$=1.5 Equation 1 becomes

\begin{equation}
\rm X_{CO}=2.65 \frac{\sqrt{n(H_2)}}{T_{b,1-0}} K_{vir}^{-1}\,
 \left(\frac{M_{\odot}}{K\,km\,s^{-1}\,pc^2}\right),
\end{equation}

\noindent
which  we  use  in the  present  work.  For  ordinary GMCs  with  $\rm
n(H_2)$$\sim  $500\,cm$^{-3}$, $\rm  T_{b,1-0}$$\sim $10\,K,  and $\rm
K_{vir}$$\sim  $1-2 the latter  yields $\rm  X_{co}$$\sim $(3-6)\,$\rm
X_l$                (where                $\rm               X_l$=$\rm
M_{\odot}$(K\,km\,s$^{-1}$\,pc$^2$)$^{-1}$).

The optically thin limit for  CO J=1--0 line emission is often adopted
to provide a lower limit on the total molecular gas mass (e.g.  Bryant
\& Scoville 1996; Solomon et  al.  1997).  The availability of several
CO lines allows a more robust such limit without assuming LTE:

\begin{equation}
\rm \rm  X^{(thin)} _{co}=0.078
\left[1+\frac{1}{3}e^{5.5/T_{ex,10}}+\frac{5}{12}\,r_{21}+\frac{7}{27}\,r_{32} +...
\frac{2J+3}{3(J+1)^2}\,r_{J+1\,J}+...\right] \left(\frac{M_{\odot}}{K\,km\,s^{-1}\,pc^2}\right),
\end{equation}

\noindent
(see Appendix),  which makes obvious  that, unless the CO  line ratios
become significantly  larger than unity  (which they can  in optically
thin emission), most of the  contributions come from the low J levels.
The LTE expression is:

\begin{equation}
\rm  X^{(thin)} _{co}(LTE)=9.45\times10^{-3}\left[\left(\frac{T_k}{K}\right) e^{5.5/T_k}\right] 
 \left(\frac{M_{\odot}}{K\,km\,s^{-1}\,pc^2}\right),
\end{equation}

\noindent
which gives  a higher  (and less reliable)  lower limit since  the LTE
partition function  can be significantly larger than  the non-LTE one.
The optically  thin approximation can  be used also for  the realistic
case of significant  CO line optical depths as  long as large velocity
gradients keep the photon escape probability local throughout the bulk
of the molecular gas mass. Then Equation 4 is  modified to

\begin{eqnarray}
\rm \rm  X^{(\beta)} _{co} & = &  0.078 \beta^{-1} _{10}
\rm \left[1+\frac{1}{3}e^{5.5/T_{ex,10}}  +  \frac{5}{12}\left(\frac{\beta_{10}}{\beta_{21}}\right)\,r_{21}+
\frac{7}{27}\left(\frac{\beta_{10}}{\beta_{32}}\right)\,r_{32} +...\right. \\
 & + & \left. \rm ...\frac{2J+3}{3(J+1)^2}\left(\frac{\beta_{10}}{\beta_{J+1\,J}}\right)r_{J+1\,J}+...\right] \nonumber 
\rm \left(\frac{M_{\odot}}{K\,km\,s^{-1}\,pc^2}\right), 
\end{eqnarray}

\noindent
 where              $\rm             \beta             _{J+1\,J}$=$\rm
\left[1-exp(-\tau_{J+1\,J})\right]/\tau_{J+1\,J}$ is the photon escape
probability (see Appendix).  The  last equation provides a more robust
lower limit  than Equation 4, as  it accounts for  finite line optical
depths  (computed from  radiative  transfer models).   If enough  line
ratios are available, the estimate  from Equation 6 will approach that
of Equation 3 as long as  one average gas phase dominates the emergent
CO line emission.

In Figure 1  we show the $\rm X_{co}$  and $\rm X^{(thin)} _{co}$(LTE)
distributions, computed from Equations 3  and 5 and the results of one
radiative transfer models for the CO SLEDs and $^{13}$CO lines for all
the LIRGs in our sample  (see Paper\,I).  Most $\rm X_{co}$ values are
$\la   $1.5\,$\rm  X_l$,  with   $\rm  \langle   X_{co}\rangle  $$\sim
$0.60\,$\rm X_l$ obtained for the main distribution (excluding the few
outliers  beyond  3\,$\rm  X_l$).   The  $\rm  X^{(thin)}  _{co}$(LTE)
distribution gives similar results, indicating that most of the states
compatible  with  the  CO  SLEDs  have optically  thin  or  moderately
optically thick  CO J=1--0 line  ($\rm \tau _{10}$$\sim  $1--3).  This
has been noted in the past in the context of a two-phase ISM model and
attributed to the high  temperatures of a turbulent ``envelope'' phase
in molecular clouds (Aalto et al.  1995).


 The $\rm X_{co}$ range obtained  using our one-phase models of global
CO SLEDs  is similar to  that reported by  DS98 and Yao et  al.  2003.
Moreover in Figure 2 the distribution of $\rm \sqrt{n(H_2)}/T_{b,1-0}$
demonstrates  that unlike  often stated,  the effects  of  density and
temperature do not  always cancel out but are  responsible for much of
the $\rm  X_{co}$ variations over different  ISM environments (compare
Figure  2,   with  the  $\rm   X_{co}$  distribution  in   Figure  1).
Nevertheless  the  average dynamical  state  of  the  gas (i.e.   $\rm
K_{vir}$) remains the most important factor affecting the average $\rm
X_{co}$.   This  is   shown  in  Figure  3  where   the  $\rm  X_{co}$
distributions for virial/near-virial and unbound average gas dynamical
states are  shown.  In the  merger-driven starbursts of  ULIRGs highly
non-virial  gas velocities  can be  caused by  significant  amounts of
stellar mass  concomitant with a continous  molecular gas distribution
as noted by DS98 (see also Solomon et al.  1997), albeit without using
radiative transfer  models to  constrain $\rm K_{vir}$.   Strong tidal
fields  acting on  GMC  cloud envelopes  and/or  a diffuse  intercloud
molecular  gas distribution  whose velcity  field traces  the combined
potential of the surviving dense GMC  cores as well as stars will both
raise the average $\rm  K_{vir}$, and further reduce the corresponding
$\rm X_{co}$ (Equation~3).

  In  Figure 4 the  distribution of  $\rm X^{(\beta)}  _{co}$ computed
  from Equation 6 is shown which, unlike that deduced from Equation 3,
  uses the  CO line ratios  explicitely.  This makes it  more suitable
  when fully-sampled CO SLEDs from  J=1--0 up to high-J (e.g.  J=6--5,
  7--6)  are   available,  and  when  the  turbulent   models  of  the
  hierarchical density structures  of molecular clouds (e.g. Ossenkopf
  2002) become refined enough to provide global CO and $^{13}$CO SLEDs
  as well as the corresponding $\rm \beta _{J+1,J}$ values per density
  ``sub-phase''.    For   well-sampled   $^{13}$CO  SLEDs   the   $\rm
  X^{(\beta)}   _{co}$   expression   (using  the   appropriate   $\rm
  [^{13}CO/H_2]$  abundance)  is  even   more  useful  as  $\rm  \beta
  _{J+1,J}$($^{13}$CO)$\sim $1.  In our current study using the output
  $\rm \beta_{J+1\,J}$ values from  our one-phase models in Equation 6
  yields an $\rm  X^{(\beta) } _{co}$ distribution similar  to that of
  $\rm X_{co}$ from Equation 3.

 Interestingly  the  $\rm  X_{co}$ distributions,  while  concentrated
 within $\sim $(0.3--1.5)\,$\rm X_l$, do extend out to Galactic values
 $\sim $(2.5-6)\,$\rm X_l$.  These  are found for disk-dominated LIRGs
 (e.g.  NGC\,157  and Mrk\,1048), (U)LIRGs  with unexpectedly ``cold''
 CO ratios (e.g.  IRAS\,05189-2524) where cool ($\sim $15\,K) gas with
 low/modest  densities ($\sim  $(10$^2$--10$^3$)\,cm$^{-3}$) dominates
 their average ISM state.  Surprisingly, {\it Galactic $X_{co}$ values
 can  be found also  in some  ULIRGs,} whose  ``hot'' CO  ratios (e.g.
 IRAS\,08572+3915) indicate a dominant warm ($\sim $(100--150)\,K) and
 dense ($\sim $(3$\times  $10$^4$--10$^6$)\,cm$^{-3}$) phase, or other
 evidence (e.g.  multi-J HCN lines) suggest a massive dense, gas phase
 (e.g.  Mrk\,231).   Thus large $\rm  X_{co}$ values are  possible for
 both high  and low excitation gas  phases and thus cannot  be used as
 indicators  of Galaxy-type  ISM  conditions in  distant star  forming
 disks (e.g.   Daddi et  al.~2010b). This also  suggests that  in many
 local ULIRGs  and similar systems  at high redshifts {\it  the widely
 adopted  $ X_{co}$$\sim  $(0.8-1)\,$ X_l$  values  understimate their
 molecular gas mass.}

\subsection{Eddington-limited star formation and a  minimum  molecular gas mass}

 Recent studies of SF feedback suggest a maximum $\epsilon_{g,*}$=$\rm
 L^{(*)}_{IR}/M_{*}(H_2)$$\sim  $500\,$\rm  (L_{\odot}/M_{\odot})$ for
 the dense and warm gas $\rm  M_{*}(H_2)$ near SF sites in galaxies as
 a  result of strong  radiation pressure  from the  nascent O,  B star
 clusters onto the concomitant dust  of the accreted gas fueling these
 sites (Scoville  2004; Thompson et al.  2005;  Thompson 2009).  Thus,
 provided that average dust  properties (e.g.  its effective radiative
 absorption coefficient  per unit  mass) remain similar  in metal-rich
 star-forming systems such  as LIRGs, a near-constant $\epsilon_{g,*}$
 is expected among them.  A value of $\epsilon _{g,*}$$\sim $500\,$\rm
 (L_{\odot}/M_{\odot})$ is actually measured in individual SF sites of
 spiral disks  such as  M\,51 and entire  starbursts such  as Arp\,220
 (Scoville     2003),     while     $\sim    $($440\pm     100$)\,$\rm
 (L_{\odot}/M_{\odot})$ is  obtained for CS-bright  star-forming cores
 in  the  Galaxy (Shirley  et  al.   2003).   Further evidence  for  a
 near-constant   $\rm  \epsilon_{g,*}$  is   the  tight   linear  $\rm
 L_{IR}$-HCN(1-0) correlation  found for individual GMCs  up to entire
 ULIRGs (Wu  et al.  2005)  with HCN J=1--0  used as a dense  gas mass
 tracer.  However the intermitency expected for galaxy-sized molecular
 gas reservoirs (i.e.  at any given epoch of a galaxy's evolution some
 dense gas regions  will be forming stars while  others will not) will
 lower  the global  $\rm \epsilon  _{g,*}$  to $\sim  $1/3-1/2 of  the
 Eddington value (Andrews \& Thompson~2011).

  Similar $\rm \epsilon_{g,*}$ values can be obtained without explicit
  use  of the  Eddington limit  (and the  detailed dust  properties it
  entails),  but  from  the  typical $\rm  L_{*}/M_{new,*}$  in  young
  starbursts where  $\rm M_{new,*}$ is the  mass of the  new stars and
  $\rm L_{*}$  their bolometric luminosity  ($\sim $$\rm L^{(*)}_{IR}$
  for  the  deeply  dust-enshrouded  SF  sites).   For  $\rm  \epsilon
  _{SF,c}$=$\rm M_{new,*}/[M_{new,*}+M_{*}(H_2)]$ as the SF efficiency
  of the dense gas regions where the new stars form it is

\begin{equation}
\rm          \epsilon_{g,*}=\frac{\epsilon         _{SF,c}}{1-\epsilon
_{SF,c}}\left(\frac{L^{(*)}_{IR}}{M_{new,*}}\right).
\end{equation}

\noindent 
For  $\rm  \epsilon  _{SF,c}$$\sim  $0.3--0.5  typical  for  dense  SF
regions,    and    $\rm   L^{(*)}    _{IR}/M_{new,*}$=(300--400)\,$\rm
(M_{\odot}/L_{\odot})$   (Downes  \&   Solomon  1998   and  references
therein),    Equation    7   yields    $\rm    \epsilon_{g,*}$$\sim
$(130--400)\,$\rm  (M_{\odot}/L_{\odot})$.   
Here we choose $\rm \epsilon_{g, *}$=250\,$\rm (L_{\odot}/M_{\odot})$,
close to the  average values from Equation 7 and  the black body limit
deduced  for compact CO  line emission  concomitant with  an optically
thick  ($\rm \tau _{100\,\mu  m}$$>$1) dust  emission (Solomon  et al.
1997).   Eddington-limited  star formation  in  LIRGs  implies {\it  a
minimum  molecular   gas  mass  $M_{SF}$=$L^{(*)}_{IR}/\epsilon_{g,*}$
fueling their observed star formation rates.}  In Figure 5 we show the
$M_{SF}$ distribution which, for the  ULIRGs in our sample, reaches up
to  $\sim $5$\times  $10$^{9}$\,$\rm M_{\odot}$  i.e.   surpassing the
total molecular gas reservoirs of typical spirals.

\subsection{Computing $\rm \bf X_{co}$ in a two-component approximation}

In some cases  where one-phase LVG models of global  CO line ratios of
  LIRGs  do   not  converge  to  a  well-defined   range  of  physical
  conditions,  a simple  two-phase model  can be  used to  examine the
  underlying  ISM  conditions   and  the  corresponding  $\rm  X_{co}$
  factors.  The total CO J=1--0 luminosity can then be expressed as

\begin{equation}
\rm L^{'} _{CO, 1-0} = L^{(h)\,'} _{CO, 1-0}+L^{(l)\,'} _{CO,1-0}
= (\epsilon _{g,*} X^{(h)} _{co})^{-1}L^{(*)} _{IR} + L^{(l)\,'} _{CO,1-0},
\end{equation}

\noindent
where (h)  and (l)  indicate the high  and the low  excitation phases,
 $\rm X^{(h)} _{co}$ is the CO-H$_2$ conversion factor for the former,
 and  $\epsilon _{g,*}$=250($\rm L_{\odot}/M_{\odot}$).   This assumes
 that   the  high  excitation   phase  fuels   Eddington-limited  star
 formation.  For any other~transition

\begin{equation}
\rm L^{'}_{CO, J+1-J} = \left[\frac{L^{(*)} _{IR}}{\epsilon _{g,*} X^{(h)} _{co}}\right] r^{(h)} _{J+1\,J} 
+ L^{(l)\,'} _{CO,J+1-J},
\end{equation}





 The molecular gas near H\,II regions and interfaces between molecular
 clouds  and  O,B  stellar  associations  is  an  obvious  choice  for
 obtaining  template high-excitation  CO SLEDs.   The  beam-matched CO
 line survey of the Orion A and B GMCs found $\rm r_{21}$=1.2--1.3 and
 $\rm R_{21}$=10 (the CO/$^{13}$CO J=2--1 ratio) in such ``hot'' spots
 (Sakamoto et  al.  1994). These ratios along  with $\rm T_{kin}$$\geq
 $100\,K (expected  from the high  CO line brightness  temperatures in
 such SF  spots) set  as constraints to  a one-phase LVG  model yield:
 $\rm T_{kin}$=(125--150)\,K, $\rm n(H_2)$=3$\times$10$^5$\,cm$^{-3}$,
 $\rm    K_{vir}$=7,    and    corresponding   line    ratios:    $\rm
 r^{(h)}_{21}$(3)=1.35,       $\rm       r^{(h)}_{32}$=1.33,      $\rm
 r^{(h)}_{43}$=1.30, $\rm r^{(h)}_{54}$=1.27, $\rm r^{(h)}_{65}$=1.25,
 and $\rm  r^{(h)}_{76}$=1.22.  For this  LVG solution set  we compute
 (Equation  3) $\rm  X^{(h)} _{co}$  $\sim $2.2\,$\rm  X_l$,  and from
 Equations  8  and  9  the   CO  line  ratios  of  the  low-excitation
 (l)-phase.  These can  then be  used  as constraints  on 1-phase  LVG
 models to obtain the average  ISM conditions and $\rm X^{(l)} _{co}$.
 For such a 2-phase decomposition the effective $\rm X^{(2-ph)} _{co}$
 factor would be

\begin{equation}
\rm \rm X^{(2-ph)} _{co} = \frac{X^{(h)} _{co}+\rho^{(l-h)} _{co}X^{(l)} _{co}}{1+\rho^{(l-h)} _{co}}
\end{equation}

\noindent
with  $\rm \rho  ^{(l-h)} _{co}$=$\rm  L^{(l)\,'} _{co,1-0}/L^{(h)\,'}
_{co,1-0}$. Using the Eddington-limit normalization for $\rm L^{(h)\,'}
_{co,1-0}$ yields

\begin{equation}
\rm X^{(2-ph)} _{co}=X^{(l)} _{co}+ \frac{L^{(*)} _{IR}}{\epsilon _{g,*} L^{'} _{CO,1-0}}
\left(1-\frac{X^{(l)} _{co}}{X^{(h)} _{co}}\right),
\end{equation}

\noindent
When lines such as higher-J CO transitions (J=4--3, 6--5) and/or heavy
rotor molecular lines (e.g.  HCN)  are available, they are used to set
constraints  on the  (h)-phase, $\rm  \rho ^{(l-h)}  _{co}$,  and $\rm
X^{(2-ph)}  _{co}$ without  assuming the  (h)-phase CO  SLED  and $\rm
X^{(h)} _{co}$ deduced from the Orion GMC star-forming ``spots''.

\section{The X$_{\bf co}$  factor in LIRGs}

The average  ISM conditions deduced  using the CO lines  available per
LIRG      (see      Paper       I)      encompasses      the      $\rm
\left[n(H_2),T_{kin},K_{vir}\right]$ parameter space where most of the
mass is  expected to  reside in ordinary  GMCs.  Thus the  narrow $\rm
X_{co}$  distribution  around   $\rm  \langle  X_{co}  \rangle  $$\sim
$0.6\,$\rm  X_l$ (Figs  1,  4), does  not  seem as  the  result of  an
excitation bias induced by  the particular molecular lines used.  This
is not to say that other  high-J CO or e.g.  HCN line luminosities are
expected to be compatible with the average ISM states deduced from our
current CO  line dataset  but rather than  that such lines  with their
higher critical  densities will probe much smaller  gas mass fractions
per typical GMC  to be of any consequence when it  comes to the global
$\rm X_{co}$.

It is tempting to consider the aforementioned picture as complete.  It
is  certainly  compatible  with  models of  supersonic  turbulence  in
ordinary  GMCs where  most molecular  gas mass  is found  at densities
$<$10$^{4}$\,cm$^{-3}$, its conditions  thus ``accessible'' to the CO,
$^{13}$CO  lines used  to constrain  them  (and $\rm  X_{co}$) in  our
study.  In the merger-driven starbursts  of ULIRGs however GMCs may be
far from ordinary.  Stripping of their outer envelopes by strong tidal
fields  and/or  the large  pressures  from  a  hot ionized  gas  phase
resulting  from   HI  cloud   collisions,  both  expected   in  merger
environments (e.g.  Solomon et  al.  1997), can dramatically alter the
M(H$_2$)-n(H$_2$)  distribution  towards   most  mass  being  at  $\rm
n(H_2)$$\geq $10$^4$\,cm$^{-3}$.  This actually occurs in the Galactic
Center (e.g. G\"usten  \& Philipp 1994), and in  ULIRGs it can involve
their  entire  molecular gas  reservoirs.   In  such  cases low-J  CO,
$^{13}$CO lines are no longer  sensitive to the average ISM conditions
and  may thus  yield  inaccurate ``corrections''  to  the global  $\rm
X_{co}$ factor with respect to its Galactically calibrated value.


\subsection{The $\rm \bf X_{co}$ in ULIRGs:  theoretical expectations 
for highly turbulent gas }
   
  The  high pressures  in the  highly  turbulent gas  disks in  ULIRGs
  (DS98) will: a)  result to much larger average  gas densities at all
  scales, and  b) ``relocate'' large  mass fractions of their  GMCs to
  densities $\rm  n(H)$$>$10$^4$\,cm$^{-3}$. The latter  can be easily
  shown  from  the  probability  distribution function  (pdf)  of  the
  density    in   supersonically    turbulent    clouds.    This    is
  well-approximated by a log-normal distribution with a dispersion:

\begin{equation}
\rm \sigma _{\rho} \approx \left[ln\left(1+\frac{3M^2}{4}\right)\right]^{1/2},
\end{equation}

\noindent
where M=$\rm \sigma_v/c_s$ is the average 1-dim Mach number (Padoan \&
 Nordlund 2002).  The mass fraction contained in cloud structures with
 overdensities x$\geq$$\rm x_{\circ}$ (x=$\rm n/\langle n \rangle$) is
 then given by

\begin{equation}
\rm f=\frac{M(x\geq x_{\circ})}{M_{tot}}=
\frac{1}{2}\left[1+erf\left(\frac{-2\ln(x_{\circ})+\sigma^2 _{\rho}}{2^{3/2} \sigma_{\rho}}\right)\right].
\end{equation}

The    high    velocity    dispersions    $\rm    \sigma    _{v}$$\sim
$(30--140)\,km\,s$^{-1}$  measured in  the molecular  disks  of ULIRGs
(DS98; see also  Swinbank et al.  2011 for  a recently discovered such
disk   at  z$\sim   $2.3)  versus   those  in   spiral   disks  ($\sim
$(5--10)\,km\,s$^{-1}$)  correspond to  M(ULIRGs)$\sim $(3--30)$\times
$M(spirals).  For M(spiral)=10 then $\rm \sigma _{\rho}$(spirals)$\sim
$2,   while  $\rm   \sigma  _{\rho}$(ULIRGs)$\sim   $2.55-3.33,  which
dramatically extends  the gas density  pdf expected in  ULIRGs towards
high values where  much of the gas mass will now  lie.  Indeed for the
GMCs in  spiral disks  only $\sim $3\%  of the  mass will be  found at
overdensities of  x$>$$\rm x_{\circ}$=500 (=5$\times$10$^4$\,cm$^{-3}$
for typical GMC with  $\rm \langle n\rangle$=100\,cm$^{-3}$) while for
ULIRGs this  is $\sim $45\% (Figure  6).  This difference  can be even
larger since the {\it average} molecular gas density in disks is lower
($\rm \langle  n \rangle $$\sim $(100-500)\,cm$^{-3}$)  than in ULIRGs
where it can reach up to at least $\sim $10$^{4}$\,cm$^{-3}$ (Greve et
al.    2009).   Thus  gas   at  n$\geq   $5$\times  $$10^4$\,cm$^{-3}$
corresponds to  an overdensity of  $\rm x_{\circ}$=100-500 (containing
$\la  $10\%  of  gas  mass)   for  GMCs  in  spirals,  and  only  $\rm
x_{\circ}$=5 (and 90\%  of the mass) for GMCs  in ULIRGs.  The typical
CO SLED available  for LIRGs in our sample  (J=1--0, 2--1, 3--2) would
then be  insensitive to the state  of the molecular gas  in ULIRGs and
thus unable to constrain  the corresponding $\rm X_{co}$ factor, while
it would  remain adequate  for such  a task for  the molecular  gas in
isolated spirals.

 Interestingly the  few cases of near-Galactic $\rm  X_{co}$ in ULIRGs
were  found when either  higher-J CO  lines were  available or  when a
dense and  warm gas  phase was  massive enough to  be obvious  even in
low-J CO  SLEDs (Paper I, and  section 4).  The $\rm  X_{co}$ for such
dense  gas  can be  high,  approaching  and  even surpassing  Galactic
values.  Indicatively, for  n=5$\times $10$^4$\,cm$^{-3}$ and warm gas
with  a   thermalized  CO  J=1--0  line   $\rm  T_{b,1-0}$$\sim  $$\rm
T_{kin}$$\sim  $(100-150)\,K,  it  is:  $\rm  X_{co}$$\sim  $(4-6)$\rm
K^{-1}  _{vir}$\,$\rm X_l$  (Eq.  3), which  for self-gravitating  gas
($\rm K_{vir}$=1) corresponds to Galactic~values.

Moreover, for the highly  turbulent, high-pressure ISM environments in
ULIRGs the  notion of molecular gas reservoirs  reducible to ensembles
of  dicrete GMCs may  no longer  apply, with  much of  the gas  in the
pre-merger   GMCs   redistributed   in   continous  disks   of   $\sim
$(100-300)\,pc  diameter  (DS98;  Sakamoto  et al.   2008).   A  warm,
low-density, molecular gas phase, with large $\rm K_{vir}$ ($>$10) can
then  be generated,  {\it  alongside}  the denser  one,  by the  tidal
disruption  of  GMC ``envelopes''  in  merger  environments, the  high
pressures turning  intercloud CNM  HI into H$_2$,  and the  effects of
supersonic turbulence  driven at  the largest scales  (e.g.  Ossenkopf
2002).   Such a  diffuse phase,  even if  containing little  mass, can
easily dominate the  emergent global CO J=1--0, 2--1  line emission of
ULIRGs, yielding  ``cold'' CO(2-1)/(1-0)  ratios and low  $\rm X_{co}$
factors (a result  of its low densities, high  temperatures, and large
$\rm  K_{vir}$ values).   These low  $\rm X_{co}$  values,  {\it while
appropriate for  the diffuse gas mass, may  not be so for  the bulk of
the molecular gas that now resides at much higher densities.}

 Thus in  the very  turbulent molecular gas  reservoirs of  ULIRGs a
combination of: a) most of their mass residing at high densities (with
potentialy  large $\rm  X_{co}$ factors),  and b)  the existence  of a
diffuse,  warm, and  unbound  phase  with little  mass  (and low  $\rm
X_{co}$)  dominating low-J  CO  SLEDs (and  the  average $\rm  X_{co}$
determined  from  them),  can  easily  result  to  {\it  a  systematic
underestimate  of  their  total  molecular  gas  mass.}   All  current
observational  studies likely suffer  from this  bias, expected  to be
most prominent  in ULIRGs with  high molecular gas  surface densities.
This  includes the  present  study as  mostly  low-J CO  SLEDs (up  to
J=3--2) are typically available per LIRG.  The only exceptions are the
few ULIRGs for which, a highly excited CO J=3--2 line, CO J=4--3, 6--5
lines, and/or HCN  lines from the literature allowed  the revealing of
their massive dense  gas reservoirs (see Section 4).   Studies of $\rm
X_{co}$ in  similar systems  at high redshifts  (e.g.  Tacconi  et al.
2008) will  be  similarly affected.  Only  molecular  SLEDs with  high
critical densities  can overcome such  an $\rm X_{co}$ bias  in ULIRGs
and see  how many  actually lie on  the high  end of the  $\rm X_{co}$
distributions shown in Figures 1 and~4.


\subsubsection{Numerical simulations and the $\rm X_{co}$ factor: important caveats, and  ways forward}
 
Recently  GMC-sized numerical  simulations that  include H$_2$  and CO
formation/destruction and  radiative transfer in  MHD turbulent models
where used to  explore $\rm X_{co}$ and its  dependance on average ISM
conditions (Shetty  et al.  2011a,b).  These  simulations also include
chemistry and  thermodynamical effects but not  SF-driven heating (via
photons or CRs)  or turbulent heating which could  dominate in ULIRGs.
Currently such  studies explore Galactic-type GMCs  whose $\rm X_{co}$
values  they  found  similar  to  the one  observed.   Extending  such
numerical  models  to  ISM   conditions  expected  in  ULIRGs  is  not
straightforward as  GMC boundary conditions such  as ambient radiation
fields, surface  pressure, external gravitational field  and its tidal
terms (important for  galactic centers of spirals and  ULIRGs) are set
constant in GMC-sized simulations or, in the case of tidal fields, are
omitted altogether.  Nevertheless in  any realistic setting of GMCs in
galaxies these  boundary conditions  do change, sometimes  faster than
internal cloud evolutionary timescales, and especially so in ULIRGs.

Galaxy-sized numerical  studies of $\rm  X_{co}$ in disks  and mergers
recently   presented   by  Narayanan   et   al.    2011  address   the
aforementioned  important  point of  coupled  GMC-galaxy evolution  by
tracking  the evolution  of  GMC  boundary conditions,  but  do so  by
necessarily  adopting some  subresolution methods  to  follow in-cloud
physics and keep the  problem computationally tractable.  They recover
Galactic $\rm  X_{co}$ values for  disks, but $\sim $5-10  times lower
ones  in   mergers,  with   a  large  spread   that  makes   a  single
ULIRG-appropriate $\rm X_{co}$  factor impractical.  Nevertheless, for
ULIRGs in  particular, these  simulations do not  have (and  could not
have)   the   resolution   necessary   to  track   gas   with   n$\geq
$10$^4$\,cm$^{-3}$ (where most of  their gas mass resides).  Resolving
the  all-important  kinematic state  of  such  a  dense gas  component
(self-gravitating or not?)  within their compact gas reservoirs ($\sim
$(100-300)\,pc) is currently impossible with galaxy-sized simulations,
as is the explicit tracking of the large turbulence-regulated range of
densities expected  in ULIRGs.  There the {\it  average} densities can
surpass $\sim $10$^4$\,cm$^{-3}$, while  the high levels of turbulence
will  maintain $\ga  $60\% of  the mass  at  n$\geq $10$^5$\,cm$^{-3}$
(Figure  6).    With  the  $\rm  T_{kin}$  values   computed  in  such
simulations  likely  to  be  lower  limits (as  CR  and  mechanical
SNR-driven heating  are omitted), gas densities  and the corresponding
kinematic states remain the only parameters affecting the average $\rm
X_{co}$  in metal-rich environments  that can  push its  values either
way.  Indeed while additional heating can only raise $\rm T_{kin}$ and
thus  the CO  J=1--0 brightness  temperature (reducing  $\rm X_{co}$),
high  densities  ($>$10$^4$\,cm$^{-3}$)  and  self-gravitating  states
($\rm K_{vir}$$\sim $1) will act to raise $\rm X_{co}$.

GMC-sized simulations of the  much more turbulent and denser molecular
gas  in ULIRGs  are  necessary for  computing  the corresponding  $\rm
X_{co}$ but  these must now use  new initial conditions  with: a) much
higher  volumed-averaged   gas  densities  (current   ones  use  $\sim
$150\,cm$^{-3}$),  b) higher  velocity dispersions  (current  ones use
$\rm \sigma _v$$\sim $2.4\,km\,s$^{-1}$),  and c) higher background CR
energy  densities  (and thus  non-negligible  CR  gas heating).   Such
simulations  and their emergent  line intensities  can be  compared to
much richer  molecular SLEDs  available now than  in the  past.  These
include the  typical CO lines from  J=1--0 up to  J=4--3 observed from
the ground, rising up to J=13--12 using the Herschel Space Observatory
(van  der  Werf  et   al.   2010),  further  complemented  by  multi-J
observations  of heavy  rotor molecules  like HCN,  HCO$^+$,  CS (e.g.
Greve  et al.  2009)  as sensitivities  vastly improve  in the  era of
ALMA.   Thus, rather  than simply  trust the  $\rm X_{co}$  yielded by
GMC-sized  simulations as  is currently  done, these  simulations will
yield $\rm X_{co}$  values as a by-product of  the much more difficult
task of reproducing relative  strengths of molecular line luminosities
with critical densities ranging from $\sim $100\,cm$^{-3}$ up to $\sim
$10$^7$\,cm$^{-3}$.  These encompass  the entire range of hierarchical
structures  in turbulent  GMCs,  from a  highly turbulent  low-density
possibly non  self-gravitating ``GMC-envelope'' phase  at large scales
up  to  self-gravitating  compact  dense  gas  cores  with  dissipated
supersonic  turbulence (e.g.   Ossenkopf  2002).  GMC-sized  numerical
simulations  exploring  a  dense   grid  of  ULIRG-type  GMC  boundary
conditions  and determining  the corresponding  $\rm X_{co}$  grid can
then be  used to inform galaxy-sized numerical  simulations where such
boundary  conditions  are tracked,  while  resolving the  hierarchical
structures of individual GMCs (and thus determining $\rm f(x_{\circ})$
from the simulations) remains out of reach.

\subsection{High $\rm \bf X_{co}$ in ULIRGs:  past observational evidence ignored?}

  As discussed in the previous sections, by relying on low-J CO SLEDs,
  most observational studies of $\rm  X_{co}$ may have imparted a bias
  towards  low values  in ULIRGs.   Nevertheless past  observations of
  heavy rotor molecules such as HCN J=1--0 have given early hints of a
  large  scale   M(H$_2$)-n(H$_2$)  redistribution  in   these  merger
  systems, which  placed large fractions  ($\ga $50\%) of  their total
  molecular    gas   mass    at   much    higher    densities   ($\geq
  $10$^4$\,cm$^{-3}$)  than in  isolated gas-rich  disk  systems (e.g.
  Solomon  et al.   1992;  Gao  \& Solomon  2004).   However a  single
  transition of a  heavy rotor molecule cannot set  constraints on the
  gas  temperature and  density (let  alone on  its  average dynamical
  state) useful  enough to  constrain the corresponding  $\rm X_{mol}$
  factor well.  Thus these early  estimates of the dense molecular gas
  mass in ULIRGs  are necessarily very uncertain as  they relied on an
  $\rm X_{HCN}$  factor derived for  n$\sim $$\rm n_{crit}$[HCN(1-0)],
  self-gravitating gas, and HCN J=1--0 brightness temperatures typical
  of the  IR color temperatures of  SF galaxies (e.g.   Gao \& Solomon
  2004).   Indicatively for $\rm  n(H_2)$=5$\times $10$^4$\,cm$^{-3}$,
  $\rm  T_{b,1-0}$(HCN)=40\,K and $\rm  K_{vir}$=1 Equation  3 applied
  for HCN  J=1--0 yields $\rm  X_{HCN}$=15\,$\rm X_{l}$, which  can be
  higher still for higher gas  densities, as indicated by multi-J HCN,
  CS and HCO$^+$ line observations of Arp\,220 and NGC\,6240 and where
  $\rm X_{HCN}$$\sim $(20-35)\,$\rm X_l$ (Greve et al. 2009).

  For such  high $\rm X_{HCN}$ values and  $\rm L^{'} _{HCN,1-0}/L^{'}
  _{co,1-0}$=1/4-1/6 observed  in ULIRGs  (e.g. Solomon et  al.  1992;
  Gao \& Solomon 2004; Gracia-Burillo et  al. 2012) it is easy to show
  that $\rm  X^{(2-ph)} _{co}$ (Equation 10)  will yield near-Galactic
  or higher  values in such  systems.  Indeed, assuming that  the only
  contribution  to  the  (h)-phase  in  Equation  10  comes  from  the
  HCN-bright gas  (i.e.  the HCN emission from  the low-excitation (l)
  phase is neglible) we can determine $\rm \rho^{(l-h)} _{co}$ from

\begin{equation}
\rm \rho^{(l-h)} _{co}=\frac{L^{'} _{co,1-0}}{L^{(h)'}_{co,1-0}}-1
=\frac{r^{(obs)} _{CO/HCN}}{r^{(h)} _{CO/HCN}}-1,
\end{equation}
  
\noindent
where $\rm r^{(obs)} _{CO/HCN}$=$\rm L^{'} _{co,1-0}/L^{'} _{HCN,1-0}$
in ULIRGs(=4-6),  and $\rm r^{(h)}  _{CO/HCN}$ that of  the (h)-phase.
The latter can be found using a radiative transfer model for densities
typical  of HCN-bright gas,  an assumed  gas temperature,  and setting
$\rm   K^{(h)}   _{vir}$=1.    For   n=10$^5$\,cm$^{-3}$,   and   $\rm
T_{kin}$=100\,K using  an LVG  model for HCN  and CO lines  we compute
$\rm T_{b,HCN  1-0}$=41\,K and $\rm T^{(h)}  _{b,co 1-0}$=96\,K.  Thus
$\rm r^{(h)} _{CO/HCN}$=2.34,  while $\rm X_{HCN}$$\sim $20\,$\rm X_l$
and $\rm  X^{(h)} _{co}$$\sim $8.6\,$\rm X_l$,  with $\rm \rho^{(l-h)}
_{co}$$\sim  $0.71-1.56.  Setting  $\rm X^{(l)}  _{co}$=0.5\,$\rm X_l$
(Figs.  1,  4) for the low-excitation (CO-bright  but HCN-dark) phase,
we  obtain   $\rm  X^{(2-ph)}  _{co}$$\sim   $(4-5)\,$\rm  X_l$,  i.e.
Galactic values. These  are now due to much  higher fractions of dense
HCN-bright gas than in the  Galaxy, with large $\rm X_{HCN}$ (and $\rm
X^{(h)} _{co}$).  If  such $\rm X^{(2-ph)} _{co}$ values  are the norm
in  the  high-pressured turbulent  gas  disks  of  ULIRGs, {\it  their
  dynamical  masses would be  dominated by  the molecular  gas}, quite
unlike isolated  SF~spirals.  Indeed  for all the  ULIRGs in  the DS98
study whose  dynamical mass is reasonably  well-constrained (with both
CO 1-0 and 2-1 imaging)  Galactic $\rm X_{co}$ values would yield $\rm
M_{gas}/M_{dyn}$$\sim   $0.6-1,  and   compatible  with   the  obvious
limitation of  $\rm M_{gas}/M_{dyn}$$\leq $1  within the observational
uncertainties of  $\rm M_{dyn}$ estimates. Regarding the  latter it is
is  worth pointing  out that  low S/N  for extended  CO  line emission
regions  in  current  interferometric  maps  as  well  as  dynamically
unrelaxed,  non-circular, molecular  gas motions  (expected  in strong
mergers)   typically   cause    {\it   underestimates}   rather   than
overestimates of  the actual $\rm M_{dyn}$.  Surprisingly  such a bias
was  deduced  even for  gas-rich  but  otherwise  isolated disks  with
numerical  simulations   showing  an  observationally-determined  $\rm
M_{dyn}$ to  be typically  be $\sim $30\%  less than the  actual value
(Daddi et  al. 2010b).  Needless to  say this state  of affairs becomes
worse still of CO-imaged systems at high redshifts.

Multi-J observations  of heavy rotor molecules and/or  high-J CO lines
in  ULIRGs can  produce  dense  gas mass  estimates  via a  (radiative
transfer model)-constrained $\rm X_{mol}$ factor. However the few such
studies available (Papadopoulos 2007; Papadopoulos et al.  2007; Krips
et al.   2008; Gracia-Carpio et  al.  2008; Greve  et al. 2009)  are a
testimony of how  hard it is to overcome  the degeneracies inherent in
radiative transfer models of  optically thick line emission from heavy
rotor molecules.   The $\rm n$-$\rm T_{kin}$  solution degeneracies of
the Large Velocity Gradient  (LVG) radiative transfer models typically
used  in such  studies  translate to  a  wide range  of $\rm  X_{HCN}$
factors, ranging  from $\sim $10\,$\rm  X_l$ and reaching up  to $\sim
$(50-90)\,$\rm   X_l$   (Krips   et   al.   2008;   Gracia-Carpio   et
al. 2008). Tellingly  even the two studies with  the largest number of
dense  gas  tracer  lines   available  per  ULIRG,  that  of  Mrk\,231
(Papadopoulos et al.   2007) and of Arp\,220, NGC\,6240  (Greve et al.
2009) cannot overcome such degeneracies,  and yield LVG solutions that
produce   $\rm    X_{HCN}$$\sim   $(10-20)\,$\rm   X_l$    and   $\sim
$(17-37)\,$\rm  X_l$ respectively.   Nevertheless  most $\rm  X_{HCN}$
values  deduced in  such studies  are  high enough  to yield  dominant
amounts  of dense gas  mass in  ULIRGs, with  a near-Galactic  or even
larger than Galactic $\rm X^{(2-ph)} _{co}$.


 This apparent contradiction of  the low $\rm X_{co}$ factor advocated
for ULIRGs  by past low-J CO  line studies (e.g. Solomon  et al. 1997;
DS98;  Yao et  al.  2003),  and the  larger effective  $\rm X^{(2-ph)}
_{co}$ implied by heavy rotor  molecular line emission (mostly HCN) in
such galaxies has not been much  noted in the literature.  In the rare
cases where  this discrepancy was  discussed, the argument went  for a
downward revision of  $\rm X_{HCN}$ by $\sim $1/5  rather than a boost
of the  widely adopted ULIRG-value of $\rm  X_{co}$$\sim $1\,$\rm X_l$
(e.g.  Papadopoulos et al.  2007).   It must be noted however that the
possibility  of HCN-based  molecular  gas mass  estimates raising  the
effective  $\rm  X_{co}$  in   ULIRGs  is  mentioned  in  the  seminal
interferometric work by DS98.

\subsubsection{What could lower $\rm X_{co}$  in ULIRGs}

Lower $\rm X_{HCN}$ values than those reported in existing studies are
possible if  higher gas temperatures and/or unbound  gas motions ($\rm
K_{vir}$$>$1) prevail for the bulk of the dense gas in ULIRGs.  In the
most  detailed  such study  Greve  et al.   (2009)  do  find warm  and
non-virial  LVG  solutions for  the  HCN-bright  gas  in Arp\,220  and
NGC\,6240    with   $\rm    n(H_2)$=3$\times$$10^5$\,cm$^{-3}$,   $\rm
T_{kin}$=(45-120)\,K   (Arp\,220)    and   $\rm   T_{kin}$=(60-120)\,K
(NGC\,6240) while $\rm K_{vir}$$\sim  $2, 5 for Arp\,220 and NGC\,6240
respectively.   Choosing $\rm T_{kin}$=100\,K  as indicative  for both
systems (for  a given density  and $\rm K_{vir}$ high  $\rm T_{kin}$'s
yield    small    $\rm   X_{mol}$'s),    we    find   $\rm    T_{b,HCN
1-0}$(Arp\,220)$\sim $60\,K,  $\rm T^{(h)} _{b,co 1-0}$(Arp\,220)$\sim
$96\,K  and $\rm T_{b,HCN  1-0}$(NGC\,6240)$\sim $40\,K,  $\rm T^{(h)}
_{b,co  1-0}$(NGC\,6240)$\sim  $86\,K  (for their  corresponding  $\rm
K_{vir}$ values).  Then from  Equation 3: $\rm X_{HCN}$$\sim $12\,$\rm
X_l$, $\rm  X^{(h)} _{co}$$\sim $7.6\,$\rm X_l$ for  Arp\,220 and $\rm
X_{HCN}$$\sim  $7.3\,$\rm X_l$,  $\rm  X^{(h)} _{co}$$\sim  $3.4\,$\rm
X_l$     for     NGC\,6240.      Thus     using     $\rm     r^{(obs)}
_{CO/HCN}$=5.9(Arp\,220), 12.5(NGC\,6240),  and the LVG-computed: $\rm
r^{(h)} _{CO/HCN}$$\sim $1.6(Arp\,220),  2.15(NGC\,6240), we find $\rm
\rho ^{(l-h)} _{co}$$\sim $2.68(Arp\,220), 4.81((NGC\,6240). Then from
Equation 10  and the aforementioned numbers (and  setting $\rm X^{(l)}
_{co}$=0.5\,$\rm  X_l$) it  is: $\rm  X^{(2-ph)} _{co}$(Arp\,220)$\sim
$2.4\,$\rm  X_l$ and  $\rm X^{(2-ph)}  _{co}$(NGC\,6240)$\sim $1\,$\rm
X_l$.

In the aforementioned  example the warm ``end'' of  a degenerate range
of  LVG solutions  to a  multi-J  heavy rotor  molecular line  dataset
yields an $\rm X_{co}$$\sim $(1/5)$\rm X_{co,Gal}$ for NGC\,6240 while
reducing   that   of  Arp\,220   to   $\sim  $(1/2)$\rm   X_{co,Gal}$.
Nevertheless such  sets of solutions, with $\rm  K_{vir}$$>$1 and $\rm
T_{kin}$ significantly  higher than  $\rm T_{dust}$ are  not typically
considered optimal  in most studies.  Indeed $\rm  K_{vir}$$\sim $1 is
often  used as a  constraint on  the LVG  solution range  of molecular
lines  tracing dense  gas (e.g.  HCN,  CS) while  solutions with  $\rm
T_{kin}$$\sim $$\rm  T_{dust}$ are considered as  more appropriate for
such a gas component since gas-dust thermal coupling is expected to be
strong (e.g.  see  the dense gas model in  Mrk\,231 by Papadopoulos et
al.  2007).   On the other  hand recent theoretical  and observational
work  has   shown  that  in   ULIRGs  large  temperatures   with  $\rm
T_{kin}$$>$$\rm T_{dust}$ are possible even  with most of the gas mass
at high  densities (n$\geq $10$^5$\,cm$^{-3}$) because  of dominant CR
and/or  turbulent  heating,  both  powered  by the  large  SNR  number
densities in these galaxies  (Papadopoulos 2010; Paper\,I; Rangwala et
al.~2011).    Nevertheless,  while   high   $\rm  T_{kin}$   certainly
contributes in lowering $\rm X^{(h)} _{co}$ of the dense gas phase, it
is the high $\rm K_{vir}$ values that are mostly responsible for this,
a result noted also by  previous studies (e.g. DS98).  Indeed, setting
$\rm K_{vir}$=1 for the aforementioned LVG solutions for the dense gas
phase in  Arp\,220 and NGC\,6240  brings their $\rm  X^{(2-ph)} _{co}$
factors  back  up  to  $\sim  $4.5\,$\rm  X_{l}$(Arp\,220)  and  $\sim
$3.3\,$\rm X_l$(NGC\,6240).

Clearly good  constraints on the dynamical  state of the  dense gas in
ULIRGs is of  utmost importance since while $\rm  K_{vir}$$\sim $1 may
be  a  safe assumption  for  the  SF-fueling  dense gas  component  of
individual GMCs in  typical spirals, it may not be  so for the extreme
ISM   environments   of    ULIRGs.    Strong   tidal   fields   and/or
``bottom''-stirred ISM by strong  SF feedback (via multiple SNR shocks
and the radiative  feedback implied by Eddington-limit-regulated SFRs)
could induce $\rm K_{vir}$$>$1 even  for the dense gas in such systems
(see  Paper\,I), lowering  $\rm X_{HCN}$  and hence  the corresponding
$\rm X^{(2-ph)} _{co}$ factor.
  

\subsection{Cold SF-quiescent molecular gas and large $\rm \bf X_{co}$  in LIRGs}

    Gas-rich  disks  containing a  starburst  in  their central  $\sim
    $(1-2)\,kpc, but also large  amounts of SF-quiescent ISM at larger
    galactocentric  radii   are  often   found  in  LIRGs   with  $\rm
    L_{IR}$$<$10$^{12}$L$_{\odot}$,  and such configurations  are even
    expected for weakly perturbed systems (e.g.  Maiolino et al.  1997
    and   references   therein).   Large   amounts   of  cold   ($\sim
    $(10-15)\,K)  molecular   gas  beyond  a   warm  component  ($\sim
    $(35-40)\,K) confined within  a star-forming central $\sim $1\,kpc
    has been shown as a general feature of LIRGs using CO J=2--1, 1--0
    lines  (Papadopoulos  \&  Seaquist  1998).  Sensitive  dust  submm
    continuum  and CO J=1--0  imaging in  individual LIRGs  found cold
    dust and  concomitant molecular gas with low  CO J=1--0 brightness
    extending out to radii of  at least $\sim $3\,kpc (Papadopoulos \&
    Seaquist 1999;  Papadopoulos \& Allen 2000).   Submm continuum and
    HI imaging  studies of nearby spirals also  suggest cold molecular
    gas as  a general feature  of their disks at  large galactocentric
    distances (Thomas et al.  2004).

 Extended,  SF-quiescent, molecular  gas  reservoirs in  the disks  of
 LIRGs  can be  undetected because  of low  CO J=1--0  line brightness
 and/or  lack of  CO in  the lower  metallicity environments  found at
 large  galactocentric  distances.    Two  SF-idle  GMCs  with  n$\sim
 $10$^2$\,cm$^{-3}$  and  potentially  very low  kinetic  temperatures
 ($\sim  $5\,K) have  been  found  in M\,31  (Allen  \& Lequeux  1993;
 Loinard et  al.  1995; Allen  et al.  1995).  For  $\rm K_{vir}$$\sim
 $1-2 (i.e.  self-gravitating or nearly so) such clouds will have $\rm
 T^{(l)}  _{co,1-0}$$\sim  $1.4\,K and  a  corresponding $\rm  X^{(l)}
 _{co}$$\sim $19\,$\rm X_l$.   In the global CO SLEDs  of LIRGs such a
 cold  SF-idle  component, even  if  massive,  it  will be  completely
 inconspicious, outshined  by the  much more CO-luminous  SF molecular
 gas where  typical $\rm  T^{(h)} _{co,1-0}$$\geq $15\,K.   The global
 $\rm X_{co}$ of a LIRG, computed  using models of its global CO SLED,
 will then invariably  be biased by the average  ISM properties of its
 central  starburst,  often  having   a  genuinely  low  $\rm  X^{(h)}
 _{co}$$\sim $(1/3-1/5)$\rm X_{co,Gal}$  factor (e.g.  Papadopoulos \&
 Seaquist 1999;  Papadopoulos \& Allen  2000).  Thus its  adoption for
 the  entire  LIRG can  much  underestimate  the  contribution from  a
 massive cold,  SF-quiescent component  where a Galactic  $\rm X_{co}$
 applies, as indicated  by submm, H\,I and CO  J=1--0 imaging for such
 galaxies.

  We can compute  an indicative $\rm X^{(2-ph)} _{co}$  in such a LIRG
 for  a  given $\rm  \rho  ^{(l-h)}  _{co}$  by setting  $\rm  X^{(l)}
 _{co}$=$\rm   X_{co,Gal}$  for   the  cold,   extended,  SF-quiescent
 component,  and  $\rm X^{(h)}  _{co}$=(1/5)$\rm  X_{Gal,co}$ for  the
 warm, nuclear,  star-forming one.  Using the  results by Papadopoulos
 \& Seaquist (1998) where ``stacked'' CO(2--1)/(1--0) ratios for LIRGs
 were  used to  determine a  generic  warm versus  cold molecular  gas
 distribution in their disks (their  Figure 6) we set $\rm L_{h}$$\sim
 $0.9\,kpc  as  the  diameter  of  the  warm  SF  component  and  $\rm
 L_{s}$$\sim $20\,kpc as that of the entire CO-bright gas distribution
 in the ``average'' LIRG.   With a warm/cold CO brightness temperature
 ratio of $\rm  t_b$=$\rm T^{(h)} _{co,1-0}/T^{(l)} _{co,1-0}$$\sim $3
 it   is:  $\rm  \rho   ^{(l-h)}  _{co}$=(1/$\rm   t_b$)$\times  $$\rm
 [(L_s/L_h)^2-1]$$\sim  $164. Thus  $\rm X^{(2-ph)}  _{co}$$\sim $$\rm
 X^{(l) }_{co}$=$\rm  X _{co,Gal}$, i.e.   dominated by the  cold disk
 component.   Given that $\rm  \rho ^{(l-h)}  _{co}$ can  very greatly
 among LIRGs, it is instructive  to also obtain $\rm X^{(2-ph)} _{co}$
 in a LIRG where the relative  distributions of the warm SF versus the
 cold  SF-idle  gas are  well  known.   For  the Sy2  NGC\,1068:  $\rm
 L_{h}$$\sim $2.7\,kpc  and $\rm L_{s}$$\sim  $6\,kpc (Papadopoulos \&
 Seaquist  1999), while  $\rm t_b$$\sim  $5. Thus  $\rm  \rho ^{(l-h)}
 _{co}$$\sim   $0.79   and   $\rm  X^{(2-ph)}   _{co}$$\sim   $0.6$\rm
 X_{co,Gal}$, which is $\sim  $3x higher than $\rm X^{(h)}_{co}$ (that
 would be deduced from its global CO~SLED). 



\subsection{Molecular gas inventories in LIRGs: critical observations}

     Highly turbulent molecular gas in the merger-driven starbursts of
    ULIRGs,  and extended,  cold,  SF-quiescent gas  in  the disks  of
    isolated or  slightly perturbed  LIRGs represent the  two extremes
    where current CO line studies may systematically underestimate the
    total molecular  gas mass.   In both cases  global CO  SLEDs often
    show  a low-excitation component,  seen as  small CO  J=1--0, 2--1
    line  flux ``excess''  (often with  a subthermal  CO (2--1)/(1--0)
    ratio) on top of the dominant CO line emission of a warm and dense
    SF gas phase with high CO line excitation to much higher J levels.
    In  ULIRGs  such  a  diffuse  low-excitation  phase  is  typically
    unbound, warm, and contains little  mass. In the disks of LIRGs it
    consists  of SF-quiescent, gravitationally  bound (or  nearly so),
    Galaxy-type  GMCs extending  beyond a  central starburst.   In the
    latter case spatially resolving  the CO SLEDs and/or dust emission
    of the nuclear  SF region versus the extended  cold disk can yield
    proper molecular gas mass estimates not dominated by the (usually)
    low $\rm X _{co}$ of the CO-bright nuclear SF region.  In practice
    CO  and $^{13}$CO  J=1--0,  and  J=2--1 imaging  of  those SF  and
    SF-idle areas of LIRGs is  adequate to determine the nature of the
    residing  ISM  (SF:  $\rm  r_{21}$$\sim $1,  $\rm  R_{10,21}$$\sim
    $10-15,   SF-quiescent:    $\rm   r_{21}$$\sim   $0.4--0.6,   $\rm
    R_{10,21}$$\sim $3-6,  the corresponding $\rm  X_{co}$ values, and
    the areas over which they apply.

In the upcoming era of ALMA  such spatial separations of SF and non-SF
ISM in LIRG disks along  with estimates of the corresponding molecular
line  ratios and  $\rm  X_{co}$ values  will  be straightfoward.   For
ULIRGs though the low-density and  the high density gas components can
be concomitant or very closely associated, especially if the former is
the  outcome  of  the  disruption  of GMC  outer  layers.   Thus  high
resolution CO observations, even with  ALMA, may be unable to separate
their distributions  and their  corresponding CO line  ratios, leaving
their $\rm  X_{co}$ factors still  uncertain.  In such  galaxies, with
the  bulk  of the  molecular  gas  mass  at n$\ga  $10$^4$\,cm$^{-3}$,
observations of  rotational transitions of heavy  rotor molecules such
as HCN are of  paramount importance, even without spatial information.
For example  a high global HCN/CO J=1--0  brightness temperature ratio
of  $\sim $1/4-1/6 would  immediately indicate  an unsually  high $\rm
M_{dense}/M_{tot}(H_2)$, contrasting the $\sim $10$\times $ lower such
ratio  in  disk  GMCs.   Then  multi-J  observations  of  heavy  rotor
molecules can  be used to  determine the corresponding  $\rm X_{mol}$,
$\rm X^{(h)} _{co}$ factors and eventually $\rm M_{tot}(H_2)$.

High  resolution interferometric  imaging of  at least  two rotational
transitions of  a heavy rotor  molecule (e.g. HCN J=1--0,  3--2) along
with  at  least one  of  its  isotopologues  (e.g. H$^{13}$CN  J=1--0)
remains an invaluable resource for better determining the distribution
and mass of the dense gas in the compact disks of ULIRGs.  This can be
achieved  using radiative  transfer models  of the  emergent  HCN line
emission as a function of position within these disks, as done earlier
for the  CO J=1--0,  J=2--1 interferometric study  by DS98.  The focus
will now  be on  the dense gas  where much  of the molecular  gas mass
resides, while  its all-important  average dynamical state  (i.e.  the
$\rm K_{vir}$) will be determined as part of the modeling.  Imaging of
the rare isotopologues can much reduce the radiative transfer modeling
degeneracies  affecting  $\rm  K_{vir}({\bf r})$,  and  yield  improved
constraints   on  the   dense  gas   surface  densities   $\rm  \Sigma
_{dense}({\bf r})$ of  the gas disks  in ULIRGs.  Their total  gas mass
can  then be  determined from  integrating the  resulting  $\rm \Sigma
_{dense}({\bf r})$ over the best-fit disk models.

Here we  must note that  any molecular line observations  that involve
the   $\rm  [C/^{13}C]$   isotope  ratio   such  as   CO/$^{13}$CO  or
HCN/H$^{13}$CN, while necessary for reducing LVG modeling degeneracies
and  better constraining  $\rm K_{vir}$,  they involve  the additional
assumption of  the $\rm [C/^{13}C]$ abundance ratio  (which we assumed
to be 50 in our LVG models).  The latter can be particularly uncertain
in ULIRGs  where their large  CO/$^{13}$CO line ratios have  also been
attributed  to a  higher  $\rm [C/^{13}C]$  abundance  than in  spiral
disks, a result of early starburst ages and/or accretion of relatively
unprocessed (by star-formation)  molecular gas (Henkel \& Mauersberger
1993). If enhanced $\rm [C/^{13}C]$  abundances are indeed the norm in
ULIRGs  this will  have  the  general effect  of  {\it increasing  the
  various X$_{co}$, X$_{HCN}$ factors} since the deduced $\rm K_{vir}$
values per gas phase can now be lower.  This will be so simply because
larger  $\rm [C/^{13}C]$ abundances  rather than  low CO  line optical
depths (and thus  high $\rm K_{vir}$, see Equation 11  in Paper 1) can
be  also responsible for  the observed  high CO/$^{13}$CO  line ratios
in~ULIRGs.

\subsubsection{Molecular gas mass estimates of ULIRGs: the promise of Herschel}

In ULIRGs a massive, warm, and  dense gas phase can have a luminous CO
SLED that remains prominent  up to very high-J rotational transitions.
Since J=1--0, 2--1 transitions can have significant contributions from
a low-excitation  diffuse gas  component containing small  fraction of
the total molecular gas mass,  this leaves only J=3--2 and higher-J CO
lines as useful  probes of the bulk of the molecular  gas mass and its
properties.  The  SPIRE/FTS aboard  the HSO can  provide access  to CO
lines from J=4--3  up to J=13--12 for local  luminous ULIRGs (e.g. van
der Werf et al. 2010; Rangwala  et al. 2011) and thus yield a critical
dataset  for obtaining better  total molecular  gas mass  estimates in
such systems.

Once  a fully-sampled global  CO SLED  from J=1--0  up to  J=13--12 is
available, a  multi-component analysis can decompose it  into a series
of gas  components, constraining their  properties, and using  them to
obtain  the  corresponding  $\rm  X_{co}$ factors.   Such  multi-phase
models  could eventually  be produced  by theoretical  advancements in
GMC-sized numerical simulations (see 3.1.1).  Thus HSO observations of
high-J CO lines will allow better inventories of molecular gas mass in
ULIRGs since much  of that mass resides in a  dense, and presumably SF
and  warm  phase.  We  note  however  that  serious degeneracies  will
remain, especially regarding the all-important dynamical state of each
gas  component  upon  which  the  corresponding  $\rm  X_{co}$  factor
strongly   depends.   Multi-J   observations  of   the   much  fainter
(especially in ULIRGs) $^{13}$CO isotopologue lines, necessarily using
the  much larger mm/submm  telescopes available  from the  ground (and
thus limited by  the atmosphere up to J=6--5,  7--6) are important for
reducing those degeneracies, and better constraining $\rm K_{vir}$, or
equivalently  (for  a  given  density  and  temperature),  the  escape
probabilities $\rm  \beta _{J+1\,J}$ per gas component  that enter the
expressions of  the $\rm X_{co}$ factors  (Equations 3 and  6).


\section{Probing the extremes: individual (U)LIRGs, their molecular gas, and  \rm X$_{\bf CO}$}

Several  (U)LIRGs  in our  sample  merit  an  individual study  either
because a larger than average  number of available CO lines permits it
(see Table  7 in Paper\,I),  and/or because very high  line excitation
make their CO SLEDs  irreducible to superpositions of star-forming and
(non-SF)  molecular gas  (see also  Paper\,I for  a discussion  on the
implication about  ISM power sources).  Finally there  are LIRGs whose
global  CO(J+1-J)/(1-0) ratios  suggest the  cold  SF-quiescent clouds
found at large galactocentric distances  in the Galaxy and the disk of
M\,31.  Such globally ``cold'' yet star-forming galaxies are very few,
as expected  for an IR-selected  (and thus SFR-selected)  LIRG sample,
and  thus   also  deserve  a   closer  look  as  they   represent  the
low-excitation range  in LIRGs.  The detailed model(s)  per galaxy and
the  associated discussion  are in  the Appendix,  while here  we 
summarize the most important~findings.


The extreme range  of CO SLEDs found for  the molecular gas reservoirs
of LIRGs  discussed in Paper\,I  is now marked by  individual objects.
On  the  high  end   are  galaxies  like  IRAS\,00057+4021,  Arp\,299,
IRAS\,12112+305  and others  whose extremely  high CO  line excitation
implies   large   amounts   of   very   dense   ($\sim   $(1-3)$\times
$(10$^4$-10$^6$)\,cm$^{-3}$   {\it  and}   (often)  very   warm  ($\rm
T_{kin}$$\ga   $100\,K)   gas.    The   large   $\rm   f_{d}$=M(n$\geq
$10$^4$\,cm$^{-3}$)/M$\rm  _{tot}$(H$_2$)  deduced  for such  (U)LIRGs
independently recovers, using mid/high-J CO lines, a well-known result
for  merger-driven starbursts  obtained using  the HCN/CO  J=1--0 
ratio as  an $\rm f_{d}$ proxy  (Gao \& Solomon  2004).  These earlier
studies (see also Gracia-Carpio et al.  2008; Krips et al. 2008) found
$\sim  $10 times higher  dense gas  mass fractions  in ULIRGs  than in
isolated  spirals.  Isofar as  the dense  molecular gas  phase closely
tracks  the  dense  and  warm   gas  associated  with  SF  sites,  the
mid-J/high-J  CO lines  of  ULIRGs  are also  expected  to show  clear
indications of  a high  $\rm f_d$,  as it is  indeed found,  and quite
unlike  what  is  expected   from   typical  GMCs  ($\rm
f_{d,GMC}$$\sim $0.02-0.03).

Furthermore,    the   very   warm    and   strongly    unbound   ($\rm
K_{vir}$$\geq$20)   states  uncovered  for   the  massive   dense  gas
reservoirs of some  LIRGs can make their global  CO SLEDs surpass even
those  expected for  SF ``hot''-spots  in the  Orion A  and  B clouds.
While in  some distinct cases this  can be due to  strong AGN feedback
(e.g.   Mrk\,231),  for  all  other  galaxies with  such  extreme  ISM
conditions       (e.g.        IRAS\,00057+4021,      IRAS\,08572+3915,
IRAS\,23365+3604)  the cause is  unclear.  It  remains to  be explored
whether the much higher SFR  {\it densities} of the compact SF regions
of ULIRGs (e.g.  Sakamoto et  al.  2008) can create such extraordinary
ISM  states  where  turbulence-injection  by SNRs  no  longer  remains
confined in small regions but encompasses much more molecular gas mass
(see   discussion  in   Paper\,I).    This,  along   with  CRs   (also
SNR-generated)  may  set  up   powerful  global  mechanisms  that  can
volumetrically heat large amounts of gas, unhindered by the large dust
extinctions and  high average gas  densities that will keep  PDRs very
localized around SF~sites.

On  the low excitation  end only  a few  cases of  low global  CO line
ratios are found  (e.g. IRAS\,05189--2524, IRAS\,03359+1523).  This is
expected for  IR-selected (and thus SFR-selected)  galaxies (see Dunne
et  al.   2000   on  the  limitations  of  such   samples  in  finding
(cold-ISM)-dominated systems).  In Arp\,193 a low-excitation ISM state
and  the lack  of large  amounts  of dense  and warm  gas is  actually
suggested by HCN rather than  CO lines. Indeed despite the presence of
a young merger with substantial IR luminosity and a significant HCN/CO
J=1--0 line ratio, its global HCN line emission is consistent with the
absence of  a dense gas phase,  possibly a case of  strong SF feedback
momentarily dispersing its dense gas supply (Papadopoulos 2007).  Such
IR-luminous/(dense-gas)-deficient galaxies will  be rare (see discussion
in B.17)  and thus  valuable for studying  the effects of  SF feedback
onto the dense ISM where  the initial conditions of star formation are
set.  The  case of IRAS\,05189--2524 on  the other hand  stands out as
one where a massive cold molecular gas-rich disk is implied but unlike
other LIRGs  (e.g.  I\,Zw\,1)  there are no  morphological indications
whatsover  for  the presence  of  such a  disk  in  this very  compact
ULIRG/AGN system.  Moreover its  ``warm'' CO(6-5)/(3-2) ratio, warm IR
``colors'', and  compact size in cm, near-IR,  and optical wavelengths
would argue for the presence of  only a warm, dense, SF gas phase, and
thus this  object represents a cautionary tale  about such conclusions
drawn for similar high-z ULIRG/AGN  systems using only high-J CO lines
(e.g.  Tacconi et al. 2006).

We note  that massive cold molecular  gas disks are  implied for other
LIRGs  as well  (e.g. I\,Zw\,1,  VII\,Zw\,031, NGC\,7469)  but without
strongly affecting  their global CO SLEDs or  CO/$^{13}$CO line ratios
which remains  dominated by the  warm SF phase.  This  simply mirrors,
for molecular lines, a  result well-known for dust continuum emission,
though in practice global CO  SLEDs are somewhat more sensitive to the
presence of  cold diffuse  gas than  the global dust  SEDs are  to the
concomitant cold dust mass (Papadopoulos \& Allen~2000).

\subsection{Effects of average ISM conditions on the X$_{\bf co}$}

 In  Table 1 we  tabulate the  total molecular  gas mass  estimates as
 produced by Equations 3,  10, the corresponding $\rm X_{co}$ factors,
 the  minimum molecular  gas mass  implied for  Eddington-limited star
 formation  (section~2.3),   and  the  mass   of  the  high-excitation
 (h)-phase  when a  2-component model  is  used to  interpret the  CO,
 $^{13}$CO lines (see Appendix). A mere inspection of the third column
 demonstrates   than   in   several   galaxies  ($\sim   $40\%)   {\it
   significantly  larger $X_{co}$ factors  than the  so-called (U)LIRG
   values of  $\sim $(0.6-1)\,$ X_l$  may apply.}  For ULIRGs  this is
 due to large and even  dominant fractions of their molecular gas mass
 being at much higher densities than in disk-dominated LIRGs while for
 the  latter because extended  cold molecular  disks (with  a Galactic
 $\rm  X_{co}$) can  contain  much  of the  molecular  gas mass  while
 remaining incospicious in  the global CO SLEDs used  to constrain the
 global  $\rm X_{co}$.  For  ULIRGs this  low-mass bias  includes some
 very well-known systems such as  Mrk\,231 and Arp\,220.  For lower IR
 luminosity galaxies  underestimates of  their total molecular  gas by
 the $\rm X_{co}$ deduced  from one-phase radiative transfer models of
 their  CO SLEDs  can  be important  in  disk-dominated systems  (e.g.
 NGC\,7469), but also in  seemingly compact star-forming galaxies that
 nevertheless have ``cold'' CO  SLEDs and Galactic $\rm X_{co}$ values
 (IRAS\,05189--2524).

The  computed $\rm X^{(2-ph)}  _{co}$ and  $\rm M^{(2-ph)}  _{tot}$ in
Table\,1 also make  clear that in the case  of ULIRGs global molecular
line SLEDs can reveal the aforementioned mass bias, provided that they
include  mid/high-J  CO  and/or  heavy  rotor  molecular  lines  (e.g.
HCN). For  less IR-luminous galaxies with nuclear  starbursts and cold
SF-quiescent gas-rich  disks global  SLEDs cannot easily  identify the
presence of the latter except in a few cases (e.g.  IRAS\,05189-2524).
This becomes  possible only when additional  spatial information (e.g.
CO 1-0, submm  dust continuum, or cm imaging) is  available for such a
disk (e.g.  NGC\,7469).  In the  absence of such information even good
one-phase  models of  global CO  SLEDs for  a  (starburst)+(cold disk)
system may be unable to reveal the disk component leaving the deduced
average  $\rm   X_{co}$  dominated   by  the  starburst   phase  (e.g.
IRAS\,02483+4302). In the era of ALMA routine CO multi-J and $^{13}$CO
line imaging will provide much better molecular gas mass estimates for
disk-dominated LIRGs  with strong ISM excitation  gradients (see 3.4).
For  ULIRGs on  the other  hand  multi-J observations  of heavy  rotor
molecules (e.g.   HCN, CS) and their rare  isotopologues are necessary
for confirming  the high total  $\rm X_{co}$ and molecular  gas masses
implied by our  results in Table 1, by  constraining the all-important
dynamic state of the massive dense gas phase as discused in 3.2.1.

\section{Conclusions}

In this work our large  CO, $^{13}$CO line survey of Luminous Infrared
Galaxies (LIRGs)  detailed in Paper\,I (Papadopoulos et  al.  2011) is
used to examine the impact of the wide range of average ISM conditions
found for these  galaxies on their total molecular  gas mass estimates
via  the  so-called   $\rm  X_{co}$=M(H$_2$)/$\rm  L^{'}  _{CO,  1-0}$
factor. Our  sample includes some  of the most prominent  local ULIRGs
(e.g.  Arp\,220, Mrk\,231, IRAS\,17208--0014), often used as templates
for merger-driven  starbursts at  high redshifts, as  well as  less IR
luminous  disk-dominated galaxies.  We  find that  one-phase radiative
transfer models  of the  global CO, $^{13}$CO  line ratios  yield $\rm
\langle            X_{co}\rangle            $$\sim           $(0.6$\pm
$0.2)\,M$_{\odot}$\,(K\,km\,s$^{-1}$\,pc$^2$)$^{-1}$,  similar to that
obtained  by past studies.   The average  gas temperature  and density
strongly influence  $\rm X_{co}$, but the gas  average dynamical state
is   the  most   important   influencing  factor   with  unbound   gas
corresponding to low $\rm X_{co}$ values while self-gravitating gas to
higher ones.

Nevertheless            higher            $\rm            X_{co}$$\sim
$(2-6)\,\,M$_{\odot}$\,(K\,km\,s$^{-1}$\,pc$^2$)$^{-1}$   values   are
deduced for  (U)LIRGs whenever adequate  molecular line data  exist to
determine   the  mass   contribution  of   gas  at   densities  n$\geq
$10$^4$\,cm$^{-3}$ (high-J  CO lines from our survey  and/or HCN lines
from  the  literature).    Theoretical  expectations  for  the  highly
turbulent  molecular gas  in the  merger-induced starbursts  of ULIRGs
indicate  that, with  most  of the  gas  at such  high densities,  the
aforementioned large  $\rm X_{co}$ values  maybe the rule  rather than
the exception in such systems.  Past observational studies were unable
to  determine this, yielding  instead much  lower $\rm  X_{co}$ values
(and often considered as ULIRG-appropriate standard ones), because the
molecular lines used (mostly CO  J=1--0, 2--1) could not constrain the
properties of the dominant (i.e.  the dense) gas phase in ULIRGs.  Our
results indicate that only high-J CO lines and multi-J observations of
heavy rotor  molecules (e.g.   HCN, CS) can  overcome this  mass bias,
placing the  Herschel Space Observatory  and ALMA front and  center in
the quest for  improved molecular gas mass estimates  in ULIRGs in the
local  and the distant  Universe.  Of  particular importance  are good
constraints on the dynamical state  of the dense molecular gas in such
systems (self-gravitating or not?)   since strongly unbound states for
this phase seem to be the only possible way that $\rm X_{co}$ in ULIRGs
could  be  much  lower  than a  Galactic  value.   Observations  of
high-density  tracing  rare  isotopologues  (e.g.   high-J  $^{13}$CO,
H$^{13}$CN)   will  be   of  crucial   importance  in   yielding  such
constraints.    On   the   theoretical  front,   GMC-sized   numerical
simulations of the turbulent molecular gas with ULIRG-type of boundary
conditions are needed  in order to obtain the  dense gas mass fraction
and  its temporal  evolution.  This  in turn  can  inform galaxy-sized
simulations of  merger systems which recently  have included molecular
gas, but cannot track the dynamical state and mass of the dense gas.

We also find  LIRGs where underestimates of their  total molecular gas
are  the result  of a  massive, cold,  gas-rich, disk  existing beyond
their central  starburst. In  such cases the  global CO SLED,  and the
$\rm  X_{co}$ factor  determined  from it,  remains  dominated by  the
central starburst and its  often low $\rm X_{co}$. Spatially resolving
the CO,  $^{13}$CO line and dust  continuum emission of  such disks in
LIRGs  with  strong  ISM  excitation  gradients  is  indispensible  in
accounting for their total molecular gas mass. In such cases, provided
that adequate  resolution is  employed to separate  the cold  gas disk
from the nuclear starburst, even low-J CO and $^{13}$CO lines (J=1--0,
2--1,  3--2)  are  adequate  to  yield  much  improved  molecular  gas
mass~estimates.

Finally our  study unfolds the  wide range of  the global CO  SLEDs of
LIRGs presented in Paper\,I over a subsample of individual systems. We
find  galaxies whose  extreme CO  SLEDs indicate  ISM  conditions that
surpass those expected for star-forming regions and suggest ISM energy
sources  other  than  photons  from  PDRs  (as  already  discussed  in
Paper\,I).  These  can be AGN,  extreme turbulence, and/or  very large
cosmic ray energy  densities. Moreover we find LIRGs  where extreme SF
feedback may  have momentarily  fully disrupted their  dense molecular
gas  reservoirs, and LIRGs  where large  amounts of  cold SF-quiescent
molecular gas  are present  despite the absence  of an  extended disk,
their compact  near-IR/cm emission  size, and ``warm''  CO (6-5)/(3-2)
ratios.  The  latter type of  objects can be particularly  worrying if
encountered at  high redshifts where such  characteristics can readily
lead  towards  large  underestimates  of  their  total  molecular  gas
mass. The wide range of  the average ISM conditions and the intriguing
possibilities  that   may  lie  behind  it,  make   our  subsample  of
individually-studied  LIRGs an  excellent target  for future  ALMA and
NOEMA (for  the northern objects) molecular  line imaging observations
leading towards  a complete picture  about ISM energetics, AGN  and SF
feedback on the molecular gas in galaxies.

\acknowledgments We  would like to thank the  referee Santiago Burillo
for his comments, and particularly for bringing into our attention two
important issues namely, the uncertainties of [CO/$^{13}$CO] abundance
ratio  in ULIRGs,  and the  reliability of  dynamical  mass estimates,
which resulted  in our  corresponding clarifications included  in this
paper. Padelis P. Papadopoulos would like to dedicate this long-coming
work to  his 10-month  old son $\Lambda  \epsilon \omega \nu  i \delta
\alpha$-$X\rho ^{´}\eta  \sigma \tau o$, for  late night inspirations,
and to his wife $\rm  M\alpha \rho\gamma \alpha \rho i\tau \alpha$ for
her  enduring support.  The  project was  funded also  by the  John S.
Latsis  Public Benefit  Foundation.  The  sole responsibility  for the
content lies with its authors.

\newpage

\appendix

\section{The  X$_{\bf CO}$ factor in an LVG setting}

The   CO  J=1--0   line   luminosity  $\rm   L^{'}  _{CO}$(1--0)   (in
K\,km\,s$^{-1}$\,pc$^2$, see Equation 5, Paper\,I) can be re-expressed
as

\begin{equation}
\rm L^{'} _{CO(1-0)}=\int _{\Delta V} \int_{A_s(V)} T_{b,1-0}(\vec{r},V)\,da\,dV=
\int _{\Delta R_s} \int_{A_s(R)} \left[T_{b,1-0}(\vec{r},R)\left(\frac{dV}{dR}\right)_{\vec{r},R}\right] \,da\,dR
\end{equation}

\noindent
where  any given velocity  V is  assumed to  correspond uniquely  to a
source  surface $\rm  A_s(V)$, emitting  at $\rm  T_b(\vec{r},V)$.  In
turn this  can be parametrized by  a ``depth-in-the-source'' parameter
R(V) so  that these iso-velocity  surfaces completely scan  the entire
source volume for  a range $\rm \Delta R_s$  (which corresponds to the
FWZI of the source velocity field), and without any radiative coupling
(the LVG  assumption).  The latter implies that  the line luminosities
emanating from these surfaces simply add up.  Thus we can write

\begin{equation}
\rm L^{'} _{CO(1-0)}=\langle T_{b,1-0}\left(\frac{dV}{dR}\right)\rangle \Delta V_s
= \langle T_{b,1-0}\rangle \left(\frac{dV}{dR}\right) \Delta V_s,
\end{equation}

\noindent
where $\langle..\rangle$ denotes averaging over the entire source volume $\rm \Delta V_s$,
and for an assumed constant  velocity gradient. Thus for

\begin{equation}
\rm \frac{dV}{dR}=K_{vir}\times \left(\frac{dV}{dR}\right)_{vir}\sim 
0.65\sqrt{\alpha}\,K _{vir}\, \left(\frac{\langle n(H_2)\rangle}{10^3\,cm^{-3}}\right)^{1/2}\,
km\,s^{-1}\,pc^{-1},
\end{equation}

\noindent
and $\rm M(H_2)$=$\rm \mu m(H_2) \langle n(H_2) \rangle \Delta V_s$ ($\mu $=1.36 accounts
for the He mass) Equations A1, A2, and A3 along with subsituting astrophysical units
yield for the $\rm X_{CO}$ factor

\begin{equation}
\rm X_{CO}=\frac{\mu m(H_2) \langle n(H_2)\rangle K_{vir}^{-1}}{\langle T_{b,1-0}\rangle}
\left(\frac{dV}{dR}\right)^{-1} _{vir}=
\frac{3.25}{\sqrt{\alpha}}\frac{\sqrt{\langle n(H_2)\rangle}}{\langle T_{b,1-0}\rangle}
 K^{-1} _{vir}\,\left(\frac{M_{\odot}}{K\,km\,s^{-1}\,pc^2}\right).
\end{equation}

\noindent
The last expression (used in  the main text with the averaging symbols
omitted  for simplicity)  for $\alpha$=1.5  becomes identical  to that
derived  by Solomon et  al.  1997  (their Equation  21 for  f=1).  For
virialized gas  motions ($\rm  K_{vir}$=1), and typical  conditions in
Galactic  GMCs with  $\rm  \langle n(H_2)\rangle$=(100-500)\,cm$^{-3}$
and  $\rm  T_{b,1-0}$=10\,K   yields  $\rm  X_{CO}$$\sim  $(3-6)\,$\rm
M_{\odot}(K\,km\,s^{-1}\,pc^2)^{-1}$, in good  accord with the average
Galactic value and its uncertainties.  Finally, expression A4 is valid
for any optically thick molecular  line emission used as a mass tracer
of a  particular gas  phase (e.g.  HCN  J=1--0 tracing dense  gas with
$\rm n(H_2)$$>$10$^4$\,cm$^{-3}$),  while multi-J observations  of the
particular  molecule and its  isotopologues can  be used  to constrain
$\rm \langle n(H_2)\rangle$, $\rm T_{b,1-0}$, and $\rm K_{vir}$.

\subsection{The optically thin approximation for X$_{\bf CO}$}

From  standard  formalism  the   integrated  line  luminosity  for  an
optically  thin  CO  J+1$\rightarrow$J  line,  omitting  the  CMB  for
simplicity, is given by

\begin{equation}
\rm L_{J+1,J}=N_{J+1}\,A_{J+1\,J}h \nu_{J+1,J}, 
\end{equation}

\noindent 
where $\rm  N_{J+1}$ is the  total number of  CO molecules at  the J+1
state  and  $\rm  A_{J+1\,J}$  is  the  Einstein  coefficient  of  the
J+1$\rightarrow   $J  transition.    This  line   luminosity   can  be
re-expressed   in   terms   of   $\rm  L^{'}   _{J+1\,J}$   (in   $\rm
L_l$=K\,km\,s$^{-1}$\,pc$^2$ units, see A1) using

\begin{equation}
\rm L_{J+1,J}=\frac{8\pi k_B\nu ^3 _{J+1,J}}{c^3}\,L^{'} _{J+1,J}, 
\end{equation}

\noindent
which combined with A\,5 yields

\begin{equation}
\rm N_{J+1}=\frac{8\pi k_B \nu^2 _{J+1,J}}{h c^3 A_{J+1\,J}} L^{'} _{J+1\,J}.
\end{equation}

\noindent
The total molecular gas mass  is then given by

\begin{equation}
\rm M(H_2)=\left(N_0+N_1+N_2+N_3+N_4....\right)R_{CO}\,\mu\, m_{H_2},
\end{equation}

\noindent
where $\rm  R_{CO}$=$\rm [H_2/CO]$=$10^{4}$  is the CO  abundance and
$\mu $=1.36 accounts for He  mass.  The molecule population at the J=0
level obviously  cannot be  estimated from a  line transition,  and we
thus     compute     it     from    $\rm     N_{0}$=$\rm     (g_0/g_1)
exp[E_1/(k_BT_{ex,10})] N_1$ where  $\rm g_{J}$=$\rm 2J+1$ denotes the
J-state  degeneracy  factor,  $\rm  E_{1}/k_B$$\sim$5.5\,K,  and  $\rm
T_{ex,10}$   is  the   excitation   temperature  of   the  CO   J=1--0
transition. Substituting this in A8 yields

\begin{equation}
\rm M(H_2)=\left(1+\frac{1}{3}e^{5.5/T_{ex,10}}+\frac{N_2}{N_1}+
\frac{N_3}{N_1}+....\right)R_{CO}\,\mu m_{H_2} N_1,  
\end{equation}

\noindent 
which after substituting the expressions from A7 yields

\begin{equation}
\rm X^{(thin)} _{co}=\frac{8\pi k_B \mu m_{H_2} \nu^2 _{10} R_{CO}}{hc^3 A_{10}}\times
\left[1+\frac{1}{3}e^{5.5/T_{ex,10}}+
\left(\frac{\nu_{21}}{\nu_{10}}\right)^2\frac{A_{10}}{A_{21}}r_{21}+
\left(\frac{\nu_{32}}{\nu_{10}}\right)^2\frac{A_{10}}{A_{32}}r_{32}+...\right]
\end{equation}

\noindent
where $\rm r_{J+1\,J}$ are the $\rm L^{'} _{CO}$(J+1-J)/$\rm L^{'} _{CO}$(1-0) ratios
observed (Equations 5, 6).  The Einstein coefficients scale as a function of J
as 

\begin{equation}
\rm A_{J+1\,J}=3 \frac{(J+1)^4}{2J+3}\,A_{1\,0}
\end{equation}

\noindent
while $\rm \nu_{J+1\,J}$=$\rm (J+1)\nu_{10}$. Thus equation A10 finally becomes

\begin{equation}
\rm  X^{(thin)} _{co}=\frac{8\pi k_B \mu m_{H_2} \nu^2 _{10}R_{CO}}{hc^3 A_{10}}\times
\left[1+\frac{1}{3}e^{5.5/T_{ex,10}}+\frac{5}{12}\,r_{21}+\frac{7}{27}\,r_{32} +..
\frac{2J+3}{3(J+1)^2}\,r_{J+1\,J}+...\right].
\end{equation}

\noindent
Substituting the various physical constants and introducing astrophysical units
yields,

\begin{equation}
\rm \rm  X^{(thin)} _{co}=0.078
\left[1+\frac{1}{3}e^{5.5/T_{ex,10}}+\frac{5}{12}\,r_{21}+\frac{7}{27}\,r_{32} +..
\frac{2J+3}{3(J+1)^2}\,r_{J+1\,J}+...\right] \left(\frac{M_{\odot}}{K\,km\,s^{-1}\,pc^2}\right).
\end{equation}

\noindent
In the literature the LTE  approximation is often used by setting $\rm
\sum _{k=0} N_k$=$\rm Z_{LTE}$/g$_1$N$_1$ and $\rm Z_{LTE}$$\sim $$\rm
2 (k_BT_k/E_1)$=$\rm 2 [T_{k}/(5.5\,K)]$ in Equation A8, which then
yields

\begin{equation}
\rm  X^{(thin)} _{co}(LTE)=9.45\times10^{-3}\left[\left(\frac{T_k}{K}\right) e^{5.5/T_k}\right] 
 \left(\frac{M_{\odot}}{K\,km\,s^{-1}\,pc^2}\right).
\end{equation}

For  optically thick  CO emission  where the  optical depths  arise in
small gas cells with respect to molecular cloud sizes CO line emission
remains effectively  optically thin  (i.e. traces the  entire emitting
gas mass) but with $\rm A_{ik}$$\rightarrow
$$\rm  \beta_{ik}A_{ik}$  in all  the  previous  equations, with  $\rm
\beta_{ik}$=$\rm  \left[1-exp(-\beta_{ik})\right]/\tau_{ik}$ being the
photon escape  probability (for  a spherical cloud).   In such  a case
Equation A10 can  be trivialy modified for finite  CO line optical depths
to yield

\begin{eqnarray}
\rm \rm  X^{(\beta)} _{co} & = & 0.078\beta^{-1} _{10}
\left[1+\frac{1}{3}e^{5.5/T_{ex,10}}+\frac{5}{12}\left(\frac{\beta_{10}}{\beta_{21}}\right)\,r_{21}+
\frac{7}{27}\left(\frac{\beta_{10}}{\beta_{32}}\right)\,r_{32} +..\right. \\
 & + & \left. \rm ..\frac{2J+3}{3(J+1)^2}\left(\frac{\beta_{10}}{\beta_{J+1\,J}}\right)r_{J+1\,J}+..\right] \nonumber 
\rm \left(\frac{M_{\odot}}{K\,km\,s^{-1}\,pc^2}\right). 
\end{eqnarray}

\newpage

\section{Models of individual LIRGs}

Here we  describe the radiative  transfer models in  selected galaxies
from our  sample.  The  details of the  Large Velocity  Gradient (LVG)
radiative transfer code used, and  the parameter space explored by its
$\rm [T_{kin},  n, K_{vir}]$ variables  can be found in  Paper\,I.  In
the several cases where a  one-phase gas LVG model manifestly fails to
reproduce the  available line ratios we use  a two-phase approximation
(see  2.4) to better  represent the  underlying average  molecular gas
properties,  and  compute  the  corresponding $\rm  X^{(2-ph)}  _{co}$
(Equation 10).   The (h)-phase properties and its  $\rm X^{(h)} _{co}$
are constrained  using CO J=4--3,  3--2 and $^{13}$CO lines,  with the
6--5  transition placing  constraints as  a  lower limit  of the  true
emergent  line  luminosity  (see  Paper\,I),  with  higher  CO  J=6--5
luminosities  typically yielding  even denser  gas LVG  solutions with
larger  $\rm   X^{(h)}  _{co}$.   The  ``residual''   CO  line  ratios
(determined  from Equations  8, 9)  are used  as inputs  into  our LVG
radiative transfer code in  order to constrain the physical properties
and  corresponding $\rm  X^{(l)} _{CO}$  factor of  the low-excitation
(l)-phase.  In the cases where  a two-phase ISM model is necessary but
inadequate CO  line observations exist to constrain  the (h)-phase, we
assume  one based  on the  model defined  by the  Orion star-formation
``hot-spots'', their  SLED, and molecular gas  mass normalization (see
2.4).  Finally in the very few cases where heavy rotor molecular lines
(mostly HCN transitions)  exist, they are often used  to determine the
(h)-phase.  For all the LIRGs  studied, their measured CO line ratios,
computed SF-powered IR luminosities  $\rm L^{(*)} _{IR}$, and the dust
temperatures   used   to  constrain   the   LVG   model  (i.e.    $\rm
T_{kin}/T_{dust}$$\ga $1) can be found in Paper\,I.


\subsection{IRAS\,00057+4021}

There are only  a few studies of this LIRG, and  the only available CO
J=1--0 interferometric map shows  a very compact nearly face-on source
with most of the gas in  a $\la 1''$ ($\sim $864\,pc) core (DS98). Its
luminous CO J=4--3 line emission  cannot be accounted by one-phase LVG
models,  with   the  best   such  model:  $\rm   T_{kin}$=90\,K,  $\rm
n(H_2)$$\sim$$3\times  10^2$\,cm$^{-3}$, $\rm  K_{vir}$=1, reproducing
well  the observed  $\rm r_{21}$,  $\rm r_{32}$  and the  $\rm R_{21}$
lower limit,  but yielding $\rm r^{(lvg)} _{43}$=0.48,  which is $\sim
$1/3 of the observed value. Moreover the fact that $\rm r_{43}$$>$$\rm
r^{(h)} _{43}$(Orion) (see 2.4) and $\rm r_{43}$$>$$\rm r_{32}$, makes
obvious  that  a superposition  of  low-excitation (SF-quiescent)  and
high-excitation  Orion-type SF  ``hot''  spots cannot  produce such  a
highly-excited global  CO SLED, and that even  extreme GMCs consisting
solely of such  ``hot'' spots are inadequate.  Thus  the {\it average}
state of  the molecular  gas in IRAS\,00057+4021  ``out-excites'' even
the SF  regions in Orion A and  B.  The existence of  such ULIRGs, and
the implications for an  ISM energy input other than far-UV/optical/IR
photons from SF sites (i.e.   turbulent and/or CR heating) has already
been noted in Paper\,I (see also Rangwala et al. 2011).

Using $\rm  r_{43}$-$\sigma$=0.89 along with the other  CO line ratios
still does  not yield  an average ISM  state that can  reproduce them.
The best LVG solution  ranges then are: $\rm T_{kin}$=(35-65)\,K, $\rm
n(H_2)$=$10^3$\,cm$^{-3}$,   $\rm   K_{vir}$$\sim   $1-4,   and   $\rm
T_{kin}$=(70-150)\,K,  $\rm  n(H_2)$=$3\times  10^2$\,cm$^{-3}$,  $\rm
K_{vir}$$\sim  $1.  These  reproduce the  observed $\rm  r_{32}$, $\rm
r_{21}$   and  the   $\rm  R_{21}$   lower  limit,   but   yield  $\rm
r_{43}$(lvg)$\sim  $0.45-0.50, i.e.   still  significantly lower  than
$\rm  r_{43}$-$\sigma$.   A   2-phase  model  using  an  Orion-defined
(h)-phase  (and SLED)  gives  $\rm r^{(l)}  _{21}$=0.91, $\rm  r^{(l)}
_{32}$=0.53, $\rm  r^{(l)} _{43}$=0.83  (Equations 8, 9),  and remains
unable to fit  these residual line ratios, failing  to account for the
J=4--3 line ($\rm r^{(l)}  _{43}$(lvg)$\sim $0.3-0.4).  Using only the
J=3--2,      J=4--3     transitions      as      constraints     ($\rm
r_{43/32}$=$2.23\pm0.97$)  of  the   (h)  phase  and  restricting  LVG
solutions with $\rm T_{kin}$$\geq
$$\rm  T_{dust}$(=37\,K)  and   $\rm  K_{vir}$$\geq  $1,  yields  $\rm
T_{kin}$$\ga $95\,K, $\rm  n(H_2)$=$3\times 10^4$\,cm$^{-3}$, and $\rm
K_{vir}$$\sim $20.  The fit yields no upper limit on $\rm T_{kin}$, as
expected  since the  maximum  possible $\rm  r_{43/32}$ (attained  for
optically    thin    thermalized    CO   lines)    is    (4/3)$^2$$\rm
e^{-E_{32}/(k_BT_{kin})}$=1.77$\rm       e^{-16.6/T_{kin}}$$\rightarrow
$1.77  for $\rm  16.6/T_{kin}$$\rightarrow $0.   Using the  lower $\rm
r_{43/32}$=1.26 value (=(observed)$-\sigma$) as  input in the LVG code
yields identical results.  Thus {\it  a very warm, dense, and strongly
kinematically stirred  gas phase is present  in IRAS\,00057+4021.}  An
estimate  of its mass  can be  obtained from  the measured  $\rm L^{'}
_{CO(4-3)}$=($5.84\times  10^9$)\,$\rm   L_l$,  and  $\rm  L^{(h-ex)'}
_{CO(1-0)}$  =$\rm  1/\langle  r^{(h-ex)}  _{43}\rangle  $$\times$$\rm
L^{'} _{CO(4-3)}$, where $\rm \langle r^{(h-ex)} _{43}\rangle $=3.4 is
the  average CO(4--3)/(1--0)  ratio  computed from  the  range of  LVG
solutions compatible with the observed one.  Thus for $\rm L^{(h-ex)'}
_{CO(1-0)}$=($1.72\times   10^9$)\,$\rm   L_l$,   and   $\rm   X^{(h)}
_{co}$$\sim  $0.92\,$\rm   X_l$  (computed  from   the  LVG  solutions
compatible  with  $\rm  r_{43/32}$)  it  is  $\rm  M_{h-ex}(H_2)$$\sim
$1.6$\times   $10$^9$\,$\rm  M_{\odot}$,   comparable  to   the  total
(HI+H$_2$)  gas   mass  of  the   Milky  Way  and  the   minimum  $\rm
M_{SF}(H_2)$=$\rm   L^{(*)}   _{IR}/\epsilon_{g,*}$$\sim   $1.2$\times
$$10^9$\,M$_{\odot}$ expected  for ``fueling'' star  formation in this
LIRG. In Table 1 we  comprehensively list all these gas mass estimates
for IRAS\,00057+4021,  along with $\rm  M_{total}(H_2)$ computed using
1-phase and 2-phase  LVG models (Eqs 3 and 10).   We must note however
that for IRAS\,00057+4021  and most (but not all)  LIRGs where 2-phase
models are used, the large degeneracy that exists for LVG solutions of
the ``residual'' CO line ratios of the (l)-phase (cool and near-virial
versus warm and strongly gravitationally unbound states) translates to
a large range of $\rm X^{(l)} _{co}$, and thus $\rm X^{(2-ph)} _{co}$,
values  (Table  1). This  in  effect is  the  low-J  CO SLED  degeracy
described in  3.4, which  the global CO  SLEDs available in  our study
cannot ``break''.

The  high-excitation  (h)-phase  in  IRAS\,00057+4021  contains  $\sim
 $(20-50)\% of $\rm  M_{tot}(H_2)$ (Table 1), which is  at least $\sim
 $10$\times  $ larger  than  what would  be  expected if  its ISM  was
 reducible  to  typical  Galactic  GMCs.  Moreover  its  $\rm  r^{(h)}
 _{21}$(lvg)$\sim $2.5, $\rm  R^{(h)} _{21}$(lvg)$\sim $23 ratios (for
 $\rm  T_{kin}$=100\,K, $\rm n(H_2)$=3$\times$$  10^4$\,cm$^{-3}$, and
 $\rm K_{vir}$=22) are higher than those in the Orion SF ``hot-spots''
 near HII regions  and O,B star associations (Sakamoto  et al.  1994).
 This further corroborates the  irreducibility of the observed CO SLED
 of IRAS\,00057+4021  to ordinary photon-powered ISM  states.  One the
 other hand strong mechanical feedback from SNR-driven shocks could in
 principle yield such  highly excited CO SLEDs (Arikawa  et al.  1999;
 Bolatto et  al.  2003),  but just like  radiative feedback from  O, B
 stars, such SNR-driven ``hot''  spots involve only small fractions of
 a typical  GMC mass.   Indicatively, in the  Galactic SNRs  W\,44 and
 IC\,443 a warm, dense,  and very kinematically stirred component with
 $\rm  r_{21}$$\sim $1.3--4 (Seta  et al.   1998) involves  only $\sim
 $1\% of the impacted GMCs,  with negligible effect on their global CO
 line ratios ($\rm \langle  r_{21}\rangle _{GMC}$$\sim $0.6).  We must
 also note however that IRAS\,00057+4021  is a Seyfert\,2 and its AGN,
 if  X-ray luminous,  can induce  such extreme  conditions to  a large
 molecular  gas  mass,  if  distributed  close to  it  (Schleicher  et
 al.~2010).



%

\subsection{IRAS\,00509+1225 (I\,Zw\,1)}

This LIRG hosts an optically and X-ray luminous QSO in the center of a
molecular  gas-rich  disk,  the   site  of  a  vigorous  circumnuclear
starburst (Barvainis et al.  1989;  Eckart et al.  1994 and references
therein).  The latter  has been imaged interferometrically (Schinnerer
et al. 1998; Staguhn et al.  2004) revealing a disk of $\sim
$$8^{´´}$-$12^{´´}$  ($\sim$12-14\,kpc) with  cold ISM  but increasing
ISM excitation  towards its central $\sim  $$1.5^{´´}$ ($\sim $2\,kpc)
SF circumnuclear ring.

In terms of global CO  line excitation while its $\rm r_{21}$=$0.84\pm
0.24$ ratio  can be  (barely) compatible with  that of  a SF-quiescent
disk like  the Milky Way ($\sim  $0.6), its luminous  J=3--2 line with
$\rm r_{32}$=$1.16\pm  0.40$ is clearly  dominated by much  higher ISM
excitation, typical of a  SF phase (even $\rm r_{32}$-$\sigma$=0.76 is
2.5x higher  than $\rm r_{32}$=0.3  typical for SF-quiescent  ISM).  A
one-phase LVG  model yields an  acceptable fit within  the measurement
uncertainties, with $\rm  r_{21}$(lvg)$\sim $1, $\rm r_{32}$(lvg)$\sim
$0.90-0.95,  and $\rm  R_{10}$$\sim $8,  $\rm R_{21}$$\sim  $5-7 while
$\rm K_{vir}$$\sim $1-2 (a result  of the modest $\rm R_{10}$ and $\rm
R_{21}$).  The  best one-phase  LVG solutions yield  $\rm X_{co}$$\sim
$1.5\,$\rm  X_l$ (Equation  3), and  a total  molecular gas  mass $\rm
M(H_2)$$\sim $8.5$\times $10$^9$\,M$_{\odot}$.   The minimum dense gas
mass fueling  an Eddington-limited SF  is $\rm M_{SF}$(H$_2$)=4$\times
$10$^8$\,M$_{\odot}$,   which   only    $\sim   $5\%   of   the   $\rm
M_{tot}$(H$_2$)  and typical for  ordinary SF  GMCs found  in galactic
disks.   Thus  I\,Zw\,1 stands  as  an example  of  a  LIRG with  star
formation occuring  in an isolated  gas-rich disk, fueled  by ordinary
molecular clouds.

The reasonable 1-phase  LVG fit of the observed  global CO line ratios
in  I\,Zw\,1 make a  2-phase decomposition  unnecessary.  Nevertheless
the  available observational  information on  the presence  of  a cold
molecular gas-rich disk and  a circumnuclear ring starburst allow such
a decomposition as a useful test on the limitations of global CO SLEDs
and their fit  by a single average ISM state,  to yield reliable total
molecular gas mass estimates.  Using the CO J=3--2, 6--5 and $^{13}$CO
J=1--0 lines to constrain the  (h)-phase we find the best LVG solution
range  at  $\rm   T_{kin}$=(65-75)\,K,  n=$10^4$\,cm$^{-3}$  and  $\rm
K_{vir}$=13 that yield  $\rm X^{(h)} _{co}$=0.65\,$\rm X_l$.  Adopting
a Galactic $\rm X^{(l)}  _{co}$=5\,$\rm X_l$ for the computed residual
CO  J=1--0 emission,  which we  attribute  to the  extended gas  disk,
yields $\rm  X^{(2-ph)} _{co}$=1.65\,$\rm X_l$  (Equation 10), similar
to that  computed from  the one-phase LVG  model.  We  caution however
that the CO J=6--5 line  luminosity is highly uncertain and along with
it the current 2-phase decomposition.

\subsection{NGC\,828}

This  is a  disturbed spiral  galaxy with  a prominent  dust  lane and
H$\alpha $ emission in its  center and two sources symmetricaly around
the center  source along  the major axis  (Hattori et al.   2004). VLA
imaging of its cm continuum emission show it elongated along the disk,
as would be expected for star formation, with no signs of AGN activity
such as a radio core or jets  (Parma et al. 1986).  Its single dish CO
J=1--0 detection  showed abundant molecular gas (Sanders  et al. 1986)
while subsequent  interferometric CO  1-0 imaging with  OVRO recovered
all  the  single  dish  flux  and  showed  it  to  be  extended  $\sim
$10$''$$\times $20$''$ ($\sim $7.5kpc$\times $3.6kpc) with the longest
dimension lying along  the optical disk (Wang et  al. 1991).  The same
study  obtained  a  dynamical  mass of  $\rm  M_{dyn}$$\sim  $4$\times
$10$^{10}$\,M$_{\odot}$ within  a CO disk radius of  3.9\,kpc (for the
adopted cosmology).  However at  $\sim $4$''$ (1.4\,kpc) south-east of
its  nucleus the  CO-derived velocity  field deviates  from that  of a
normal rotating  spiral, possibly  indicating the presence  of another
molecular gas  concentration brought in as  part of a  merger (Wang et
al. 1991).

The available CO lines are  consistent with a star-forming disk with a
well-excited  CO J=3--2 line  ($\rm r_{32}$=0.70$\pm  $0.15, Paper\,I)
but  whose modest $\rm  R_{10}$=10$\pm $2  and $\rm  R_{21}$=12$\pm $3
ratios indicate that any merger activity has not disturbed the average
dynamical  state of  the molecular  clouds. This  is reflected  in the
one-phase  LVG solutions  found for  $\rm T_{kin}$$\geq  $35\,K which
invariably  have  $\rm  K_{vir}$$\sim  $2-4.  These  solutions  remain
highly    degenerate    however    with   $\rm    T_{kin}$=(35-50)\,K,
n=$10^3$\,cm$^{-3}$, and $\rm K_{vir}$$\sim  $4 almost as good as $\rm
T_{kin}$=(55-150)\,K, n=3$\times
$$10^2$\,cm$^{-3}$, and  $\rm K_{vir}$$\sim$2. The  corresponding $\rm
X_{co}$$\sim   $(0.8-1.1)$\rm    X_l$,   yields   $\rm   M_{tot}$$\sim
$(4.6-6.3)$\times  $10$^{9}$\,M$_{\odot}$,   while  $\rm  M_{SF}$$\sim
$3$\times    $10$^8$\,M$_{\odot}$   is    necessary    to   fuel    an
Eddington-limited star formation.

Thus the dense and warm gas fueling SF sites in this low IR luminosity
galaxy will amount to only $\sim $5\%-6.5\% of the total molecular gas
mass,  consistent  also with  its  small  HCN/CO  J=1--0 ratio  of  $\rm
r^{(obs)} _{HCN/CO}$=0.022 (Gao \& Solomon 2004).  Nevertheless global
CO SLED  can still be easily  dominated by even small  amounts of star
forming gas whose  often modest $\rm X_{co}$ may  be much smaller than
that of an incospicious SF-quiescent cold gas reservoir.  For NGC\,828
there  is apriori  knowledge of  an extended  molecular gas  disk with
$\sim  $8\,kpc  diameter.   However  a  2-phase fit  with  an  assumed
(h)-phase SLED  and mass normalization (see 2.4)  leaves the (l)-phase
properties   largely  undetermined   with  $\rm   X^{(l)}  _{co}$$\sim
$(0.7-2.5)\,$\rm X_l$ ($\sim $$\rm  X^{(2-ph)} _{co}$ since the (l)-phase
CO J=1--0 luminosity dominates), and $\rm M_{tot}$$\sim $(4-14)$\times
$10$^9$\,M$_{\odot}$.

\subsection{IRAS\,02483+4302}

This two-nuclei system is  a gas-poor/gas-rich merger of an elliptical
 (nucleus  A) going  through the  disk of  a former  spiral containing
 nucleus B  (Kollatschny et  al.  1992) and  all of the  molecular gas
 detected  in a CO  J=1--0 interferometer  map (DS98).   The optically
 bright nucleus  A hosts an  AGN with a  Sy2 spectrum, and  is located
 3.8$''$ ($\sim $3.76\,kpc)  to the west of nucleus  B.  The latter is
 where  star formation  occurs in  this system,  ``activated''  by the
 merger  event which  also likely  triggered the  QSO activity  in the
 gas-poor merger progenitor.
 
 Its CO line ratios up to J=3--2 can be well-fitted by a wide range of
conditions  with even a  cold $\rm  T_{kin}$=15\,K but  highly unbound
($\rm  K_{vir}$=22) phase  yielding a  good fit.   Most  solutions are
found at warmer temperatures though with $\rm T_{kin}$=(35-55)\,K, and
$\rm     T_{kin}$=(90-120)\,K,     low     densities     of     n$\sim
$(3$\times$10$^2$-10$^3$)\,cm$^{-3}$, and strongly unbound states $\rm
K_{vir}$=7-40.  Such  a wide degeneracy of  LVG solutions representing
the  average conditions of  the molecular  gas in  IRAS\,02483+4302 is
typical  for  LIRGs where  $\rm  r_{J+1,J}$$<$1,  and no  CO/$^{13}$CO
ratios are available  to constrain line optical depths  (and thus $\rm
K_{vir}$).  For this LIRG a lower limit of $\rm R_{10}$$\ga $13 is not
enough  to eliminate  the  aforementioned degeneracy,  though it  does
preclude SF-quiescent  {\it and} virialized  states.  For for  all the
LVG solutions  compatible with its  global line ratios we  obtain $\rm
X_{co}$=(0.25--0.70)\,$\rm   X_l$  (Equation   3).    Restricting  the
admissible range of LVG  solutions with $\rm T_{kin}/T_{dust}$$\ga $1,
and $\rm  K_{vir}$$\la $20  yields $\rm X_{co}$$\sim  $0.45\,$\rm X_l$
and a total molecular gas mass $\rm M(H_2)$$\sim$1.6$\times
$$10^9$\,M$_{\odot}$,  similar to  that obtained  using  the optically
thin,  LTE,  approximation   $\rm  M^{(LTE)}  _{thin}$$\sim  $2$\times
$10$^9$\,M$_{\odot}$ (Equation 5 for  $\rm T_{kin}$=55\,K) a result of
the low/moderate  optical depths of  the LVG solutions used  to deduce
$\rm X_{co}$ in this LIRG.

The minimum gas mass  for an Eddington-limited star formation in
IRAS\,02483+4302     is    $\rm     M_{SF}$(H$_2$)$\sim    $1.4$\times
$$10^9$\,M$_{\odot}$, which is $\sim  $90\% of its total molecular gas
mass (for  $\rm X_{co}$=0.45\,$\rm X_l$).  This is  rather puzzling as
the   SF   phase,   with    its   expected   high   densities   ($\geq
$10$^{4}$\,cm$^{-3}$)  and  near  gravitationally bound  states  ($\rm
K_{vir}$$\sim  $1-5) can  easily have  CO lines  thermalized  and with
substantial  optical depths up  to at  least J=3--2.   Thus if  the SF
phase indeed dominates the total molecular gas mass in this ULIRG, one
would expect  $\rm M^{(LTE)}  _{thin}$$\ll $M(H$_2$) rather  than $\rm
M^{(LTE)} _{thin}$$\sim $M(H$_2$) as the latter implies small/moderate
optical depths for  the bulk of the molecular  gas mass.  Moreover the
physical  states  compatible  with   the  global  CO  line  ratios  of
IRAS\,02483+4302 while  highly degenerate, are hardly  indicative of a
dense SF gas phase dominating its total molecular gas mass. It is also
telling  that all  one-phase LVG  solutions fail  to reproduce  the CO
J=6--5  line  luminosity of  this  LIRG,  indicating  the presence  of
another,  potentially  massive, gas  component  whose  higher CO  line
excitation becomes prominent beyond J=3--2.  Using only its CO J=3--2,
6--5,  and the upper  limits on  the $^{13}$CO  J=1--0, 2--1  lines as
constraints  of  the (h)-phase  in  a  2-phase  model still  yields  a
significant    range   of   solutions    though   all    have   n$\sim
$(10$^4$-10$^5$)\,cm$^{-3}$ with $\rm T_{kin}$$\geq $60\,K, typical of
a dense  and warm  SF-related phase.  Most  (but not all)  have virial
states   ($\rm  K_{vir}$$\sim  $1)   with  $\rm   X^{(h)}  _{co}$$\sim
$(1.5-2.5)\,$\rm  X_l$.  From the  corresponding $\langle  \rm r^{(h)}
_{65}\rangle  _{lvg}$$\sim  $0.90   and  $\rm  \langle  X^{(h)}  _{co}
\rangle_{lvg}$$\sim   $2.2\,$\rm  X_{l}$   we   obtain  $\rm   L^{(h)}
_{co,1-0}$=1.5$\times      $10$^9$\,$\rm      L_l$,      and      $\rm
M_{h-ex}$=3.3$\times $10$^9$\,$\rm M_{\odot}$.   Higher CO J=6--5 line
luminosities will only  make this mass larger (both  via a higher $\rm
L^{(h)}  _{co,1-0}$ and a  higher deduced  $\rm \langle  X^{(h)} _{co}
\rangle_{lvg}$).

The ``residual'' global  CO line ratios of the  (l)-phase in this LIRG
cannot  distinguish between a  cold, SF-quiescent  gas at  near virial
dynamical states (and Galactic $\rm  X^{(l)} _{co}$) versus a warm and
highly unbound phase expected  in the highly turbulent environments of
mergers (and with  a low $\rm X^{(l)} _{co}$ because  of the high $\rm
K_{vir}$).  Thus for $\rm  \rho ^{(l-h)} _{co}$=1.43 computed from our
2-phase  model, $\rm  X^{(h)} _{co}$=2.2\,$\rm  X_l$ and  $\rm X^{(l)}
_{co}$=(0.5-2.5)\,$\rm    X_l$,     we    obtain    $\rm    X^{(2-ph)}
_{co}$=(1.2-2.4)\,$\rm  X_l$.   Higher,  near-Galactic, values  remain
possible if high resolution CO imaging were to reveal the (l)-phase as
an extended cold, SF-quiescent gas reservoir, something that global CO
SLEDs cannot easily do.


\subsection{IRAS\,03359+1523}

Two interacting galaxies $\sim  $10$''$ ($\sim $6.9\,kpc) apart can be
discerned in optical images, with only the eastern source being bright
in radio wavelengths  (Condon et al.  1990; Goldader  et al. 1997) and
containing most ($\ga  $75\%) of the CO J=3--2  emission (Leech et al.
2010,  Paper\,I).  This  system has  the lowest  CO(3--2)/(1--0) ratio
(=0.18$\pm$0.05)  indicating  the  lowest-excitation  global  CO  line
excitation  in our sample,  though we  cannot exclude  the possibility
that  significant  CO  J=3--2  flux  was  ``missed''  by  the  2-point
observations of this two-nuclei system.

The best LVG solution  indicates SF-quiescent gas ($\rm T_{kin}$=15\,K
n=3$\times  $10$^2$\,cm$^{-3}$ and  $\rm  K_{vir}$$\sim $2),  yielding
$\rm r_{32}$$\sim  $0.22 and $\rm  R_{21}$$\sim $18 which  is somewhat
higher than  the observed value (=$12\pm 3$).   The corresponding $\rm
X_{co}$$\sim  $2.5\,$\rm  X_l$  gives  $\rm  M_{tot}$$\sim  $2.2$\times
$10$^{10}$\,M$_{\odot}$,   while  the   LTE  approximation   for  $\rm
T_{kin}$=15\,K     gives     $\rm    M^{(LTE)}     _{thin}$=1.8$\times
$10$^{9}$\,M$_{\odot}$, which is $\sim  $10$\times $ smaller than $\rm
M_{tot}$  as expected  given the  large CO  line optical  depths ($\tau
_{10}$$\ga  $14) of  the corresponding  LVG solution.   The SF-related
molecular      gas       mass      $\rm      M_{SF}$(H$_2$)=8.2$\times
$10$^{8}$\,M$_{\odot}$  amounts  to   only  $\sim  $4\%  of  M(H$_2$),
consistent with the  low average CO line excitation  of this system
and ensembles of ordinary GMCs.

\subsection{VII\,Zw\,031}


This  galaxy has  one of  the highest  CO J=1--0  luminosities  in our
sample  and while early  ground-based images  were suggestive  of an
elliptical  system (Sanders  \& Mirabel  1996), near-IR  NICMOS images
clearly show  a spiral disk  with very bright asymmetric  arms tightly
around  its nucleus  over scales  of R$\sim  $(1-1.4)\,kpc (half-light
radii),  with numerous  star clusters  on  its disk  (Scoville at  al.
2000).  This  is one of  the few LIRGs  where there is no  evidence of
what  has triggered its  starburst activity,  namely neither  a nearby
companion system  (indicating an early merger stage),  nor tidal tails
(signs  of  an  advanced  or  post-merger  system).   High  resolution
interferometric  imaging  indicates a  nearly  face-on  source with  a
rapidly rotating gaseous ring on scales of few hundred parsecs (DS98).

    The  low-J  CO SLED  with  $\rm  r_{21}$=0.72$\pm  $0.12 and  $\rm
r_{32}$=0.43$\pm  $0.19 indicate rather  low average  line excitation,
compatible  even with  SF-quiescent ISM  ($\rm  r_{21}$$\sim $0.5-0.6,
$\rm r_{32}$$\sim $0.25-0.30).  A radiative transfer model of only the
CO J=1--0,  2--1, 3--2, and  $^{13}$CO J=1--0, 2--1 lines  yields $\rm
T_{kin}$=(30-65)\,K,    n=3$\times    $10$^2$\,cm$^{-3}$   and    $\rm
K_{vir}$$\sim $2  (near-virial).  The corresponding  $\rm X_{co}$$\sim
$1.25\,$\rm    X_l$     factor    yields    $\rm    M(H_2)$=1.5$\times
$10$^{10}$\,$\rm  M_{\odot}$  while  the  minimum molecular  gas  mass
needed to  fuel an Eddington-limited  star formation in this  LIRG is:
$\rm M_{SF}(H_2)$$\sim  $2$\times $10$^9$\,M$_{\odot}$.  On  the other
hand a luminous CO  J=4--3 with $\rm r_{43}$=1.46$\pm $0.45 (Paper\,I)
is highly excited,  and the reason why one-phase  LVG solutions, while
reproducing  the observed  $\rm  r_{32,21}$ and  $\rm R_{10,21}$  line
ratios  within  their   measurement  uncertainties,  they  yield  $\rm
r^{(lvg)}  _{43}$$\la $0.20-0.26,  i.e.  $\sim  $6-7 times  lower than
observed.  The  CO J=6--5  line luminosity with  $\rm r_{65}$=0.22$\pm
$0.07  is  also much  higher  than  predicted  by these  models  ($\rm
r^{(lvg)} _{65}$=0.01-0.02).  Interestingly the $^{13}$CO J=1--0, 2--1
lines,    with   modest    $\rm   R_{10,21}$$\sim    $10    and   $\rm
r_{21}$($^{13}$CO)$\sim  $0.72  ratios,  also indicate  a  low/average
excitation state.

 Thus  only the  CO J=4--3  and J=6--5  line luminosities  signify the
 presense of another potentially  massive warmer and denser gas phase,
 demonstrating   once  more   the   importance  of   high-J  CO   line
 measurements.  Moreover,  the $\rm r_{43}$$>$$\rm  r_{32}$ inequality
 sets the global CO SLED of VII\,Zw\,031 apart from those reducible to
 a mixture of  a dense and warm  gas phase (with a CO  SLED typical of
 Orion hot-spots),  and a cooler,  diffuse one associated  with non-SF
 gas and a low-excitation  SLED.  As in IRAS\,00057+4021, this implies
 large  amounts of hot  {\it and}  dense gas,  maintained in  a strongly
 unbound dynamical state as to  keep the average optical depth of even
 high-J CO  lines below unity  (see also discussion in  Paper\,I).  We
 note however  that in VII\,Zw\,031  the CO J=3--2 line  luminosity is
 highly uncertain because of the large system temperature, though even
 a 3$\times$ stronger CO J=3--2 line would maintain the aforementioned
 inequality and  its implications for extraordinary  ISM conditions in
 this~galaxy.

We  use the  CO (4-3)/(3-2)  ratio  $\rm r_{43/32}$  to constrain  the
(h)-phase in a 2-phase model while using the CO J=6--5 luminosity only
as a  lower limit.  We  also assume the minimum  $\rm r_{43/32}$=($\rm
L^{'}   _{CO(4-3)}$-$\sigma$)/($\rm  L^{'}  _{CO(3-2)}$+$\sigma$)=1.67
value  consistent within  the  measurement uncertainties  in order  to
place limits on  the lowest possible average excitation  level for the
(h)-phase.   This will still  be high  as even  that minimum  value is
close to the theoretical maximum of (4/3)$^2$, attained only for a hot
gas phase with fully optically thin and thermalized CO lines.  We find
good solutions only for $\rm T_{kin}$$\geq $95\,K which keep improving
up to  the maximum  $\rm T_{kin}$=150\,K considered,  while n=3$\times
$10$^4$\,cm$^{-3}$  and $\rm  K_{vir}$=22.  No  good solutions  can be
found   for  lower   temperatures  as   these  have   unphysical  $\rm
K_{vir}$$<$1 values.   Over the good solution  range the corresponding
$\rm \langle  r^{(h)} _{43}\rangle  $=3.4 and $\rm  r^{(h)} _{65}$$\ga
$2.  The  corresponding $\rm \langle  X^{(h)} _{co}\rangle $=0.9\,$\rm
X_l$,  and  from  $\rm  L^{(h)'}  _{CO,1-0}$=$(1/\rm  \langle  r^{(h)}
_{43}\rangle)$$\rm  L^{'} _{CO(4-3)}$$\sim  $3.45$\times $10$^9$\,$\rm
L_l$  (we  used  the   CO(4-3)-$\sigma  $  value)  we  estimate:  $\rm
M_{h-ex}$=3.1$\times   $10$^9$\,M$_{\odot}$.   As  expected,   with  a
$^{13}$CO(2-1)/(1-0) ratio  of $\sim $0.73  (subthermal), an (h)-phase
model  cannot successfuly include  the $^{13}$CO  lines ,  while their
modest global $\rm R_{10,21}$$\sim $10 (Table 8, Paper\,I) are typical
of  disk  GMCs  rather  than   of  those  in  starbursts  (where  $\rm
R_{10,21}$$\geq  $15). This  along with  the fact  that there  is very
little  CO J=2--1  and 3--2  emission  left after  subtraction of  the
corresponding  (h)-phase  line   emission  suggest  a  cold  disk-like
(l)-phase.   Assuming  an  $\rm  X^{(l)} _{co}$=4.5\,$\rm  X_l$  (i.e.
Galactic),  and for  $\rm \rho  ^{(l-h)} _{co}$=2.34,  we  obtain $\rm
X^{(2-ph)}  _{co}$=3.4\,$\rm  X_l$  and  a  total  gas  mass  of  $\rm
M^{(2-ph)}  _{tot}$=3.9$\times $10$^{10}$\,M$_{\odot}$.   Thus despite
its  prominence  in CO  J=4--3  and  6--5  lines, the  high-excitation
component amounts to  only $\sim $8\% of the  total molecular gas mass
in VII\,Zw\,031.  Interestingly, the  CO J=2--1 interferometric map by
DS98  resolves out  half  of  the 30-m  single  dish flux,  indicating
extended emission beyond the $\sim $2$''$(2\,kpc) nuclear region where
the  bright CO  J=1--0 emission  is distributed.   Indeed in  the DS98
interferometric image  faint CO J=1--0 emission  can be seen  out to a
radius of $\sim $4$''$ (4\,kpc). New sensitive CO and $^{13}$CO J=1--0
and 2--1  interferometric observations with similar  u-v coverage (and
thus enabling  reliable line  ratio maps) can  be used to  discern the
existence of such a massive cold molecular gas disk in this intriguing
object.


\subsection{IRAS\,05189--2524}

This is Sy2 galaxy that appears to  be a late stage merger with a very
red compact nucleus  and a tidal tail, (Farrah  et al.  2003; Veilleux
et al.   2002). This (U)LIRG is  the most compact and  has the warmest
mid-IR  colors ($\rm f_{25\mu  m}/f_{60\mu m}$=0.25)  out of  a nearly
complete ULIRG  sample selected for their ``warm'' mid-IR  colors ($\rm
f_{25\mu  m}/f_{60\mu  m}$$>$0.2). Such  systems  are  are thought  to
represent  a   critical  stage  in   the  so-called  ULIRG$\rightarrow
$(optically  luminous   QSO)  transition  scenario   (Sanders  et  al.
1988b,c).   The  AGN  in  this  ULIRG  is  X-ray  luminous  with  $\rm
L_{x}$(2-10keV)$\sim  $1.3$\times  $10$^{43}$\,ergs\,s$^{-1}$  (Dadina
2007).  NICMOS  near-IR imaging reveals  an unresolved nucleus  in all
three near-IR bands,  with a half light radii  of $\sim $(100-140)\,pc
while  CO J=1--0  observations  show large  amounts  of molecular  gas
(Sanders et al.  1991) but  no interferometric CO images are available
yet for this southern ULIRG.

Unlike  all other  mergers/starbursts, IRAS\,05189--2524  has  a small
$\rm  R_{21}$=$6\pm  2$ ratio,  typical  of  the self-gravitating  (or
nearly so) GMCs in spiral disks.  As a result all LVG solutions within
the expected temperature range (i.e.  $\rm T_{kin}/T_{d}$$\ga $1) have
$\rm  K_{vir}$$\sim  $1-2, indicating  virialized  gas motions,  quite
unlike the  much higher $\rm K_{vir}$ found  in merger/starbusts.  The
observed  $\rm r_{21}$=$0.67\pm  0.15$ together  with the  modest $\rm
R_{21}$ are  actually perfectly  compatible with a  SF-quiescent phase
(where  $\rm  r_{21}$$\sim  $0.6),  although  a  SF-active  one  ($\rm
r_{21}$$\sim $0.8-1) is certainly  not excluded within the measurement
uncertainties.   The best  LVG solutions  converge  towards conditions
typical  of  Galactic GMCs  but  at  elevated  temperatures with  $\rm
T_k$=(30-55)\,K,  $\rm  n(H_2)$=$3\times  10^2$\,cm$^{-3}$,  and  $\rm
K_{vir}$$\sim   $1   and   yielding   $\rm   r_{21}$=0.86-0.91,   $\rm
r_{32}$=0.50-0.67, $\rm  r_{43}$=0.30-0.40 and $\rm  R_{21}$=9.  Using
values within the $\rm  r_{J+1,J}\pm \sigma$ range as constraints does
not change the  basic picture of low-density, somewhat  warm gas, in a
virialized or nearly so dynamical~state.  For the best solution it is:
$\rm    X_{co}$$\sim    $3.5\,$\rm    X_l$,    which    yields    $\rm
M_{tot}(H_2)$=1.37$\times$$10^{10}$\,M$_{\odot}$.

   Denser and  warmer gas must  of course be present,  associated with
  the  vigorous star  formation in  IRAS\,05189--2524 and  the minimum
  mass  needed  to  fuel  it  of  $\rm  M_{SF}(H_2)$$\sim  $3.6$\times
  $10$^{9}$\,$\rm M_{\odot}$.   The CO J=6--5 line  also indicates gas
  with  significantly higher  excitation  than that  indicated by  the
  lower-J CO lines since, for  the optimal LVG solutions for the low-J
  CO SLED segment: $\rm  r_{65}$(lvg)$\sim $0.01-0.05, much lower than
  the observed $\rm r_{65}$=$0.37\pm 0.14$. Using a 2-phase model with
  an Orion-derived SLED and  an Eddington-limit mass normalization for
  the  (h)-phase (see  2.4) yields  (l)-phase ratios  of  $\rm r^{(l)}
  _{21}$=0.45,  $\rm r^{(l)} _{32}$=0.28  and $\rm  R^{(l)} _{21}$=3.5
  with the best LVG solution being $\rm T_{kin}$=15\,K, n$\sim
  $$300$\,cm$^{-3}$  and  $\rm K_{vir}$$\sim  $2.2,  i.e.  typical  of
  SF-quiescent   GMCs  found  in   the  disk   of  the   Galaxy.   The
  corresponding  $\rm  X^{(l)} _{co}$=2.5\,$\rm  X_{l}$,  which for  a
  computed  $\rm L^{(l)'}  _{CO(1-0)}$=2.3$\times  $10$^9$\,$\rm L_l$,
  yields  a   total  mass  of  cold  SF-quiescent   molecular  gas  of
  $\sim$6$\times   $10$^9$\,M$_{\odot}$.   For   the   estimated  $\rm
  \rho^{(l-h)} _{co}$=1.43 and $\rm X^{(h)} _{co}$=2.2\,$\rm X_l$ (see
  2.4)  we find $\rm  X^{(2-ph)} _{co}$$\sim  $2.4\,$\rm X_{l}$  and a
  $\rm  M^{(2-ph)} _{tot}$$\sim $9.4$\times  $10$^9$\,M$_{\odot}$ with
  $\sim $65\% of this gas mass in a cold SF-quiescent phase.

 This  merger  ULIRG/AGN  with  its  large  molecular  gas  reservoir,
 nearly-Galactic  $\rm  X_{co}$   ($\sim$3-6  times  higher  than  the
 so-called ULIRG-values  of $\sim $(0.6-0.8)\,$\rm X_l$),  and most of
 its molecular  gas mass in a  cold reservoir, stands  as the clearest
 counter-example to standard views regarding this class of galaxies in
 the  local Universe.   Moreover, unlike  I\,Zw\,1  (another prominent
 ``warm''  LIRG  considered  midway  in a  supposed  ULIRG$\rightarrow
 $(optical  QSO)  transition)  whose  known  spiral disk  can  be  the
 repository  of its  large SF-quiescent  molecular gas  reservoir, the
 very compact  size of IRAS\,05189-2524  in cm continum,  near-IR, and
 optical  wavelengths (Condon  et al.   1990; Scoville  et  al.  2000;
 Surace et  al.  1998) seems  to preclude such a  configuration.  This
 compactness  along with  a  ``warm'' CO(6--5)/(3--2)  ratio of  $\sim
 $0.63  (i.e.   dominated  by  the  SF phase)  in  a  ULIRG/AGN  whose
 molecular gas reservoir is dominated by a cold SF-quiescent component
 offer a  cautionary tale regarding similar ULIRG/AGN  objects at high
 redshifts where such charecteristics  were often used (and still are)
 to argue for  the presence of only a dense and  warm SF gas reservoir
 (Tacconi  et al. 2006).   Future ALMA  observations of  this southern
 ULIRG  and  similar  objects  would  thus  be  very  interesting  for
 revealing the distribution of their large molecular gas reservoir and
 especially of their cold phase.

\subsection{IRAS\,08030+5243}

This  LIRG  is  one  of  the  very few  (=3)  appearing  as  a  single
undistorted  nucleus  in a  large  near-IR  (2.2$\mu  $m) and  optical
imaging  survey of  56  ULIRGs  selected from  the  IRAS 2\,Jy  sample
(Murphy  et al.   1996).   Its CO  line  ratios reveal  a low  average
excitation state but lack of $^{13}$CO observations permit a very wide
range of conditions to reproduce  its low global $\rm r_{32}$ and $\rm
r_{21}$ ratios. In  turn these conditions yield a  large range of $\rm
X_{co}$ values, from  $\rm  X_{co}$$\sim $4.6\,$\rm  X_l$ (for  $\rm
T_{kin}$=15\,K, n$\sim $300\,cm$^{-3}$  and $\rm K_{vir}$$\sim $1), to
$\rm     X_{co}$$\sim      $(1.5-1.7)\,$\rm     X_l$     (for     $\rm
T_{kin}$=(55-90)\,K),  n$\sim $100\,cm$^{-3}$  and  $\rm K_{vir}$$\sim
$1)  and also  down to  $\rm X_{co}$$\sim  $0.61\,$\rm X_l$  (for $\rm
T_{kin}$=[20\,K,  (40-50)\,K],  n$\sim  $[10$^3$, 300\,cm$^{-3}$]  and
$\rm K_{vir}$$\sim  $[13, 7]  respectively).  For a  set of  very warm
($\rm  T_{kin}$=(95-130)\,K) and  strongly unbound  ($\rm K_{vir}$=22)
diffuse gas (n=3$\times $$10^2$\,cm$^{-3}$) solutions the $\rm X_{co}$
factor  can become  as  low as  $\sim  $0.3\,$\rm X_l$.   It is  worth
remembering  that such  gas may  indeed exists  in most  ULIRGs  as an
``envelope'' phase of a much denser dense phase, a result of disrupted
GMCs during a strong merger. Its  high brightness in low-J CO lines (a
result  of  its  high  $\rm  T_{kin}$ and  $\rm  K_{vir}$  values)  is
responsible for the  low $\rm X_{co}$ deduced for  such systems (which
nevertheless  can have  most of  their molecular  gas mass  in  a much
denser phase with high $\rm X_{co}$, see discussion in 3.2).

Demonstrating the  importance of  $^{13}$CO line measurements  we note
that  the CO/$^{13}$CO  J=1--0  ratio changes  dramatically among  the
aforementioned  LVG solutions  groups, from  $\rm R_{10}$$\sim  $5 for
cold  virialized gas,  to $\rm  R_{10}$$\sim $30  for the  warm highly
non-virial ISM  states. Neverthelesss lack of  such measurements makes
it  impossible to  constrain the  LVG  solution groups  and thus  $\rm
X_{co}$ in this  ULIRG.  Values as low as  $\rm X_{co}$=0.3\,$\rm X_l$
are  likely   excluded  however  as  they   yield  $\rm  M_{tot}$$\sim
$3$\times$10$^9$\,M$_{\odot}$,        comparable        to        $\rm
M_{SF}$(H$_2$)=2.4$\times  $10$^9$\,M$_{\odot}$.  In  such a  case the
global  CO SLED  would be  that  of dense  warm SF  gas (i.e.   highly
excited) rather than of low-excitation.

\subsection{IRAS\,08572+3915}

This  is  another  prominent  ULIRG  from the  ``warm''  ULIRG  sample
(Sanders  et al.   1988c) consisting  of a  close pair  of interacting
spirals  with their nuclei  separated by  5$''$ ($\sim  $5.6\,kpc) and
clearly discernible in  cm, near-IR and optical images  (Condon et al.
1990; Scoville  et al. 2000; Surace  et al.  1998). Its  NW nucleus is
unresolved  in near-IR  and  is the  only  one detected  in CO  J=1--0
interferometer  maps  where it  appears  unresolved  with $\rm  \theta
_{co}$$\leq  $2.1$''$($\sim $2.3\,kpc) (Evans  et al.   2002).  Mid-IR
(12-25\,$\mu$m) imaging showed all  the mid-IR emission also emanating
from the NW nucleus and a  region $\la $330\,pc in diameter (Soifer et
al. 2000, and for the adopted cosmology).

The  discovery of  very luminous  CO  J=6--5 emission  towards the  NW
nucleus  (Paper\,I,  Figure 2),  corresponds  to  second highest  $\rm
r_{65}$ ratio in our entire  sample ($\sim $1), indicating extreme ISM
excitation    conditions,    with    average    gas    densities    of
$\sim$(10$^5$-10$^6$)\,cm$^{-3}$,  and LVG  fits  that keep  improving
past  $\rm  T_{kin}$=150\,K.   The  3$\sigma  $ upper  limit  on  $\rm
r_{32}$$\la $2.54 is certainly compatible with such conditions, albeit
not  imposing any  useful additional  constraints (high  $\rm T_{sys}$
because  of  $\rm  \nu  _{sky}$(3-2)  near  the  325\,GHz  atmospheric
absorption feature prevented sensitive  CO J=3--2 observations of this
system).  The high  $\rm r_{65}$ ratio measured for  the NW nucleus of
this  ULIRG  is  actually  perfectly  compatible  with  a  pure  Orion
``hot-spot''  CO  SLED  (see   2.4),  further  emphasizing  the  truly
extraordinary  levels of ISM  excitation. The  corresponding molecular
gas      mass     is      at     least      $\rm     M_{SF}(H_2)$$\sim
$3.9$\times$10$^9$\,M$_{\odot}$  (assuming  an  Eddington-limited~SF).
{\it       A      Galactic      $       X_{co}$$\sim      $(3-6)\,$\rm
  M_{\odot}$\,(K\,km\,s$^{-1}$\,pc$^2$)$^{-1}$ factor is deduced} from
the LVG solutions with $\rm T_{kin}$$\geq $65\,K, which corresponds to
$\rm M(H_2)$$\sim  $(4.8--9.6)$\times $10$^9$\,M$_{\odot}$.  Thus $\rm
M_{SF}/M_{tot}$$\sim $0.41-0.81, among  the highest such fractions for
the individualy-studied LIRGs (see Table 1).


\subsection{Arp\,55}

This is another clear merger having double nuclei as IRAS\,08572+3915,
but  more widely  separated ($\sim  $12$''$, 9.3\,kpc)  seen  in radio
continuum maps (Condon et al.  1990)  as well as optical and CO J=1--0
interferometric maps  (Sanders et al.   1988).  A tidal  tail emerging
from  the eastern  nucleus (also  the  most gas-rich  one) is  clearly
visible in H$\alpha$ emission (Hattori  et al.  2004).  The two nuclei
have  been separately  detected  in  CO J=3--2  (Leech  et al.   2010,
Paper\,I) while the NE nucleus may have been tentatively detected also
in  CO  J=6--5  (Paper\,I),   with  an  implied  high  CO(6--5)/(3--2)
brightness  temperature  ratio of  $\rm  r_{65/32}$=0.75 indicating  a
highly  excited molecular  ISM phase.   On the  other hand  the global
HCN/CO  J=1--0 brightness  temperature ratio  is $\sim  $0.03  (Gao \&
Solomon 2004), typical  of the molecular gas found  in isolated spiral
disks   and   quite   unlike   the   much  higher   such   ratios   in
merger/starbursts.

Using the global  $\rm r_{21}$ and $\rm r_{32}$,  along with the lower
limit on the  CO/$^{13}$CO J=2--1 ratio (see Paper\,I)  yields a large
range of  possible average ISM  states, all with  n$\sim $(1-3)$\times
$10$^3$\,cm$^{-3}$  and  most with  surprisingly  large $\rm  K_{vir}$
($\sim $70-126).  Restricting $\rm K_{vir}$ to values $\leq $40 yields
a much  narrower range of solutions with  $\rm T_{kin}$=(45-70)\,K and
corresponding  $\rm X_{co}$$\sim  $(0.25-0.8)\,$\rm X_l$.   This still
considerable   $\rm   X_{co}$    range   yields   $\rm   M_{tot}$$\sim
$(2.9-9.2)$\times   $10$^9$\,M$_{\odot}$,   while  $\rm   M_{SF}$$\sim
$0.9$\times $10$^9$\,M$_{\odot}$.   The latter is  $\sim $1/10--1/3 of
the total  molecular gas mass,  leaving ample room for  a SF-quiescent
molecular gas  reservoir. In Arp\,55  however SF-quiescent is  not the
cool  and  mostly gravitationally  bound  gas  found  in the  GMCs  of
isolated  disks.   Indeed the  large  temperatures  and $\rm  K_{vir}$
values needed to fit its  highly-excited CO lines and $\rm R_{21}$$\ga
$13 suggest warm and strongly  unbound gas which, unlike in most other
ULIRGs, contains most of the molecular gas mass. The low HCN/CO J=1--0
ratio  then  may  not  be  due  to  the  prevelance  of  Galactic-type
SF-quiescent  GMCs rather  than to  a  disrupted ISM  state that  has,
momentarily, a low (dense)/(total) molecular gas mass fraction.  Short
periods during which the dense gas reservoir of a LIRG can be strongly
disrupted by  SF feedback while  its IR luminosity remains  intact are
expected  in  strongly evolving  mergers  (Loenen  2009).   In such  a
scenario Arp\,55 would be one of  the few mergers ``caught'' in a very
short  act (see  also Arp\,193).   Interestingly  its HCN(3--2)/(1--0)
ratio  is  $<$0.3 (Juneau  et  al.  2009),  much  lower  than in  some
classical  (U)LIRGs  such as  Arp\,220  and  NGC\,6240  (Greve et  al.
2009).   More sensitive  $^{13}$CO  and higher-J  HCN observations  of
Arp\,55 are necessary to confirm  such a tantalizing picture of global
ISM dynamics.

\subsection{IRAS\,09320+6134 (UGC\,05101)}

Near-IR NICMOS observations show a single extremely red and unresolved
nucleus with  $\theta _s$$\leq $0.22$''$  ($\sim $170\,pc), surrounded
by a strongly perturbed spiral  structure (Scoville at el.  2000) with
a tidal  tail and  a large ring  seen in  the optical (Sanders  et al.
1988b).   The half-light source  radius at  2$\mu $m  is $\sim$450\,pc
while an interferometric CO J=3--2 image reveals a gas distribution of
$\sim $2$''$(1.5\,kpc)  in size (Wilson  et al. 2008).   Its disturbed
morphology is interpreted as the result of an interaction with another
gas-rich spiral (Sanders et al.  1988b) and even with a gas-poor dwarf
galaxy 17$''$  to the southeast  (Majewski et al.  1993).   This ULIRG
contains an AGN luminous in hard X-rays (Imanishi et al.  2003) and is
deeply   dust-enshrouded    along   most   lines    of   sight,   with
N(H)$>$10$^{24}$\,cm$^{-2}$ (Imanishi et al.~2001).

Its CO  SLED betrays the  presence of a high-excitation  gas component
already  from  the  low-J   lines  with  $\rm  r_{21}$=1.23  and  $\rm
r_{32}$=0.93   (Paper\,I).    These  ratios   along   with  the   high
CO/$^{13}$CO J=2--1 ratio of $\rm R_{21}$=18 can be well-fitted by two
ranges   of   LVG   solutions,   namely:   $\rm   T_{kin}$=(35-65)\,K,
n=3$\times$10$^3$\,cm$^{-3}$,     $\rm    K_{vir}$=22,     and    $\rm
T_{kin}$=(100-130)\,K,  n=10$^3$\,cm$^{-3}$,   $\rm  K_{vir}$=4.   The
corresponding  average  $\rm  X_{co}$  factors are  $\rm  X_{co}$$\sim
$0.4\,$\rm  X_l$ and $\rm  X_{co}$$\sim $0.6\,$\rm  X_l$ respectively.
We note that a small range of solutions with $\rm T_{kin}$=(25-30)\,K,
n=$10^4$\,cm$^{-3}$ and $\rm K_{vir}$=40  also exists but are unlikely
to be good representation of the average ISM state in this ULIRG since
$\rm   T_{kin}/T_{dust}$$<$1  (considered  unlikely,   see  Paper\,I).
Moreover their very large $\rm  K_{vir}$ make it even less likely that
these modest  gas temperatures can  correspond to such  highly unbound
dynamical states  where the molecular gas can  be significantly heated
by the  dissipated supersonic turbulence  (see Paper\,I). Nevertheless
adopting this  narrow cooler  gas solution range  would not  make much
difference  when it  comes to  the corresponding  $\rm  X_{co}$ factor
which is $\sim $0.5\,$\rm X_l$.  The corresponding total molecular gas
mass     then     is     $\rm     M_{tot}$$\sim     $(1.9--2.9)$\times
$10$^9$\,M$_{\odot}$     while    $\rm     M_{SF}$$\sim    $1.6$\times
$10$^9$\,M$_{\odot}$.

The large fraction of SF  gas mass in UGC\,05101 ($\sim $(55-84)\%) is
 further corroborated by the high HCN/CO J=1--0 brightness temperature
 ratio of $\sim  $0.2 (Gao \& Solomon 2004)  (for typical spiral disks
 this  ratio is $\sim  $0.02-0.03) isofar  as HCN  J=1--0 is  a linear
 proxy for  dense gas mass  and most of  that gas is involved  in star
 formation. This dense gas reservoir is likely confined into the small
 (near-IR)-bright nucleus fueling  a compact starburst.  Such compact,
 $\sim$100\,pc-sized   gas-disk/starburst  configurations   have  been
 revealed for Arp\,220  (Sakamoto et al. 2008; Mathushita  et al 2004)
 where     extremely     high     extinctions     ($\rm     N(H)$$\geq
 $10$^{25}$\,cm$^{-3}$) correspond to  significant dust optical depths
 even at submm wavelengths (Sakamoto  et al. 2010, Papadopoulos et al.
 2010a).   These  can be  responsible  for  Compton-thick  AGN, and  a
 surpression of high-J CO lines in such systems.  This may be the case
 also for  UGC\,05101 as indicated  by its low CO(6-5)/(1-0)  yet high
 HCN/CO J=1--0 ratio (see Figure 11, Papadopoulos et al. 2010b).

  Despite the good one-phase fit  obtained for the global CO SLED (and
  $\rm R_{21}$) of this ULIRG, the observed CO line emission may still
  be  dominated by  a phase  that  does not  contain the  bulk of  its
  molecular gas  mass.  Indeed, neither  the modest densities  nor the
  potentially highly  unbound average dynamical states  implied by the
  one-phase LVG  fits would be typical of  the HCN-bright star-forming
  phase.   As discussed  in 3.2  the  $\rm X_{co}$  obtained from  the
  global low-J CO SLED may thus  be suitable only for a gas phase that
  does not contain much of the molecular gas mass.  Using the observed
  $\rm r^{(obs)} _{CO/HCN}$=5 in the  context of a 2-phase model along
  with  $\rm  r^{(h)}  _{CO/HCN}$=2.34  (see  3.2)  yields  $\rm  \rho
  ^{(l-h)}  _{co}$=1.136  (Eq. 14).  In  the  absence  of multi-J  HCN
  observations  that  could  constrain   $\rm  X_{HCN}$  we  set  $\rm
  X_{HCN}$=10\,$\rm X_l$, the smallest  value produced by such studies
  (e.g.  Papadopoulos  2007; Gracia-Carpio  et al.  2008).   Thus $\rm
  X^{(h)}  _{co}$=$\rm   (r^{(h)}  _{CO/HCN})^{-1}$$\rm  X_{HCN}$$\sim
  $4.3\,$\rm X_l$, and for $\rm X^{(l)} _{co}$=0.5\,$\rm X_l$ it would
  be   $\rm   X^{(2-ph)}  _{co}$=2.3\,$\rm   X_l$,   which  is   $\sim
  $(4-6)$\times $  higher than that deduced from  one-phase LVG models
  of the  low-J CO  SLED. Multi-J observations  of HCN or  other heavy
  rotor  molecules (e.g.   CS, HCO$^+$)  can confirm  such potentially
  high masses of molecular gas at high densities.


\subsection{NGC\,3310}

The  young  and intense  star  formation  activity  of this  UV-bright
 galaxy is on  par with M\,82, and  the likely result of  a recent merger
 with  a smaller  galaxy (Balick  \&  Heckman 1981;  Concelice et  al.
 2000;  Elmegreen et  al.   2002). Its  gas  and dust  are warmed  and
 disrupted  by  the presence  of  the  starburst  in its  central  1-2
 kpc. Tidal  HI tails, unusually large HI  velocity dispersions ($\sim
 $40\,km\,s$^{-1}$) for a spiral  galaxy (Kregel \& Sancisi 2001), and
 large  velocity offsets (up  to 150\,km\,s$^{-1}$)  between molecular
 cloud  and adjacent  H\,II  regions (Kikumoto  et  al. 1993)  provide
 further evidence for a dynamically disturbed gas typical of mergers.

This galaxy  has been  extensively studied  by Zhu et  al.  2009  as a
template of  a highly-excited ISM,  indicated by $\rm  r_{21, 32}$$>$1
ratios and  warm IR ``colors'' ($\rm  S_{100\mu m}/S_{60\mu m}$=1.28).
Despite its  low IR  luminosity ($\rm L^{(*)}  _{IR}$$\sim $2.7$\times
$10$^{10}$\,L$_{\odot}$,  Paper\,I) its  global $\rm  r_{21}$=1.47 and
$\rm r_{32}$=1.21 ratios surpass those of Orion cloud SF ``hotspots'',
and  are among  the highest  observed  in our  sample.  Its  proximity
allowed  spatially   resolved  studies  of  its   molecular  gas  with
single-dish  telescopes  obtaining  beam-matched observations  of  the
CO(3--2)/(1--0) ratio (see  Zhu et al. 2009 for  details).  These high
global CO  line ratios  of NGC\,3310  can be fitted  well by  a single
phase provided $\rm T_{kin}$$\geq  $35\,K ($\sim $$\rm T_{dust}$), but
remain highly  degenerate over the $\rm [T_{kin},  n, K_{vir}]$ domain
up to  the highest  temperature searched (150\,K).   Interestingly for
$\rm  T_{kin}$=(35-50)\,K  a  class  of  dense  (n$\sim  $(1-3)$\times
$10$^4$\,cm$^{-3}$)  {\it and}  strongly  unbound ($\rm  K_{vir}$$\sim
$20-40)  states are  possible, though  only higher-J  and/or $^{13}$CO
line  observations  could  reduce  the  wide  degeneracy  of  the  LVG
solutions and designate such  phases as those most likely representing
the   actual  ISM   conditions.   Other   solution  ranges   are: $\rm
T_{kin}$=(70-90)\,K,    n=3$\times    $10$^3$\,cm$^{-3}$   and    $\rm
K_{vir}$=22  while   near-virial  solutions  can  be   found  at  high
temperatures         $\rm         T_{kin}$=95\,K,        (140-145)\,K,
n=(1-3)$\times$10$^4$\,cm$^{-3}$ and $\rm  K_{vir}$=4, 2.2.  The large
degeneracy  of  possible  average   ISM  conditions  translates  to  a
significant  one for  the  $\rm X_{co}$  factors,  ranging from  $\sim
$0.4\,$\rm   X_l$   (for  the   highly   unbound   states  with   $\rm
K_{vir}$=22-40)  and   up  to  $\sim  $(1.2-2.2)$\rm   X_l$  (for  the
near-virial states with $\rm K_{vir}$=4, 2.2).

For   $\rm  X_{co}$=(0.4-2.2)\,$\rm   X_l$   the  corresponding   $\rm
M_{tot}$=(0.3-1.5)$\times     $10$^{8}$\,M$_{\odot}$     while    $\rm
M_{SF}$=1.1$\times  $10$^8$\,cm$^{-3}$. The latter  seems to  rule out
the lowest $\rm X_{co}$ values while  even for the highest ones it is:
$\rm M_{SF}$/$\rm M_{tot}$$\sim $0.73, i.e.  most of the molecular gas
in  NGC\,3310   is  directly  involved  in  star   formation  (for  an
Eddington-limited  SF  process), leaving  little  room  for a  massive
low-excitation  molecular   gas  reservoir.   Interestingly   the  LVG
solutions corresponding  to the highest $\rm X_{co}$  values are those
indicative  of a  SF phase  with n$\sim  $3$\times $10$^4$\,cm$^{-3}$,
$\rm  K_{vir}$=2.2 and large  $\rm T_{kin}$=(140-145)\,K.   We caution
though that all this pertains only  to the inner $\sim $40$''$ of this
galaxy  where  the starburst  takes  place,  a  massive molecular  gas
reservoir with low  densities, highly unbound gas motions,  and low CO
line  excitation  can still  exist  beyond  it,  a situation  actually
encountered in  M\,82 in the  form large scale molecular  gas outflows
(Weiss et al. 2005).


\subsection{IRAS\,10565+2448}

This is a ULIRG with H\,II region-type line ratios (Armus, Heckman, \&
Miley  1989) implying  an optical  spectrum dominated  by  young stars
rather  than an  AGN.  Nevertheless  a weak  contribution from  an AGN
seems  necessary in  order  to  fit its  IR/submm  dust continuum  SED
(Farrah et al.   2003). Early r-band (6550\AA) imaging  has shown this
object  to  be a  possible  triple  merger  galaxy system  (Murphy  et
al. 1996),  while near-IR imaging with  the HST NICMOS  camera shows a
luminous  primary galaxy  interacting with  a much  fainter  one $\sim
$8$''$ ($\sim $6.7\,kpc) to the southeast (Scoville et al.  2000). The
primary galaxy is  compact with a half-light source  diameter at 2$\mu
$m of  $\sim $650\,pc,  and the only  source where luminous  CO J=1--0
line  emission   is  detected  (DS98)  with  a   diameter  of  1.5$''$
(1.25\,kpc). Moreover this ULIRG  along with VII\,Zw\,031 and Arp\,193
are  the only  galaxies  where  kinematic models  of  the CO  emission
indicate a  rotating ring rather  than a filled disk  gas distribution
(DS98).

Its low-J CO lines (J=1--0,  2--1, 3--2) indicate a well-excited low-J
CO  SLED  ($\rm  r_{21}$=1.06,   $\rm  r_{32}$=0.80)  expected  for  a
starburst while the large $\rm R_{10}$=15 and $\rm R_{21}$$\ga $18 are
typical for merger systems. Two ranges of one-phase LVG solutions that
can   be  found   for  $\rm   T_{kin}/T_{dust}$$\geq  $1   (with  $\rm
T_{dust}$=40\,K) namely: $\rm [T_{kin}, n, K_{vir}]$=$\rm [(40-50)\,K,
3\times10^3\,cm^{-3},  7]$   and  $\rm  [T_{kin},   n,  K_{vir}]$=$\rm
[(80-140)K, 10^3\,cm^{-3},  4]$, with the best ones  obtained for $\rm
T_{kin}$=45\,K and  115\,K respectively.   Both also reproduce  the CO
J=6--5 line luminosity with $\rm r^{(lvg)} _{65}$$\sim $0.15 (observed
value: 0.18$\pm $0.055).

  The resulting $\rm X_{co}$ factors are: $\rm X_{co}$=0.75\,$\rm X_l$
(for   $\rm   T_{kin}$=45\,K)    and   0.60\,$\rm   X_l$   (for   $\rm
T_{kin}$=115\,K),  yielding  a  total   molecular  gas  mass  of  $\rm
M_{tot}$=(3.8-4.80)$\times  $10$^9$\,M$_{\odot}$.   The  minimum  such
mass  needed  for  Eddington-limited  SF  is  $\rm  M_{SF}$=2.6$\times
$10$^9$\,M$_{\odot}$, which is $\sim  $54\%-68\% of $\rm M_{tot}$.  As
in UGC\,05101 the aforementioned  large fraction of molecular gas mass
directly   involved   in  star   formation   in  IRAS\,10565+2448   is
corroborated by  a high HCN/CO J=1--0 brightness  temperature ratio of
$\rm r_{HCN/CO}$$\sim  $0.17 (Gao \&  Solomon 2004).  However  all the
1-phase LVG solutions obtained are hardly representative of a dominant
dense  and  self-gravitating  gas   phase  but  more  typical  of  the
``cloud-envelope''  warm and  diffuse gas  often found  in the  ISM of
mergers. Moreover  all one-phase LVG  fits fail to reproduce  the $\rm
R_{21}$$\ga $18  lower limit, with $\rm R^{(lvg)}  _{21}$=12 being the
highest  value obtained  (for the  $\rm T_{kin}$=115\,K  solution).  A
2-phase model with  an assumed (h)-phase aleviates this  and yields an
(l)-phase  with  $\rm  T_{kin}$=65\,K, n=$10^3$\,cm$^{-3}$,  and  $\rm
K_{vir}$=13  that   has  $\rm   R_{21}$$\sim  $20  and   $\rm  X^{(l)}
_{co}$=0.4\,$\rm  X_l$.   The  computed  $\rm  X^{(2-ph)}  _{co}$$\sim
$0.75\,$\rm X_l$ is within the range of the one-phase values, a result
of the low  $\rm X^{(l)} _{co}$, the large  luminosity contribution of
the  (l)-phase ($\rm  \rho^{(l-h)} _{co}$$\sim  $4.5), and  the modest
$\rm X^{(h)} _{co}$=2.2\,$\rm X_l$ of the (h)-phase (see~2.4).

Setting constraints on the (h)-phase  and $\rm X^{(h)} _{co}$ by using
 the  observed CO  J=3--2, 6--5  and $^{13}$CO  J=1--0  lines (assumed
 emanating mostly from that phase), does not change the aforementioned
 picture,  at least when  it comes  to the  values of  $\rm X^{(2-ph)}
 _{co}$.  Indeed  while the corresponding LVG models  do recover dense
 gas solutions (n$\sim  $10$^4$\,cm$^{-3}$), substantial $\rm K_{vir}$
 values  ($\sim$13) keep  the corresponding  $\rm X^{(h)}  _{co}$ well
 below Galactic.   In this model the (l)-phase  contribution, and thus
 the influence of  $\rm X^{(l)} _{co}$ on the  $\rm X^{(2-ph)} _{co}$,
 is  negligible.  We  must stress  however that  other high-J  CO line
 observations, and crucially multi-J HCN observations are necessary to
 confirm this  picture. If luminous  such lines were to  be discovered
 for this  ULIRG, they could  substantially raise $\rm  X^{(h)} _{co}$
 and thus $\rm X^{(2-ph)} _{co}$ as discussed in 3.2.

\subsection{Arp\,299}

This is a  spectacular merger of two galaxies,  IC\,694 and NGC\,3690,
with luminous CO J=1--0 emission detected in the nuclei of both (which
are  22$''$  --  4.4\,kpc,  apart),  and the  interface  between  them
(Sargent et  al.  1987;  Sargent \& Scoville  1991).  A long  HI tidal
tail is  also seen  extending out to  $\sim $180\,kpc (Hibbard  \& Yun
1999).   Along  with  the   Antennae  galaxy  this  (U)LIRG  has  been
considered an early template for mergers, documenting their ability to
rapidly funnel  the gas in the  disks of the  progenitors into compact
regions (Sargent \& Scoville 1991).  Indeed the two nuclei in Arp\,299
contain $\sim  $80\% of its  molecular gas reservoir,  fueling intense
star-formation,  while the  nucleus of  IC\,694 contains  also  an AGN
(Sargent \& Scoville 1991 and  references therein), and is the primary
source  of  the  bolometric  luminosity  of  this  spectacular  system
(Charmandaris et al. 2002).  The most detailed molecular line study of
this  template  merger system  is  the  interferometric  study of  CO,
$^{13}$CO and  HCN J=1--0  line emission by  Aalto et al.   1997 which
finds most of the HCN-bright emission located in compact regions ($\la
$310\,pc) in  the two  nuclei of the  merging galaxies,  and unusually
large  CO/$^{13}$CO  line ratio  variations  (from  $\sim  $60 in  the
IC\,694 nucleus,  to $\sim$5-10 in  its disk).  Actually the  state of
the  gas in  the IC\,694  nucleus with  the largest  HCN/CO  {\it and}
CO/$^{13}$CO ratios  {\it exemplifies the effects  of highly turbulent
  environments found in mergers,} with  much of the molecular gas mass
``resettled'' at high densities  (boosting the HCN/CO ratio) while the
violent disruption of  GMCs creates a diffuse warm  and highly unbound
``envelope''  phase  dominating  the  low-J  CO  lines  (boosting  the
CO/$^{13}$CO ratios).

While we have no CO  J=6--5 measurements for this galaxy (too extended
for  the  narrow JCMT  beam  at 690\,GHz)  our  CO  J=3--2 and  J=4--3
measurements indicate  high CO line excitation. A  one-phase LVG model
of the global CO  (3--2)/(1--0), (4--3)/(1--0) and CO/$^{13}$CO J=1--0
line   ratios   gives   the   best  fit   for   $\rm   T_{kin}$=30\,K,
n=10$^4$\,cm$^{-3}$  and  $\rm  K_{vir}$=40 which  indicates  strongly
unbound gas motions. Other average  gas states are also possible (e.g.
$\rm   T_{kin}$=40\,K,    n=3$\times   $10$^3$\,cm$^{-3}$   and   $\rm
K_{vir}$=22) but  all have  $\rm K_{vir}$$\ga $22.   The corresponding
$\rm   X_{co}$  factor   over   the  LVG   solution   range  is   $\rm
X_{co}$=(0.35-0.42)\,$\rm X_l$,  with the largest  value corresponding
to   the  dense   gas  solution.    These  yield   $\rm  M_{tot}$$\sim
$(1-1.2)$\times   $10$^{9}$\,M$_{\odot}$   while   $\rm   M_{SF}$$\sim
$1.9$\times  $10$^{9}$\,M$_{\odot}$, which  clearly favors  the higher
$\rm X_{co}$ values  and indicates that $\sim $100\%  of the molecular
gas mass  is associated with SF  sites. Larger values  of $\rm X_{co}$
(and $\rm M_{tot}$)  remain possible however if most  of the molecular
gas mass in  Arp\,299 is contained in the  HCN-bright phase with large
$\rm X_{HCN}$ factors.  Multi-J CO, $^{13}$CO and HCN, H$^{13}$CN line
imaging would be  ideal in determining this as  a function of position
within a  template merger whose  ISM conditions seem to  encompass the
full range possible, from quiescent disks to compact starburst nuclei.

\subsection{IRAS\,12112+0305}

This ULIRG consists  of a strongly interacting pair  of galaxies whose
nuclei are $\sim  $2.9$''$ ($\sim $4\,kpc) apart (Carico  et al. 1990;
Scovile et  al.  2000), while  tidal tails and ``plumes''  are visible
both in  optical (Surace  et al.  2000)  and near-IR (Scoville  et al.
2000) wavelenghts.   CO J=1--0  interferometry finds  the  bulk ($\sim
$75\%) of  the emission emanating from  the NE nucleus  of this system
(Evans et  al. 2002), which is  unresolved with a size  of $\la $2$''$
($\la $2.7\,kpc),  and also  the only one  tentatively detected  in CO
J=6--5  (see Paper\,I).   Thus  our one-phase  LVG radiative  transfer
modeling pertains only to the  NE nucleus of this strongly interacting
system.  The observed $\rm r_{32}$=1.58 ratio is the second highest in
our sample  (the highest  one found in  another double  nuclei system:
IRAS\,22491-1808)  and  indicates   the  presence  of  high-excitation
molecular  gas.  This  surpasses  the corresponding  ratio  of the  SF
``hot-spots''   in  the   Orion  molecular   cloud,  while   the  $\rm
r_{32}$$>$$\rm   r_{21}$  inequality  is   another  indication   of  a
qualitatively different  CO SLED, irreducible  to a mixture of  SF and
SF-quiescent gas (see Paper\,I).

All one-phase LVG solutions  for IRAS\,12112+0305 indicate average gas
states    with    n$\sim   $(3$\times    $10$^4$-10$^{6}$)\,cm$^{-3}$.
Nevertheless  lack  of  $^{13}$CO  line  observations  leaves  serious
degeneracies with  e.g.  $\rm [T_{kin}(K),  n(cm^{-3}), K_{vir}]$=[60,
  10$^5$,  13],  [125,  10$^4$,   1]  being  nearly  equivalent.   The
corresponding  $\rm  X_{co}$$\sim  $(1.5-5.3)\,$\rm  X_l$,  with  {\it
  Galactic  values of $\sim  $(4-5)\,$ X_l$  obtained for  the warmest
  (=(120--150)\,K) and  densest (=(10$^5$-10$^6$)\,cm$^{-3}$) states}.
The presense of  such high densities is corroborated  by a high HCN/CO
J=1--0 ratio  of $\rm r_{HCN/CO}$=0.16 (HCN from  Gracia-Carpio et al.
2008, CO  from Paper\,I), though  without offering any  constraints on
the  temperatures of  the  dense gas  or  its all-important  kinematic
state.   However  even the  warmest  and  densest one-phase  solutions
remain  rather poor  fits of  the ``hot''  $\rm r_{32}$$\sim  $1.6 and
comparatively ``cooler''  $\rm r_{21}$$\sim $0.9  ratio.  Indicatively
such  a solution (n=10$^6$\,cm$^{-3}$)  with $\rm  T_{kin}$=150\,K and
$\rm K_{vir}$=4 yields $\rm  r_{32}$$\sim $$\rm r_{21}$$\sim $1 as the
high  density, and  modest  $\rm K_{vir}$  act  to keep  the CO  lines
well-thermalized but also optically thick up to high-J~levels.

A  two-phase model of  the global  CO line  ratios in  this intriguing
ULIRG cannot  use an Orion-derived  (h)-phase CO SLED as  the observed
$\rm  r_{32}$$>$$\rm   r^{(h)}  _{32}$(Orion).   Using   only  the  CO
(3-2)/(2-1)  ratio  $\rm r_{32/21}$=$1.76\pm  0.48$  to constrain  the
high-excitation (h)-phase  yields: n=3$\times $10$^4$\,cm$^{-3}$, $\rm
K_{vir}$=22, and $\rm  T_{kin}$$\ga $140\,K (for $\rm T_{kin}$=140\,K,
$\rm r^{(lvg)} _{32/21}$=1.28$\sim  $$\rm r_{32/21}$-$\sigma$), with a
fit  that  keeps  improving  past  $\rm  T_{kin}$=200\,K  (where  $\rm
r_{32/21}$=1.41).    This  is   expected  since   $\rm  r_{32/21}$$>$1
indicates a thermalized J=3--2 line (and thus n$\ga
$$\rm n_{crit,3-2}$$\sim  $10$^4$\,cm$^{-3}$), with low/modest optical
depths (ensured  in part by the  large $\rm K_{vir}$),  {\it and} warm
enough gas to populate the J=3  level.  The molecular gas mass of this
phase   can  be  estimated   from  $\rm   L^{(h)\,'}  _{CO(1-0)}$=$\rm
(1/r^{(h)} _{32})$$\rm  L^{'} _{CO(3-2)}$$\sim $4$\times $10$^9$\,$\rm
L_l$ (where $\rm r^{(h)} _{32}$$\sim $4) and the computed $\rm X^{(h)}
_{co}$$\sim $1\,$\rm  X_l$, thus $\rm  M_{h-ex}$(H$_2$)$\sim $4$\times
$10$^9$\,M$_{\odot}$,  and   similar  to   that  needed  to   fuel  an
Eddington-limited      SF     in      this      ULIRG     of      $\rm
M_{SF}$(H$_2$)$\sim$5$\times $10$^9$\,M$_{\odot}$.   For the (l)-phase
$\rm   L^{(l)\,'}  _{CO(1-0)}$$\sim  $6.1$\times   $10$^9$\,$\rm  L_l$
luminosity there are no  constraints on its corresponding $\rm X^{(l)}
_{co}$ factor.  Setting  $\rm X^{(l)} _{co}$=(0.5-5)\,$\rm X_l$ yields
$\rm X^{(2-ph)}  _{co}$=(0.7-3.4)\,$\rm X_l$ and  a corresponding $\rm
M_{tot}$$\sim  $(7.1-34.3)$\times $10$^{9}$\,M$_{\odot}$.  Multi-J HCN
and H$^{13}$CN as well as $^{13}$CO line observations are critical for
reducing  the degeneracies of  the radiative  transfer models  for the
dense and the low-excitation gas, and their $\rm X_{co}$~factors.

\subsection{Mrk\,231}

This is an archetypal ULIRG/QSO and the most IR luminous galaxy in the
Revised IRAS Bright Galaxy Survey  (RBGS; Sanders et al. 2003). It has
a compact  nucleus, surrounded by  irregular incomplete rings  of star
formation along  with a  small tidal arm  that contains  numerous blue
star-forming 'knots' (Farrah et  al. 2003 and references therein).  It
has large  amounts of molecular  gas in a compact  (0.85$''$, 700\,pc)
nearly  face-on disk  (Bryant  \& Scoville  1996;  DS98) which  allows
nearly  unobscured  view  towards  an  optically  luminous  AGN  which
classifies this object also as  a Seyfert\,1 galaxy. Mrk\,231 was also
the first ULIRG for which high-J CO lines (J=4--3, 6--5) were detected
(Papadopoulos et  al. 2007)  while recent SPIRE/FTS  observations with
the HSO  revealed luminous  high-J CO lines  up to  J=13--12 emanating
from AGN-induced X-ray Dissociated Regions (XDRs) (van der Werf et al.
2010).  Its  well-excited high-J CO  lines, the largest  HCN/CO J=1--0
ratio   ($\rm   r_{HCN/CO}$$\sim    $0.29)   observed   among   ULIRGs
(Papadopoulos et al. 2007), and  a very high CO/$^{13}$CO J=2--1 ratio
exemplify  the  extraordinary   {\it  average}  molecular  gas  states
possible in such systems (similar  to the IC\,694 core in the Arp\,299
merger  system).   Finally  the   availability  of  HCN  J=4--3,  1--0
measurements along with  the high-J CO lines allowed  a detailed study
of its dense gas phase  (Papadopoulos et al.  2007), and make Mrk\,231
a    good   case    study    for   the    effects    of   the    large
M(n$>$10$^4$\,cm$^{-3}$)/M$_{\rm  tot}$ fractions  on the  global $\rm
X_{co}$ of ULIRGs.

 A  one-phase LVG  fit predictably  fails to  converge on  any average
 state  as it  cannot accomodate  both  the high  densities needed  to
 excite the  CO J=3--2,  4--3, and 6--5  lines {\it and}  maintain the
 very low  optical depths needed  for reproducing the very  large $\rm
 R_{10}$=47  ratio.  Indicatively  the best  one-phase  LVG solutions:
 $\rm T_{kin}$=(115-150)\,K,  n$\sim $(300-10$^3$)\,cm$^{-3}$ and $\rm
 K_{vir}$=7-40    yield   $\rm    r^{(lvg)}    _{21}$$\sim   $0.83-1.1
 (obs=0.89$\pm    $0.12),   $\rm   r^{(lvg)}    _{32}$$\sim   $0.5-0.7
 (obs=0.72$\pm   $0.13),    $\rm   r^{(lvg)}   _{43}$$\sim   $0.22-0.3
 (obs=0.8$\pm $0.2) and $\rm r^{(lvg)} _{65}$$\sim $0.03 (obs=0.42$\pm
 $0.12,  but  not  used  for   the  LVG  fit),  while  $\rm  R^{(lvg)}
 _{21}$=25-32 (obs=47$\pm $16).  Thus the average ISM state deduced by
 a one-phase  LVG model can be  responsible for up to  J=3--2 (and the
 $^{13}$CO  J=2--1) line  emission but  severely underpredicts  the CO
 J=4--3,   and  6--5  line   luminosities.   The   corresponding  $\rm
 X_{co}$$\sim  $(0.25-0.45)\,$\rm  X_l$ is  low  but  within what  was
 thought as  the ULIRG-appropriate range (for Mrk\,231  DS98 give $\rm
 X_{co}$$\sim $(0.7-0.8)\,$\rm X_l$).   These yield $\rm M_{tot}$$\sim
 $(1.8-3.2)$\times   $10$^9$\,M$_{\odot}$,    but   a   minimum   $\rm
 M_{SF}$=5.5$\times  $10$^9$\,M$_{\odot}$  clearly  favors the  higher
 $\rm  M_{tot}$  (and  $\rm  X_{co}$)  among  those  computed with the
 one-phase model.

 For a 2-phase model we use the results by Papadopoulos et al. 2007 of
 warm ($\sim  $(85-140)\,K), diffuse gas  ($\sim $300\,cm$^{-3}$) that
 is  highly  unbound  ($\rm  K_{vir}$$\gg$1),  and  dense  gas  ($\sim
 $(1-3)$\times     $10$^4$\,cm$^{-3}$,    $\rm    T_{kin}$=(40-70)\,K)
 responsible for  the bright CO  J=6--5, HCN J=1--0, 4--3  lines.  The
 latter  phase is  virialized with  $\rm  X_{HCN}$$\sim $(10-25)\,$\rm
 X_l$ (for our  numerical factor in Equation 3 rather  than the one in
 Papadopoulos et  al.  2007) and $\rm  r^{(h)} _{CO/HCN}$$\sim $2-2.54
 (high value  corresponding to the lower $\rm  X_{HCN}$ values).  Thus
 $\rm  X^{(h)}  _{co}$=(1/$\rm  r^{(h)} _{CO/HCN}$)$\rm  X_{HCN}$$\sim
 $(3.94-12.5)\,$\rm  X_l$ while  for the  diffuse and  HCN-dark phase:
 $\rm X^{(l)} _{co}$$\sim $(0.4-1)\,$\rm  X_l$.  For an $\rm r^{(obs)}
 _{CO/HCN}$=3.45   we  then  find   $\rm  \rho   ^{(l-h)}  _{co}$$\sim
 $0.36-0.73 (yielding $\rm X^{(h)} _{co}$=(3.94-12.5)\,$\rm X_l$), and
 an $\rm  X^{(2-ph)} _{co}$$\sim  $(3-7)\,$\rm X_l$ (for  the smallest
 $\rm  X^{(l)} _{co}$).   Thus  {\it a  Galactic  $\rm X_{co}$  factor
   applies in  Mrk\,231,} a result of  a large mass of  dense gas with
 high $\rm X^{(h)} _{co}$ values,  themselves a result of the high gas
 densities {\it and} their virial dynamical~states.


\subsection{Arp\,193}

This galaxy is a template  of a young merger-induced starburst with an
age  of few$\times$10$^7$\,yr  and  near-IR imaging  showing a  highly
dust-enshrouded  nucleus, and  tidal  tails (Smith  et  al. 1995).   A
disturbed highly inclined disk is  also seen in NICMOS near-IR imaging
with reddening increasing towards the  NW within the disk (Scoville et
al.   2000), and  a similar  extinction distribution  implied by  a CO
J=1--0 interferometer  map (Bryant \& Scoville  1999).  Extended radio
continuum  emission  ($\sim  $3$''$$\times  $3.7$''$) is  seen  at  cm
wavelengths with two nuclei $\sim $1$''$ apart and a spectral index of
$\alpha  $$\sim  $0.6   ($\rm  S_{\nu}$$\propto  $$\nu  ^{-\alpha}$),
consistent with SF-related  non-thermal synchrotron emission (Clements
\& Alexander 2004). The latter study also finds significant decoupling
of  the HI  and CO-bright  H$_2$ gas  motions and  distributions.  The
young  age  of  this  disk-disk merger/starburst  is  corroborated  by
several  extra-planar  highly  luminous  star  clusters  (Scoville  et
al. 2000) and many bright H\,II regions (Clements \& Alexander~2004).

High S/N CO J=1--0 and 2--1 interferometric imaging confirmed a highly
inclined  2.8$''$$\times $0.8$''$ ($\sim  $1.3\,kpc$\times $0.37\,kpc)
gas disk whose CO emission is  best fit by a molecular ring (DS98).  A
compact southeast CO source coinciding with the secondary cm continuum
emission  peak in  Arp\,193 has  the properties  of a  giant molecular
core, and is  much warmer and denser than the rest  of the disk. Along
with a region in Mrk\,273, and the western nucleus of Arp\,220 (DS98),
it  is one  of the  most  extreme SF  regions in  the local  Universe,
forming stars very near or at the Eddington-limit set by the radiation
pressure from young  massive stars on the dust  of the dense molecular
gas        (i.e.         $\rm        L_{IR}/M_{dense}$(SF-region)$\sim
$(250-500)\,L$_{\odot}$/M$_{\odot}$).  One-phase  LVG solutions to its
CO line  ratios yield low-density solutions  with n$\sim $(1-3)$\times
$10$^3$\,cm$^{-3}$  but  large   temperature  degeneracies  with  $\rm
T_{kin}$=40\,K,   (60-75)\,K,  (80-115)\,K   and   corresponding  $\rm
K_{vir}$=22,   4  and   13   values.   Warmer   solutions  with   $\rm
T_{kin}$=(120-150)\,K and  even larger $\rm  K_{vir}$(=40) values also
exist.   Considering only the  gas states  with $\rm  K_{vir}$$\la $20
yields   $\rm   X_{co}$$\sim    $(0.35-0.75)\,$\rm   X_l$   and   $\rm
M_{tot}$$\sim $(1.6-3.5)$\times $10$^9$\,M$_{\odot}$, with the highest
$\rm X_{co}$  (and $\rm M_{tot}$)  values corresponding to  the states
with the smallest  $\rm K_{vir}$.  The minimum molecular  gas mass for
an Eddington-limited  star formation is  $\rm M_{SF}$$\sim $0.9$\times
$$10^9$\,M$_{\odot}$.

  A  young starburst  and  the  strong SF  feedback  expected in  such
  extreme SF events  could be responsible for Arp\,193  being the only
  LIRG  where,  despite  a   large  SF-powered  IR  luminosity  ($\sim
  $2.2$\times $10$^{11}$\,L$_{\odot}$) and  HCN/CO J=1--0 ratio ($\sim
  $0.044, with HCN J=1--0 from  Carpio-et al. 2008, and CO J=1--0 from
  Paper\,I),  the average state  of its  molecular ISM,  is compatible
  with  a  complete  absence   of  a  dense  molecular  gas  reservoir
  (Papadopoulos 2007).  This is indicated by its low HCN (4--3)/(1--0)
  and (3--2)/(1--0) line ratios  of $\rm r_{43}$$\la $0.12 (3$\sigma$)
  and $\rm r_{32}$=0.22$\pm $0.03 (Papadopoulos 2007; Gracia-Carpio et
  al.  2008).   Using them as constraints yields  a narrow temperature
  range  of   dense  gas  solutions   with  $\rm  T_{kin}$=(50-60)\,K,
  n=3$\times $10$^4$\,cm$^{-3}$ and $\rm K_{vir}$=14 (rather large for
  a  dense gas  phase).  The  widest range  of solutions  is  at: $\rm
  T_{kin}$=(65-95)\,K,   n=3$\times    $10$^3$\,cm$^{-3}$   and   $\rm
  K_{vir}$$\sim  $1.  Unlike the  former set  of solutions  the latter
  reproduces also the CO(3-2)/HCN(1-0)  line ratio of $\sim $17$\pm $4
  (which can be reasonably assumed  as an additional constraint on the
  denser SF gas phase\footnote{the  non-detection of CO J=6--5 in this
    LIRG is likely  due to a pointing offset of  the JCMT}). Using the
  upper  limit of  the HCN(4--3)/(1--0)  ratio as  a detection  (as to
  obtain  the   densest  possible  LVG  solutions)   only  pushes  the
  aforementioned    set    of   dense    gas    solutions   to    $\rm
  T_{kin}$=(90-105)\,K,  while the lower  density/virial ones  are now
  found at $\rm T_{kin}$=(110-150)\,K  (both sets of solutions can now
  reproduce the CO(3-2)/HCN(1-0) ratio).

For a 2-phase  model where the HCN-bright gas is  set as the (h)-phase
we  obtain $\rm X^{(2-ph)}  _{co}$$\sim $(0.70-1.75)\,$\rm  X_l$ (with
the low values corresponding  to the dense/unbound solutions). This is
is  $\sim   $(0.9-2)$\times  $$\rm  X_{co}$(DS98),   and  yields  $\rm
M_{tot}$$\sim  $(3.3-8.1)$\times $10$^9$\,M$_{\odot}$.   Thus Arp\,193
stands  as a  case  where multi-J  line  observations of  high-density
tracers such as  HCN do not significantly change  the picture obtained
by the low-J CO lines  regarding the global $\rm X_{co}$ factor.  This
may be  because this LIRG is a  rare merger, ``caught'' in  the act of
having momentarily  dispersed/consumed its dense gas  mass while still
having  large SF-powered  IR luminosities,  a brief  state  of affairs
expected  because  of  SF-feedback  during  the  evolution  of  merger
starbursts (Loenen 2009).  We  note however high-resolution CO imaging
revealing the ``excess''  CO J=1--0 line emission of  the (l)-phase to
be a  cold SF-quiescent phase can  still push the  global $\rm X_{co}$
towards near-Galactic values via a higher $\rm X^{(l)} _{co}$.

\subsection{NGC\,5135}

Optical and near-IR  {\it HST} observations reveal this  Sy2 galaxy to
be a grand design spiral with star formation across its arms but being
most intense in its central  $\sim $1\,kpc (Martini et al. 2003).  The
inner 1.6\,kpc  is also where most  of its cm  radio continuum emerges
(Condon et al.  1996).  Optical  and UV studies by Gonzalez Delgado et
al.  (1998) reveal a young  nuclear starburst with gas likely provided
by a  bar instability.  Its HCN/CO  J=1--0 line ratio  of $\sim $0.087
(Gao  \&  Solomon  2004), while  not  as  high  as  in ULIRGs,  it  is
nevertheless $\sim  $(3-4)$\times $ higher than in  spiral disks. This
indicates a significant  dense gas mass fraction, even  in the absence
of a strong merger and/or galaxy interaction. Lack of the latter makes
the large $\rm R_{10}$=26 and $\rm R_{10}$=13 measured in NGC\,5135 an
exception as these are  typical mostly in mergers.  Interestingly this
galaxy is known for powerful  gas outflows of hundreds of km/s, driven
by its young starburst (Gonzalez Delgado et al. 1988).

One-phase  models  of its  CO  line  ratios  yield $\rm  [T_{kin},  n,
  K_{vir}]$=$\rm  [(30-40)K, 3\times10^3\,cm^{-3},  22]$,  with warmer
(but  also  more  degenerate)  solutions  such  as:  $\rm  [(60-110)K,
  10^3\,cm^{-3}, 13]$  also remaining possible.  For  the entire range
of solutions  the corresponding $\rm  X _{co}$$\sim $(0.35-0.45)\,$\rm
X_l$,    similar   to   $\rm    X^{(thin)}   _{co}$(LTE)    for   $\rm
T_{kin}$=(30-70)\,K  (the  temperatures  of  the best  solutions),  as
expected for the low/modest optical  depths of the LVG solutions.  For
these    $\rm    X_{co}$    values   $\rm    M_{tot}$=(1.1-1.4)$\times
$10$^9$\,M$_{\odot}$ while $\rm M_{SF}$=3$\times $10$^8$\,M$_{\odot}$,
which corresponds to $\sim $20\%-30\% of the total molecular gas mass.
Nevertheless, with  the distribution of  the molecular gas  unknown, a
Galactic $\rm X_{co}$ for disk-distributed gas not intimately involved
with star-formation remains a distinct possibility. In such a case the
true $\rm X_{co}$ and $\rm M_{tot}$ can easily be $\sim $(10-15) times
higher, reducing $\rm M_{SF}/M_{tot}$  to only few\%, and more typical
of  spiral disks.  As described  in 3.4  CO line  imaging observations
rather than global CO SLEDs can  be used to determine the existence of
such a cold molecular gas disk.

\subsection{Mrk\,273}

This extraordinary ULIRG shows up  as a double-nuclei system in NICMOS
near-IR observations where the  nuclei are only $\sim $1$''$ (740\,pc)
apart (Scoville  et al.  2000).   A strikingly long tidal  tail ($\sim
$1$'$, 44\,kpc)  as well  as a decoupling  of stellar and  gas motions
(Tacconi  et  al.  2002)  are  the  telltale  signatures of  a  strong
merger. Its  northern nucleus contains  a compact hard  X-ray luminous
Sy2 AGN (Xia et al.  2002), but the dust emission SED of this ULIRG is
dominated by a starburst (Farrah  et al.  2003).  This galaxy actually
contains the most extreme starburst  region among all ULIRGs imaged in
the  DS98   study,  with  an   IR  luminosity  of   $\sim  $6.7$\times
$10$^{11}$\,L$_{\odot}$  emanating   from  its  core   of  only  $\sim
$0.35$''$$\times    $($<$0.2$''$)    (260\,pc$\times    $($<$149\,pc))
diameter. This  makes the compact  nuclear CO line source  of Mrk\,273
akin to  the west nucleus of  Arp\,220, and along  with Arp\,193 these
three ULIRGs host the most  prodigious SF events in the local Universe
(DS98).  For  the core of Mrk\,273  the IR brightness  is: $\rm \sigma
_{IR}$$\sim     $$\rm    L_{IR}$/($\rm     \pi     R^2    _{co})$$\sim
$10$^{13}$\,L$_{\odot}$\,kpc$^{-2}$,  i.e.   the  Eddington limit  for
radiation-pressure-regulated star  formation (Thompson 2009).   In the
framework of  an Eddington-limited SF  this makes the  compact nuclear
starburst  in Mrk\,273  a  maximal  SF event,  and  its astounding  CO
linewidth of $\sim $1060\,km\,s$^{-1}$  (DS98) may be a direct outcome
of radiation pressure becoming dynamically important as to affect bulk
gas motions.

The global  CO line ratios, while  typical of a  star-forming ISM they
are not typical of an extreme starburst as Mrk\,273 certainly is.  The
$\rm R_{21}$=$7\pm  2$ ratio  in particular is  low, found  mostly for
GMCs in  SF-quiescent spirals ($\rm  R_{21}$$\geq $20 is  observed for
extreme  starbursts).   The  compact  starburst core  of  Mrk\,273  is
surrounded  by extended  CO  line emission  ($\sim $3$''$-7$''$,  i.e.
(2.2-5.1)\,kpc,   DS98)   where   ordinary   star-forming   and   even
SF-quiescent  ISM could  be  containing significant  fractions of  the
total molecular gas mass.  The  optimum LVG solutions obtained for the
global    CO     ratios    are:    $\rm     [T_{kin},    n,    K_{vir}
]$=[30-70,3$\times$$10^2$, 1].   Apart from the  elevated temperatures
such  conditions  are  typical   of  Galactic  GMCs,  while  the  $\rm
X_{co}$$\sim  $2\,$\rm  X_l$  of   the  best  LVG  solution  (at  $\rm
T_{kin}$=45\,K)  is  higher  than  the so-called  ULIRG  value  ($\sim
$0.8\,$\rm    X_l$).     The    corresponding    $\rm    M_{tot}$$\sim
$10$^{10}$\,M$_{\odot}$  while  a minimum  mass  of $\rm  M_{SF}$$\sim
$3.3$\times    $10$^9$\,M$_{\odot}$    is    needed   to    fuel    an
Eddington-limited star formation in this system.

The total  molecular gas mass of  Mrk\,273 can be higher  still if the
mass contributions of a colder low-density and SF-quiescent phase, and
a warmer denser  SF one are accounted separately.   A clear indication
of a highly-excited SF gas phase is given by the CO J=6--5 line which,
while  not  as bright  as  in  some  other (U)LIRGs  (e.g.   Mrk\,231,
NGC\,6240), yields  a $\rm  r_{65}$ that is  $\sim $5$\times  $ higher
than that anticipated from the best LVG fit of the lower-J CO lines. A
high  HCN/CO J=1--0  ratio of  $\rm  r _{HCN/CO}$=0.14  (using the  CO
J=1--0  value from  Paper\,I and  the HCN  J=1--0 from  Carpio  et al.
2008) and    a    well-excited    HCN    J=3--2   line    with    $\rm
r_{32}$(HCN)=0.49$\pm  $0.11 (Gracia-Carpio et  al.  2008)  offer more
evidence  for  the  presence  of  a  dense  gas  phase.   A  two-phase
decomposition using only the CO 3-2, 6-5 and $^{13}$CO J=2--1 lines to
constrain the high-excitation  phase yields $\rm M_{h-ex}$(H$_2$)$\sim
$6$\times  $10$^9$\,M$_{\odot}$,  and  an  effective  $\rm  X^{(2-ph)}
_{co}$$\sim    $2.5\,$\rm    X_l$.     The    latter    yields    $\rm
M_{tot}$(H$_2$)=1.3$\times  $10$^{10}$\,M$_{\odot}$,  of  which  $\sim
$53\% correspond to the low CO brightness (l)-phase.  Thus despite its
spectacular starburst, Mrk\,273 is an example of a ULIRG that contains
also large  amounts of  cooler low CO-excitation  gas. Use of  the two
available HCN lines to determine  the propreties of the (h)-phase does
not change this picture, and can actually raise $\rm X^{(2-ph)} _{co}$
up to $\sim $(4-5.5)\,$\rm X_l$.

  

\subsection{3C\,293}

This  powerful radio  galaxy was  the first  F-R\,II source  (i.e with
edge-brightened  radio  lobes,  see  Fanaroff  \& Riley  1974)  to  be
detected in CO locally (z=0.046), revealing large amounts of molecular
gas distributed in a rotating disk of 7$''$ ($\sim $6.2\,kpc) centered
in   its  nucleus,   and  with   a   large  CO   linewidth  of   $\sim
$900\,km\,s$^{-1}$  (Evans et  al.  1999).   Its  disturbed morphology
maybe   due  to  an   interaction  with   a  companion   galaxy  $\sim
$40$''$(35\,kpc)    to    the     southwest,    while    a    powerful
optical/IR-luminous jet  interacts with the ambient ISM  (Floyd et al.
2006).    The   SF   efficiency   of   the  galaxy   is   low   ($\sim
$8\,L$_{\odot}$/M$_{\odot}$)    and     along    with    a    SFR$\sim
$(6-7)\,M$_{\odot}$\,yr$^{-1}$ it is typical of ordinary spirals.  The
modest  global  CO  $\rm  r_{21}$=0.74 and  $\rm  r_{32}$=0.44  ratios
(Paper\,I) are also consistent with a SF-quiescent ISM.  Thus the high
excitation levels of the CO J=4--3, and 6--5 discovered in this system
were  certainly  a  surprise.   Their  interpretation  as  due  to  an
AGN-driven jet-ISM interaction provided  the first known example where
``mechanical'' AGN feedback globally affects galaxy-size molecular gas
reservoirs (Papadopoulos et al.  2008, 2010b), though similar examples
with lower-power  AGN and much  smaller molecular gas  reservoirs have
been reported much earlier (Matsushita et al.~2004).

The physical  conditions compatible with the luminous  CO J=4--3, 6--5
lines were studied in detail  by Papadopoulos et al.  2010b (see their
Table  4) where  dense ($\sim  $3$\times $(10$^4$-10$^5$)\,cm$^{-3}$),
warm  $\rm  T_{kin}$$\sim   $(75-300)\,K  and  strongly  unbound  $\rm
K_{vir}$$\sim $15-50 states have  been deduced. The corresponding $\rm
X^{(h)} _{co}$$\sim $(0.8-1.5)\,$\rm  X_l$, with computed $\rm r^{(h)}
_{43}$=2-3.2 and $\rm L^{(h)\,'} _{10}$=(1/$\rm r^{(h)} _{43}$)$\times
$$\rm   L^{'}  _{43}$=(1.46-2.33)$\times  $10$^9$\,$\rm   L_l$.   This
AGN-excited  gas  will   have  $\rm  M_{h-ex}$$\sim  $(1.1-3.5)$\times
$10$^9$\,M$_{\odot}$, much higher than the $\rm M_{SF}$$\sim $5$\times
$10$^7$\,M$_{\odot}$ of  warm and dense gas expected  fueling its star
formation.  Setting  a Galactic  $\rm X^{(l)} _{co}$=5\,$\rm  X_l$ for
the remaining (l)-phase CO  J=1--0 luminosity of the SF-quiescent disk
phase yields  an effective $\rm  X^{(2-ph)} _{co}$=(2.9-3.9)\,$\rm
X_l$, which  are near-Galactic values.  Indicatively,  a one-phase fit
of only the  CO 1--0, 2--1, and 3--2 lines (any  LVG fit including the
CO J=4--3  and/or J=6--5 predictably  fails) yields maximum  values of
$\rm X_{co}$=(1.2-1.6)\,$\rm X_l$, obtained for the solutions with the
lowest $\rm K_{vir}$  values ($\sim $2-4), but values  as low as $\rm
X_{co}$$\sim $0.5\,$\rm X_l$ also~probable.



\subsection{IRAS\,14348--1447}

This is an impressive ULIRG/merger  of two gas-rich spirals with their
two nuclei  3.4$''$ (5.2\,kpc) apart,  and clear indications  of tidal
features and  strongly disrupted disks  seen in {\it HST}  optical and
near-IR  NICMOS images  (Evans et  al. 2000;  Scoville et  al.  2000).
Interferometric  imaging  of CO  J=1--0  has  shown  that both  nuclei
contain copious  amounts of molecular  gas but remain  unresolved with
size  $\la  $2.5$''$ (3.8\,kpc)  (Evans  et  al.   2000). Using  radio
continuum  measurements   to  obtain   size  estimates  for   the  two
star-forming  cores of  this ULIRG  yields very  compact  regions with
diameters of  D$\sim $200\,pc  (Condon et al.   1991), similar  to the
compact molecular gas disks found in Arp\,220 and Mrk\,231 (DS98).

The  observed  CO  J=1--0,  2--1,  3--2  and  $^{13}$CO  J=2--1  lines
(Paper\,I) are compatible with a  wide range of conditions having $\rm
T_{kin}$$\ga   $35\,K   with   the    best   ones   found   for   $\rm
T_{kin}$=(40-85)\,K,  n=10$^3$\,cm$^{-3}$, and $\rm  K_{vir}$=4. Quite
unlike most  merger/ULIRGs, all good LVG solutions  correspond to $\rm
K_{vir}$$\sim  $1-4.  Nevertheless  $\rm  X_{co}$$\sim $(0.65-1)\,$\rm
X_l$,  i.e.   much  lower   than  Galactic.   The  corresponding  $\rm
M_{tot}$$\sim  $(1.1-1.74)$\times  $10$^{10}$\,M$_{\odot}$  while  the
minimum  mass  needed  for  Eddington-limited star-formation  is  $\rm
M_{SF}$$\sim $5.6$\times $10$^9$\,M$_{\odot}$,  which amounts to $\sim
$30\%-50\% of  the total molecular gas mass.   In IRAS\,14348-1447, as
in  other ULIRGs, such  high mass  fractions of  what is  presumably a
dense and warm  gas phase are not reflected by the  global CO SLED and
$^{13}$CO   lines   available,   dominated   by   lower-density   gas.
Unfortunately neither  high-J CO nor  HCN line observations  exist for
this ULIRG  to set  independent constraints on  the mass  and physical
properties  of its dense  SF molecular  gas. A  2-phase model  with an
assumed (h)-phase CO  SLED (section 2.4), while not  necessary in this
case (as  a wide range of good  one-phase fits exist), can  be used to
assess the impact of a massive dense gas phase on $\rm X_{co}$.  Apart
from  n$\la  $10$^3$\,cm$^{-3}$   the  (l)-phase  remains  essentially
unconstrained   (as   expected),   with   $\rm   X^{(l)}   _{co}$$\sim
$(0.5-1.12)\,$\rm X_l$. For the computed $\rm \rho^{(l-h)} _{co}$=0.85
this yields $\rm X^{(2-ph)} _{co}$=(1.4-1.7)\,$\rm X_l$.

\subsection{Zw\,049.057}

This  galaxy has  a  SF-powered $\rm  L^{(*)} _{IR}$$\sim  $1.2$\times
$$10^{11}$\,$\rm  L_{\odot}$   typical  for  SF   spiral  disks  (e.g.
NGC\,7469) and is one of the lowest IR luminosity galaxies in the IRAS
{\it BGS}.   NICMOS near-IR imaging with  the {\it HST}  indeed show a
highly inclined  disk-dominated system with  a heavily dust-enshrouded
nucleus (Scoville  et al. 2000). Despite its  disk-like appearance its
near-IR emission  is better fitted by  a $\rm r^{1/4}$  rather than an
exponential  disk profile  (Scoville  et al.   2000),  while its  $\rm
L_{IR}/L_B$ ratio  is almost as  extreme as the much  more IR-luminous
Arp\,220 (Planesas et al.   1991).  Early interferometric images of CO
J=1--0  found large  molecular gas  mass within  a 1.3\,kpc-sized
region (Planesas et al.  1991, for the cosmology adopted here).

Its CO SLED remains well excited up to J=4--3 and also has substantial
J=6--5 emission  (see Paper\,I) while both its  $^{13}$CO J=1--0, 2--1
lines  are   detected  and  yield  $\rm  R_{10}$$\sim   $16  and  $\rm
R_{21}$$\sim  $24,  typical of  merger-driven  starbursts rather  than
disk-dominated  LIRGs.    One-phase  LVG  models   reproduce  all  CO,
$^{13}$CO line ratios  up to CO J=4--3 though  they still underpredict
the luminosity  of the  latter by $\sim  $1.2$\sigma$.  Unfortunately,
even  with this large  set of  lines significant  degeneracies remain.
Indicatively   the  two   best  solutions:   $\rm   \left[T_{kin},  n,
K_{vir}\right]$=$\rm   \left[35,  3\times  10^{3},   22\right]$,  $\rm
\left[105,   10^{3},   13\right]$   yield   CO   line   ratios:   $\rm
\left[r_{21},r_{32},       r_{43},        r       _{65},       R_{10},
R_{21}\right]_{lvg}$=[1.05,  0.75, 0.41, 0.05,  22, 21],  [1.05, 0.76,
0.40,  0.02, 21,  16]. Much  warmer  gas with  $\rm \left[T_{kin},  n,
K_{vir}\right]$=$\rm \left[130-140, 10^{3}, 4\right]$ is also possible
and  yields:  $\rm   \left[r_{21},r_{32},  r_{43},  r  _{65},  R_{10},
R_{21}\right]_{lvg}$=[1.06, 0.86,  0.60, 0.17, 14, 13],  now in better
agreement with the observed $\rm r_{65}$ (even if not used in the fit)
but   a  rather  worse   fit  of   $\rm  R_{21}$.    Nevertheless  the
corresponding  $\rm X_{co}$ values  span a  rather narrow  range $\sim
$(0.35-0.6)$\rm X_l$, yielding $\rm M_{tot}$$\sim $(3-5.2)$\times
$$10^8$\,M$_{\odot}$.  This is comparable to the minimum molecular gas
mass   required   for  Eddington-limited   star   formation  of   $\rm
M_{SF}(H_2)$$\sim   $4.7$\times  $10$^8$\,M$_{\odot}$,  yet   all  the
densities deduced from the LVG solutions of the global CO SLED are far
from being representative of what  is typically a much denser SF phase
with small/modest $\rm K_{vir}$ values.

A  two-phase decomposition  of  the  average ISM  state  using the  CO
J=3--2,  4--3, 6--5  and the  $^{13}$CO transitions  to  determine the
(h)-state      converges      to      $\rm      \left[T_{kin},      n,
K_{vir}\right]$=$[$110-140,10$^3$,4$]$,   with   $\rm  T_{kin}$=140\,K
corresponding to the best  solution (and $\rm X^{(h)} _{co}$=0.6\,$\rm
X_l$).  Higher  density solutions with  n=$10^4$\,cm$^{-3}$ exist over
$\rm T_{kin}$=(40-65)\,K  but represent poorer fits  and correspond to
unbound states with $\rm K_{vir}$$\sim  $13.  This is a general aspect
of all (h)-phase LVG solutions (e.g.   even if only the CO J=4--3, 6-5
and $^{13}$CO  lines are used), namely  the best fits  occur at higher
temperatures, lower  densities, and more  gravitationaly bound states.
High  density  solutions  ($\sim  $10$^4$\,cm$^{-3}$)  and  thus  more
typical for the  presumably star forming (h)-phase are  found at lower
temperatures  ($\sim $(35-70)\,K), and  strongly unbound  states ($\rm
K_{vir}$$\sim $13-40), but yield  poorer fits.  Moreover the high $\rm
K_{vir}$ of  the denser/cooler LVG solutions yield  similarly low $\rm
X^{(h)} _{co}$ (e.g.  for $\rm T_{kin}$=50\,K, n=$10^4$\,cm$^{-3}$ and
$\rm K_{vir}$$\sim $13 it is $\rm X^{(h)} _{co}$=0.7\,$\rm X_l$).

It is the combination of  weak $^{13}$CO J=1--0, 2--1 and well-excited
high-J  CO lines  that  nevertheless still  have $\rm  r_{J+1\,J}$$<$1
which  forces LVG  solutions to  a parameter  space with  low  CO line
optical depths and modest gas  densities. This is why $\rm K_{vir}$ is
high towards the  low-$\rm T_{kin}$/high-n domain (as to  keep CO line
optical  depths low), and  why the  need for  high $\rm  K_{vir}$'s is
relaxed  towards  higher  $\rm   T_{kin}$  which  now  becomes  partly
responsible for the low optical depths.  Average gas densities towards
the high $\rm T_{kin}$ regime cannot  be high as this would yield $\rm
r_{J+1\,J}$$>$1 in the optically thin domain. If we omit the CO J=6--5
line from the fit (as its luminosity can be highly uncertain) and also
leave  out  the  $^{13}$CO  J=1--0  line  (which  may  have  a  larger
contribution from  a cooler disk  phase) we obtain LVG  solutions with
densities  of  $\sim $(1-3)$\times  $10$^4$\,cm$^{-3}$  that are  more
typical of SF gas, yet with high enough $\rm K_{vir}$$\sim $13-22 that
still   yield   a   low   corresponding   $\rm   X^{(h)}   _{co}$$\sim
$(0.65-0.75)\,$\rm X_l$.  Moreover,  with an essentially unconstrained
(l)-phase, and no evidence for a  cold disk, a 2-phase gas would yield
$\rm X^{(2-ph)} _{co}$ similar to that obtained from 1-phase model.

 Thus  the molecular  gas reservoir  in this  LIRG is  expected  to be
dominated by its  SF component, yet neither the  average densities nor
the  strongly  unbound dynamical  states  deduced  from its  available
global CO  SLED are  typical of  the latter. If  this is  confirmed by
future HCN or  other heavy rotor molecular line  observations (or more
high-J  CO and  $^{13}$CO transitions)  it  may come  to resemble  the
average ISM state deduced for Arp\,193.

\subsection{Arp\,220 and NGC\,6240}

These are the two (U)LIRGs in  our sample that currently have the best
molecular line  data tracing their  dense gas (i.e.  multi-J  HCN, CS,
HCO$^+$ lines),  which are  used in a  detailed study of  its physical
conditions (Greve et  al.  2009). The presence of  a massive dense gas
phase in  both of  them is well-established  with large  HCN/CO J=1--0
line   ratios   ($\sim   $0.19(Arp\,220),   $\sim   $0.08(NGC\,6240)),
well-excited  HCN   and  CS  higher-J   lines,  yet  also   the  large
CO/$^{13}$CO line ratios  typical for merger-driven extreme starbursts
($\sim  $43-45 for  J=1--0).   The  latter implies  very  low CO  line
optical depths,  and thus runs  counter what is expected  for dominant
high-density molecular gas reservoir ($\rm \tau _{J+1,J}$(CO)$\gg $1),
necessitating  the  2-phase models  used  in  such  systems (Aalto  et
al. 1995; Papadopoulos \& Seaquist 1998; this work).

 Predictably  one-phase models  of  the CO  and  $^{13}$CO lines  (see
Paper\,I)  fail to  converge on  any good  set of  solutions  for both
galaxies,  yielding  $\rm  [T_{kin},  n,  K_{vir}]$=$\rm  [(30-90)\,K,
10^3\,cm^{-3},  40]$ (Arp\,220) and  $\rm [T_{kin},  n, K_{vir}]$=$\rm
[(15-40)\,K,  3\times10^3\,cm^{-3},  40]$  (NGC\,6240).  For  Arp\,220
even the  CO J=1--0, 2--1, 3--2  lines are impossible to  fit with one
phase  since $\rm  r_{21}$=0.67$\pm $0.07  (see Paper\,I)  indicates a
diffuse  gas  phase that  leaves  even  the  J=2--1 line  subthermally
excited,  while $\rm  r_{21}$=0.97$\pm  $0.14 marks  the emergence  of
another,  denser  and  warmer  phase  from  J=3--2  and  higher.   The
corresponding  one-phase $\rm  X_{co}$$\sim $0.3\,$\rm  X_l$  for both
systems,  a low  value  due to  the  very large  $\rm K_{vir}$  values
``forced'' by the extreme  large CO/$^{13}$CO line ratios obtained for
these systems.  This is comparable to $\rm X^{(thin)} _{co}$(LTE) (for
the  $\rm T_{kin}$=(30-40)\,K  range of  one-phase LVG  solutions), as
expected for optically thin CO line emission.

 This low  $\rm X_{co}$ factor yields  $\rm M_{tot}$$\sim $1.85$\times
$10$^9$\,M$_{\odot}$(Arp\,220)       and       $\sim       $2.5$\times
$10$^9$\,M$_{\odot}$(NGC\,6240), while the  minimum molecular gas mass
needed   for   fueling   Eddington-limited   SF   in   these   extreme
merger/starbursts       is      $\rm       M_{SF}$$\sim      $4$\times
$10$^9$\,M$_{\odot}$(Arp\,220)       and       $\sim       $1.5$\times
$10$^9$\,M$_{\odot}$(NGC\,6240).  Given  that $\rm M_{SF}/M_{tot}$$\ga
$0.6, and in  the case of Arp\,220 actually  $>$1, it becomes obvious,
even in  the absence of  heavy rotor molecular line  observations that
the true  $\rm X_{co}$ factor for  these two galaxies  must be higher.
From the analysis detailed in 3.2.1 the multi-J HCN LVG solutions in a
2-phase LVG model  yield: $\rm X^{(2-ph)} _{co}$$\sim $(2.4-4.5)\,$\rm
X_l$(Arp\,220)   and   $\rm   X^{(2-ph)}  _{co}$$\sim   $(1-3.3)\,$\rm
X_l$(NGC\,6240) with the high values corresponding to virial dynamical
states for the  dense gas. The new $\rm  M_{tot}$ estimates are: $\sim
$(1.5-2.8)$\times    $$10^{10}$\,M$_{\odot}$(Arp\,220)    and    $\sim
$(0.8-2.8)$\times  $$10^{10}$\,M$_{\odot}$(NGC\,6240). In the  case of
Arp\,220 these amount to $\sim $40\%-77\% of the dynamical mass of its
molecular gas disks ($\rm M_{dyn}$  from DS98), while for NGC\,6240 it
surpasses the total mass contained  within its central radius of r$\la
$1$''$  (480\,pc)  and  argues  for  significant  molecular  gas  mass
corresponding  to the  CO line  emission  seen beyond  its central  CO
J=2--1  peak (see  Tacconi  et al.   1999  for CO  J=2--1  for a  high
resolution  interferometry  map).   In  the latter  case  the  deduced
molecular gas mass may actually  dominate the mass distribution out to
much larger radii  than deduced by Tacconi et  al.  1999.  Multi-J HCN
line  interferometric  maps,  along   with  at  least  one  H$^{13}$CN
transition,  at  resolutions  of  $\la  $1$''$, will  be  valuable  in
determining the $\rm X_{HCN}$,  the effective $\rm X^{(h)} _{co}$, and
$\rm X^{(2-ph)} _{co}$ of these template merger systems.

\subsection{IRAS\,17208--0014}

This  classic ULIRG  is the  most  luminous in  our sample,  it has  a
compact  gas-rich  nucleus  (Planesas  et  al.  1991;  DS98),  a  very
disrupted  disk,  and  tidal  tails  indicating a  merger  (Murphy  et
al. 1996; Scoville et al.  2000). In the interferometric study by DS98
this is  the ULIRG  whose compact gas  disk ($\sim $1.4\,kpc)  has the
largest   face-on  velocity   dispersion  of   $\rm   \sigma  _V$$\sim
$150\,km\,s$^{-1}$,     indicating    a    highly     turbulent    gas
environment. This is also suggested  by the large $\rm R_{21}$$\ga $35
ratio,   expected   in  the   high-pressure,   highly  turbulent   ISM
environments of  strong mergers  (Aalto et al.  1995). Its CO  SLED is
highly excited  up to CO J=4--3  while the weak CO  J=6--5 reported in
Paper\,I is most  likely due to pointing offsets  as ZEUS measurements
at the CSO indicate a much stronger line (Stacey 2011).

No  reasonable one-phase  LVG fits  were found  for this  source while
attempts to fit  the CO J=1--0 up to J=4--3  and $\rm R_{21}$=35 yield
highly unbound states ($\rm K_{vir}$$\sim $13-70).  The best such fit:
$\rm \left[T_{kin}, n, K_{vir}\right]$=$[$70,10$^3$,13$]$, yields $\rm
r_{21}$=0.96, $\rm  r_{32}$=0.64, $\rm r_{43}$=0.32,  the latter being
$\sim $3$\times $ smaller than observed. The large $\rm L^{(*)} _{IR}$
of  this  system  corresponds  to $\rm  M_{SF}$(H$_2$)$\sim  $6$\times
$10$^9$\,M$_{\odot}$.  On the  other hand $\rm X_{co}$$\sim $0.4\,$\rm
X_l$   computed    from   the   best   LVG    solution   yields   $\rm
M_{tot}$$\sim$5.2$\times  $10$^9$\,M$_{\odot}$,   and  thus  even  the
minimum  mass expected  for a  dense  and warm  SF-fueling phase  will
dominate  $\rm  M_{tot}$, as  computed  by  a  one-phase $\rm  X_{co}$
factor.   One-phase LVG  solutions  with their  low densities,  highly
unbound dynamical states certainly do  not hint this, while their poor
fit of the  global CO line ratios of this ULIRG  suggest that at least
two main gas  phases are necessary to represent  the average ISM state
in this system.  The presence of  a massive gas phase much denser than
those implied by one-phase LVG fits is further corroborated by an $\rm
r_{HCN/CO}$=0.14 (HCN  J=1--0 from Gracia-Carpio et al.   2008, CO 1-0
from Paper\,I) and an HCN(3--2)/(1--0) ratio of $\rm r_{32}$(HCN)=0.39
(Gracia-Carpio et al.~2008).

  Using only CO J=4--3, 3--2, and $\rm R_{21}$=35 (the lower limit) to
  constrain the  (h)-phase in  a 2-phase model  yields only  warm $\rm
  T_{kin}$=(90-150)\,K            and           dense           n$\sim
  $(3$\times$10$^4$-10$^5$)\,cm$^{-3}$    gas    phases   with    $\rm
  K_{vir}$$\sim $13-22, and  $\rm X^{(h)} _{co}$$\sim $0.85-1.40 (high
  values for  the denser gas  solutions).  These solutions  yield $\rm
  r^{(h)}  _{J+1\,J}$$\sim $2.3-3  for J+1=2,3,4  indicating  CO lines
  with   small/moderate  optical   depths,  a   result  of   the  high
  temperatures and  large $\rm K_{vir}$  values which, along  with the
  high densities, are necessary for well-excited CO lines up to high-J
  levels {\it and} a high global $\rm R_{21}$ ratio.  The mass of such
  high-excitation gas can be computed from $\rm L^{(h)\,'} _{10}$=$\rm
  (1/\langle  r^{(h)}  _{43}\rangle)$$\rm L^{'}  _{43}$=(4-4.4)$\times
  $10$^9$$\rm    L_l$   which   yields    $\rm   M_{h-ex}$(H$_2$)$\sim
  $(3.4-6)$\times  $10$^9$\,M$_{\odot}$.   Thus,   as  in  some  other
  (U)LIRGs studied here (e.g.  IRAS\,00057+4021), this galaxy contains
  large reservoirs of  dense gas in thermal and  dynamical states much
  more extreme  than those expected for  dense gas in  SF regions.  As
  detailed  in  Paper\,I these  could  be  indicative  of other  power
  sources  such as  turbulent  heated regions  (THRs) or  CR-dominated
  regions  (e.g.  Papadopoulos 2010).   Future high  resolution high-J
  CO,  $^{13}$CO  imaging of  this  system  with  ALMA would  be  very
  valuable  for   confirming  and  quantifying   such  mechanisms  via
  determining   the  level   of   turbulence  present   in  its   disk
  (i.e. measuring $\rm  \sigma _z$({\bf r}) and the  average CR energy
  density permeating its ISM.

Despite  the  prominence of  the  dense  highly-excited  phase in  the
J=3--2, 4--3 lines,  the (l)-phase remains the main  contributor of CO
J=1--0  line  luminosity  in  IRAS\,17208-0014, with  $\rm  L^{(l)\,'}
_{1-0}$$\sim $(8.7-9)$\times  $10$^9$\,$\rm L_l$ ($\sim  $(66-68)\% of
the observed  one).  LVG fits of the  corresponding ``residual'' ratio
$\rm  r^{(l)}  _{21}$$\sim $0.5,  while  $\rm T_{kin}$-degenerate  (as
expected),  all  indicate   low  gas  densities  n$\sim  $(1-3)$\times
$10$^2$\,cm$^{-3}$ and a wide  range of $\rm K_{vir}$$\sim $1-70.  The
coresponding  $\rm X^{(l)} _{co}$$\sim  $(0.5-3)\,$\rm X_l$  (over the
$\rm  K_{vir}$=1-20  range   of  LVG  solutions)  remains  essentially
unconstrained,  with  the  lowest  values corresponding  to  the  most
dynamically  unbound  states  as  expected.   Thus  for  $\rm  X^{(h)}
_{co}$=1.4\,$\rm  X_l$  (densest  gas  LVG  solutions)  we  find  $\rm
X^{(2-ph)} _{co}$$\sim $(0.75--2.5)\,$\rm X_l$.

Using  the  states  compatible  with  the  HCN(3--2)/(1--0)  ratio  in
IRAS\,17208--0014 to  define the (h)-phase  leads to even  higher $\rm
X^{(2-ph)} _{co}$$\sim  $(2.5-5.9)\,$\rm X_l$, with  larger still $\rm
M_{h-ex}$ (see  Table 1).  These  correspond to a  high-density ($\sim
$(3$\times  $10$^4$-10$^5$)\,cm$^{-3}$) phase with  $\rm K_{vir}$$\sim
$1-6   (unphysical   solutions  with   $\rm   K_{vir}$$<$1  were   not
considered).  Nevertheless, for  $\rm T_{kin}$=(75-105)K, LVG solutions
with high-density gas  ($\sim $10$^5$\,cm$^{-3}$) but $\rm K_{vir}$=20
are also  possible.  These strongly unbound dense  gas solutions yield
$\rm  X^{(h)}  _{co}$$\sim  $0.8\,$\rm  X_l$ and  similarly  low  $\rm
X^{(2-ph)}   _{co}$   values    (unless   cold   (l)-phase   solutions
corresponding to a SF-quiescent disk are considered).  Observations of
more HCN  lines and,  crucially, of at  least one H$^{13}$CN  line can
determine whether such extraordinary dynamical states are possible for
the HCN-bright dense gas phase (see  3.4).

\subsection{IRAS\,22491--1808}

This is  a ULIRG  with a spectacular  morphology involving  two nuclei
separated by 1.6$''$ ($\sim $2.3\,kpc) (Carico et al. 1990), two tidal
tails and numerous  SF knots embedded in them  (Scoville at el.  2000)
indicating an advanced  merger. It is also one of  the very few ULIRGs
where an  AGN contributes  more than half  of the total  IR luminosity
(Farrah et al.  2003). In our  sample it stands out as the galaxy with
the most highly excited CO  J=3--2 line with $\rm r_{32}$=1.88, though
a large  applied correction makes  this value somewhat  uncertain (see
Paper\,I).  All  one-phase LVG solutions correspond to  dense gas $\rm
n$=(10$^4$-3$\times$10$^5$)\,cm$^{-3}$,  with the  best  fits obtained
for  $\rm T_{kin}$$\geq $40\,K.   Unfortunately lack  of more  CO line
ratios and especially at  least one $^{13}$CO line measurement permits
a  wide  range  of $\rm  [T_{kin},  n,  K_{vir}]$  states up  to  $\rm
T_{kin}$=150\,K  and  $\rm  K_{vir}$$\sim$1-20  to  fit  the  observed
ratios.   The   one-phase  $\rm  X_{co}$   values  remain  essentially
unconstrained with $\rm X_{co}$$\sim $(0.7-2.4)\,$\rm X_l$ and the low
values  associated with  strongly unbound  states  ($\rm K_{vir}$$\sim
$13-22), while the low values with states that have $\rm K_{vir}$$\sim
$1-7.   The corresponding  molecular  gas mass  is $\rm  M_{tot}$$\sim
$(6.3-22)$\times   $10$^9$\,M$_{\odot}$    while   $\rm   M_{SF}$$\sim
$5.2$\times $10$^9$\,M$_{\odot}$.

  Another extraordinary  aspect of  IRAS\,22491--1808 is that  many of
the  conditions compatible  with its  global CO  line  ratios indicate
dense gas ($\sim $3$\times $10$^5$)\,cm$^{-3}$) with temperatures $\rm
T_{kin}$=(100-130)\,K,  significantly larger  than those  of  the dust
($\rm T_{dust}$=(32-49)\,K  for emissivity  of $\beta $=1.5,  2). Such
potentially strong decoupling of gas and dust temperatures, especially
at high  gas densities where  $\rm T_{kin}$$\rightarrow$$\rm T_{dust}$
is  expected, are another  indicator of  other dominant  power sources
than PDR  photons.  Turbulent  and CR heating  are capable  of driving
large $\rm T_{kin}/T_{dust}$$>$1 inequalities  even for the very dense
gas  mass reservoirs  found in  ULIRGs (Paper\,I,  Papadopoulos 2010).
These  mechanisms can  easily dominate  when  strong gas-rich/gas-rich
spiral galaxy  interactions drive the  bulk of the molecular  gas into
compact regions in the two nuclei of the progenitors. This will inject
large amounts  of gas kinetic energy  into $\sim$100\,pc-sized regions
fueling  extreme turulence,  while high  SFR densities  will establish
extreme  CRDRs.    Interestingly  two  other   double-nuclei  systems,
IRAS\,12112+305  and IRAS\,08572+3915, also  show very  highly excited
(i.e.  $\rm r_{J+1\,J}$$>$1) CO lines.

The IRAS\,22491-1808 and similar  systems are thus excellent cases for
studying extreme  ISM conditions and  their drivers using  ALMA.  Such
observations will be crucial for: a) inferring CR energy densities (by
finding starburst region sizes), b) measuring velocity dispersions and
the turbulence levels of molecular gas disks and c) directly providing
key  molecular  line  diagnostics  for the  presense  of  CR-dominated
regions (CRDRs)  and/or turbulent-heated regions  (THRs) (Papadopoulos
2010; Bayet et al. 2011, Meijerink et al.~2011).

\subsection{NGC\,7469}

This is a well-studied Sy\,1  spiral galaxy whose AGN is surrounded by
a  nearly complete  ring of  powerful starburst  activity contributing
$\sim $2/3  of its $\rm  L_{IR}$ (Genzel et  al.  1995; Riffel  et al.
2006 and references therein).  Interferometric CO J=1--0 imaging found
its  central starburst  to be  very  gas-rich (Meixner  et al.   1990;
Tacconi  \& Genzel  1996). Its  starburst is  thought triggered  by an
interaction with a neighboring galaxy IC\,5283 that lies 83$''$ ($\sim
$27\,kpc) away  (projected distance).  A  $\rm X_{co}$$\sim $(1/5)$\rm
X_{co,Gal}$ was  inferred for its inner $\sim  $1\,kpc, and attributed
to the  effects of strong SF  feedback on molecular  clouds (Genzel et
al.  1995).  A small $\rm X_{co}$ for the starburst region of the disk
in NGC\,7469  is indeed corroborated  by CO J=3--2,  1--0 observations
and  850$\mu $m/450$\mu  $m  dust continuum  imaging (Papadopoulos  \&
Allen  2000).  However the  same observations  also revealed  faint CO
J=1--0 and 450$\mu $m dust emission extending well beyond the staburst
region,  and containing  a  massive SF-quiescent,  cold  dust and  gas
reservoir with a Galactic $\rm X_{co}$.

 These two ISM components make a good one-phase LVG fit impossible for
 the  ``cold'' $\rm  r_{21}$=0.75 and  the ``warm''  $\rm r_{43}$=0.83,
 $\rm r_{65}$=0.22 ratios observed.   The best such solution with $\rm
 T_{kin}$=70\,K,   $\rm  n(H_2)$=$10^3$\,cm$^{-3}$,   $\rm  K_{vir}$=4
 yields: $\rm r^{(lvg)}  _{21}$=0.97, $\rm r^{(lvg)} _{32}$=0.75, $\rm
 r^{(lvg)} _{43}$=0.48, $\rm r^{(lvg)} _{65}$=0.08, and $\rm R^{(lvg)}
 _{21}$=11.    The  first   two   are  higher   than  those   observed
 ($\sim$2$\sigma$  higher in  the  case of  $\rm  r_{21}$), while  for
 (4--3)/(1--0)  and (6--5)/(1--0)  the LVG-computed  ratios  are $\sim
 $1.7  and  $\sim  $2.75   times  smaller  than  observed.   The  $\rm
 X_{co}$=0.72\,$\rm X_l$ value computed from the optimal one-phase LVG
 fit  to  the  global CO  SLED  is  dominated  by  the SF  phase,  and
 corresponds  to $\rm M_{tot}$$\sim  $2.5$\times $10$^9$\,M$_{\odot}$.
 The  minimum  SF  gas  mass  expected  on  the  other  hand  is  $\rm
 M_{SF}$=6.5$\times $10$^8$\,M$_{\odot}$.

 Using only  the CO J=3--2,  4--3, 6--5\footnote{The wide beam  of the
 CSO was used to obtain the CO 6-5 measurent for this extended object}
 and  the  $^{13}$CO J=2--1  lines  to  constrain the  high-excitation
 (h)-phase in a 2-phase models  yields a typical LVG solution for $\rm
 T_{kin}$=60\,K, $\rm  n(H_2)$=$10^4$\,cm$^{-3}$, and $\rm K_{vir}$=13
 with a corresponding $\rm  X^{(h)} _{co}$=0.66\,$\rm X_l$.  From this
 solution    we    compute    $\rm    L^{(h)\,'}_{CO,1-0}$=2.51$\times
 $10$^9$\,$\rm   L_l$,   and   $\rm   L^{(l)\,'}_{CO,1-0}$=0.97$\times
 $10$^9$\,$\rm L_l$  which yield $\rm  \rho^{(l-h)} _{co}$$\sim $0.39.
 Then using the Galactic  $\rm X^{(l)} _{co}$=5\,$\rm X_l$ deduced for
 the extended CO J=1--0 and  submm continuum dust emission of the disk
 in NGC\,7469 (Papadopoulos \&  Allen 2000), we obtain $\rm X^{(2-ph)}
 _{co}$$\sim $1.88\,$\rm X_l$, which is $\sim $2.6$\times$ higher than
 that obtained by the one-phase LVG fit of the global SLED.

  Thus the  cold extended disk of NGC\,7469,  despite containing $\sim
  $3$\times $ more  molecular gas mass than its  central starburst, is
  nearly  incospicious   in  the  global  CO  SLED,   resulting  to  a
  significant underestimate  of the total  molecular gas mass  in this
  galaxy. Only  the available  spatial information about  its presence
  could  rectify this,  a  state  of affairs  mirrored  in other  such
  systems  (e.g.   NGC\,1068). For  disk-dominated  LIRGs with  strong
  excitation gradients  induced by the presense  of central starbursts
  only CO, $^{13}$CO line and  dust continuum imaging of their ISM can
  avoid such pitfals (see 3.4).   This is quite unlike ULIRGs where it
  is  the low-excitation  ``cloud-envelope''-phase  that contains  the
  smaller mass  fraction and is  concomitant or closely  follows the
  distribution of a massive much denser gas component.

\subsection{IRAS\,23365+3604}

Near-IR imaging of this LIRG  reveals tidal tails indicating a merger,
and  a single  nucleus embedded  into a  nearly face-on  disk  that is
nearly $\sim $20kpc (17.3$''$) in diameter, and shows disturbed spiral
structure (Surace et al.  2000).  The nucleus is classified as a LINER
(Veilleux et  al. 1995).   The gas dynamics  as revealed by  CO J=1--0
interferometry  (DS98)  is strongly  decoupled  from stellar  dynamics
(Genzel  et  al.  2001),  another  hallmark  of  strong mergers.   Its
molecular gas  disk is  compact with $\sim  $1.2\,kpc diameter,  and a
face-on velocity dispersion of $\sim $100\,km\,s$^{-1}$ (DS98). Its CO
SLED has  a ``cool'' CO(2--1)/(1--0) ratio (=0.75)  but a well-excited
CO J=3--2 line with a  ``warm'' CO(3-2)/(1-0) (=0.82), while its large
$\rm R_{21}$(=17) is typical  for mergers.  One-phase LVG models yield
moderately good but rather degenerate fits with $\rm \left[T_{kin}, n,
K_{vir}\right]$=[30-65, 10$^3$, 13] and [70-150,3$\times $10$^2$, 2.2]
being the  optimal solution ranges.   The best solutions  are obtained
for  $\rm  T_{kin}$=50K,  70K  with corresponding  line  ratios:  $\rm
r^{(lvg)}  _{21}$=0.88, 0.84,  $\rm  r^{(lvg)} _{32}$=0.56,0.55,  $\rm
r^{(lvg)} _{43}$=0.24,  0.27 and $\rm R^{(lvg)}  _{21}$=20, 16.  These
are compatible  with the measured  values though the  models typically
yield ``warmer'' $\rm r_{21}$  and ``cooler'' $\rm r_{32}$ ratios than
observed, in the latter case by $\sim $1.1$\sigma$.  The corresponding
$\rm X_{co}$$\sim $(1-1.3)\,$\rm  X_l$, which gives $\rm M_{tot}$$\sim
$(7.3-9.5)$\times $10$^{9}$\,$\rm  M_{\odot}$, while $\rm M_{SF}$$\sim
$2.5$\times $10$^9$\,$\rm M_{\odot}$.

The latter  represents a substantial  fraction of the  total molecular
gas mass where much higher  gas densities must be prevailing.  A clear
indication of  another highly-excited denser gas phase  is provided by
CO J=6--5 line which is  $\sim $9-16 times more luminous than expected
from the best LVG solutions. The presence of a dense gas phase able to
emit such  high-J CO lines is  also indicated by a  high HCN/CO J=1--0
line  ratio   of  $\rm  r^{(obs)}  _{HCN/CO}$$\sim   $0.10  (HCN  from
Gracio-Carpio et al.   2008, CO from Paper\,I).  Using  the CO J=3--2,
J=4--3 (upper  limit), J=6--5 and $^{13}$CO J=2--1  lines to constrain
the   high-excitation    phase   yields:   $\rm    \left[T_{kin},   n,
  K_{vir}\right]$=[40-80, 10$^4$,  40] as the  optimum solution range,
with $\rm T_{kin}$=60K being the  best.  The corresponding CO SLED has
$\rm  r^{(h)}  _{21}$=1.94,  $\rm  r^{(h)} _{32}$=1.90,  $\rm  r^{(h)}
_{43}$=1.50,  and  $\rm  r^{(h)}  _{65}$=0.45, typical  of  low/modest
optical  depths (indicatively  $\rm \tau  _{10}$$\sim $0.2,  $\rm \tau
_{32}$$\sim $2.5), a result  of the exceptionally large $\rm K_{vir}$.
A less  optimal solution  range with lower  densities and  more modest
(but still high) $\rm  K_{vir}$ values does exists $\rm \left[T_{kin},
  n,  K_{vir}\right]$=[90-150,  3$\times$10$^3$,  22], but  could  not
reproduce the  high HCN(1-0)/CO(6-5) ratio  of $\rm R_{HCN/CO65}$$\sim
$0.48$\pm $0.18 observed  for this galaxy as such  a low-density phase
would hardly emit any HCN J=1--0 ($\rm R_{HCN/CO65}$$\la $0.08).

The mass of  the dense, warm, and strongly unbound  phase can be found
from   $\rm   X^{(h)}   _{co}$$\sim   $0.45\,$\rm   X_l$   (for   $\rm
T_{kin}$=60\,K),  and  $\rm  L^{(h)\,'}  _{CO,  1-0}$=$\rm  (1/r^{(h)}
_{32})$$\rm  L^{'} _{CO,  3-2}$$\sim $3.15$\times  $10$^9$\,$\rm L_l$,
which  yields  $\rm  M_{h-ex}$$\sim $1.4$\times  $10$^9$\,M$_{\odot}$.
This  large gas  mass  with such  high  $\rm K_{vir}$  {\it and}  high
densities is rather surprising  since such strongly unbound states are
associated  with cloud-envelopes or  intercloud diffuse  molecular gas
whose  $\rm  n$$\la  $10$^{3}$\,cm$^{-3}$.   The remaining  CO  J=1--0
luminosity  of the  (l)-phase on  the  other hand  is $\rm  L^{(l)\,'}
_{CO,1-0}$$\sim $4.15$\times $10$^{9}$\,$\rm  L_l$ with an essentially
unconstrained   $\rm    X^{(l)}   _{co}$.    Setting    $\rm   X^{(l)}
_{CO}$=(0.5-5)\,$\rm X_l$ to encompass  the range from a warm, diffuse
and  unbound gas  component to  an underlying  cold, self-gravitating,
disk phase yields $\rm  X^{(2-ph)} _{co}$$\sim $(0.5-3)\,$\rm X_l$ and
$\rm M_{tot}$$\sim $(3.7-22)$\times $10$^9$\,$\rm M_{\odot}$.  The low
$\rm X^{(l)} _{CO}$ (and thus  $\rm M_{tot}$) values are more probable
because of  the negligible  CO J=2--1 luminosity  of the  (l)-phase (a
cold, self-gravitating, phase would have higher J=2--1 luminosities).

Using  the SNR-GMC  interfaces  as benchmark  entities  for dense  yet
highly  unbound (due  to the  kinetic energy  injected by  SNR shocks)
molecular  gas  phase  makes  clear  that  a  mass  fraction  of  $\rm
M_{h-ex}/M_{tot}$$\sim $0.06-0.38 exceeds that  expected per GMC for a
SNR-impacted  gas phase ($\sim  $1\% of  a GMC's  mass).  It  is worth
noting  that only in  the powerful  radio galaxy  3C\,293 galaxy-sized
molecular gas reservoirs at high densities were inferred to be at such
strongly gravitationally unbound states, likely caused by the powerful
jet-ISM  interaction.  While  no  such  AGN influence  on  the ISM  is
apparent in IRAS\,23365+3604 it is  interesting to note that it is one
of  the  3 ULIRGs  (the  others  being  Arp\,220 and  Mrk\,273)  where
molecular  gas  velocity fields  decoupled  from  those  of stars  and
ionized gas are found, with  molecular gas having $\sim$2 times larger
linewidths (Colina  et al.  2005).   High-J CO, $^{13}$CO and  HCN and
H$^{13}$CN  line  observations   are  necessary  for  confirming  such
extraordinary  dynamical states  for the  massive dense  gas reservoir
residing in IRAS\,23365+3604 and similar systems.


\begin{figure}
\epsscale{1.2}
\plotone{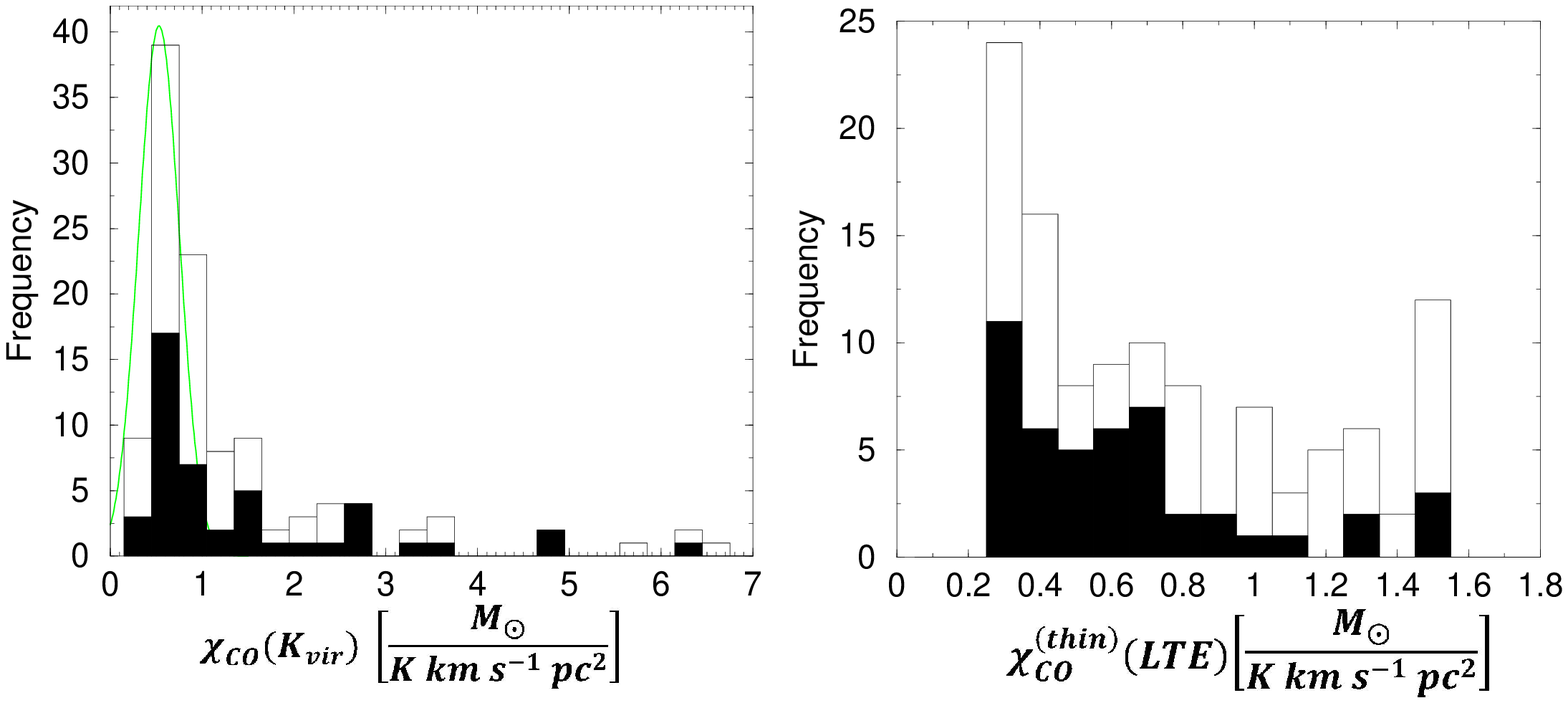}
\caption{The $\rm  X_{co}$ factors, estimated from Equations  3 and 5,
using the  results from one-phase  LVG radiative transfer  models. The
black-bar  histogram   corresponds  to  the   best and  least  degenerate
solutions obtained from these models. }
\end{figure}

\begin{figure}
\epsscale{1.0}
\plotone{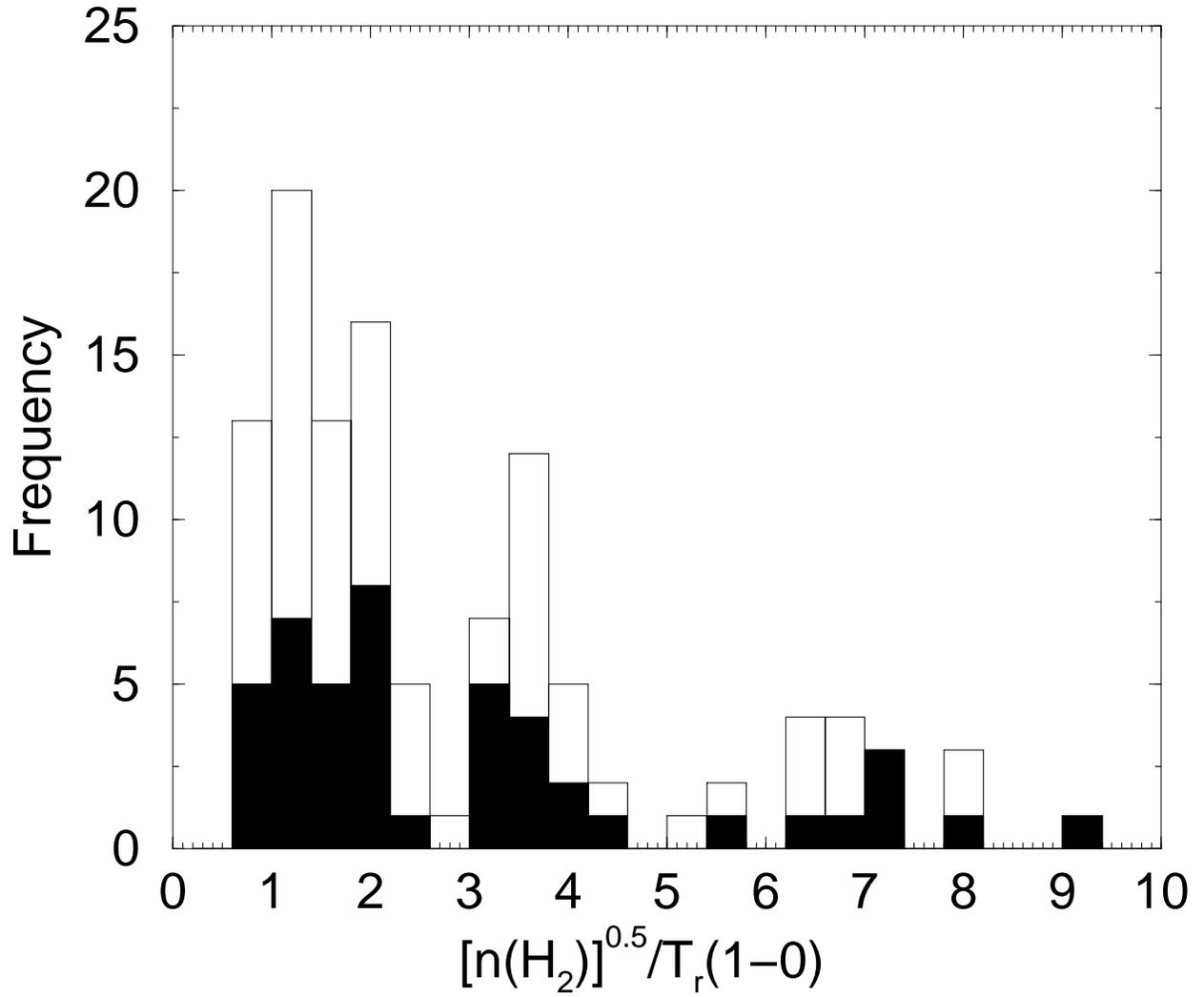}
\caption{The distribution of  the $\rm \sqrt{n(H_2)}/T_{r,1-0}$ factor for the
LVG solutions obtained for the sample (black  bars as in Figure 1).}
\end{figure}

\clearpage

\begin{figure}
\epsscale{1.0}
\plotone{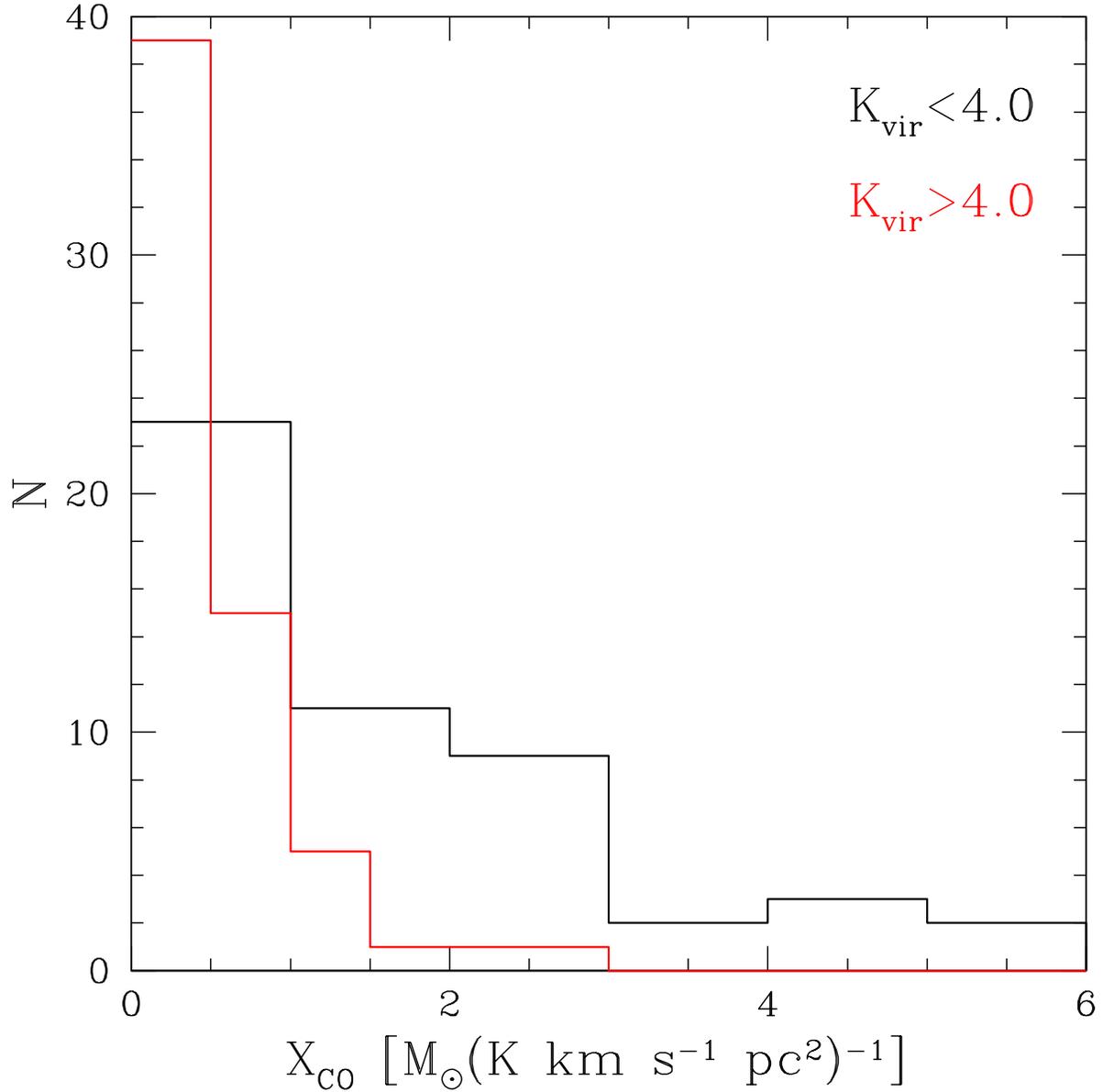}
\caption{The $\rm X_{co}$ factors, computed from Equation 3 and the
1-phase LVG solutions obtained for the CO line ratios measured in LIRGs (see Paper\,I)
for two different dynamical regimes of average gas motions:  self-gravitating
($\rm 1\la K_{vir}$$\la $4), and unbound ($\rm K_{vir}$$>$4).}
\end{figure}

\begin{figure}
\epsscale{1.0}
\plotone{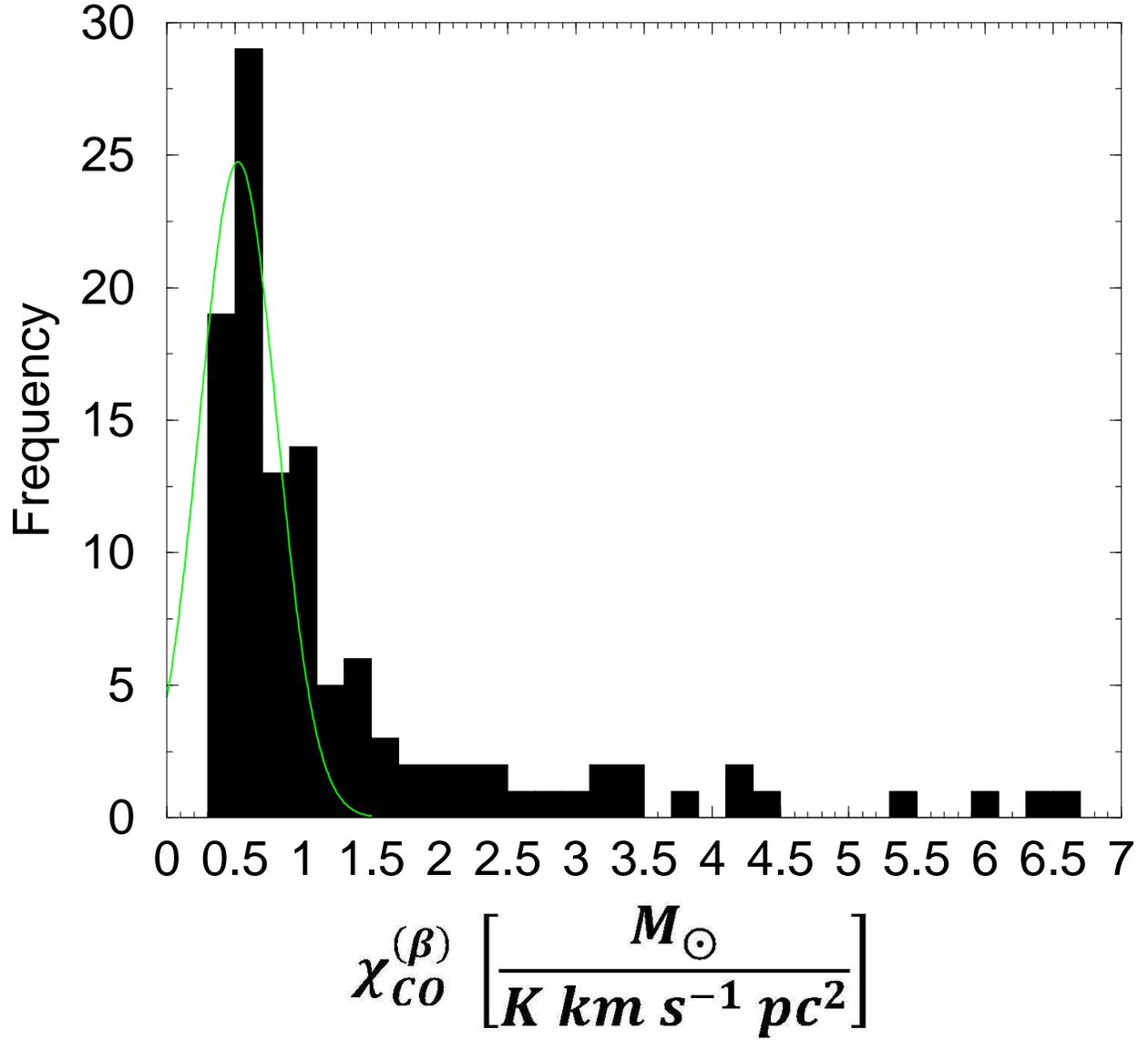}
\caption{The $\rm X_{co}$ factor, estimated from Equation 6 using the
$\beta _{J+1\,J}$ values computed from one-phase radiative transfer models. }
\end{figure}

\clearpage 

\begin{figure}
\epsscale{1.0}
\plotone{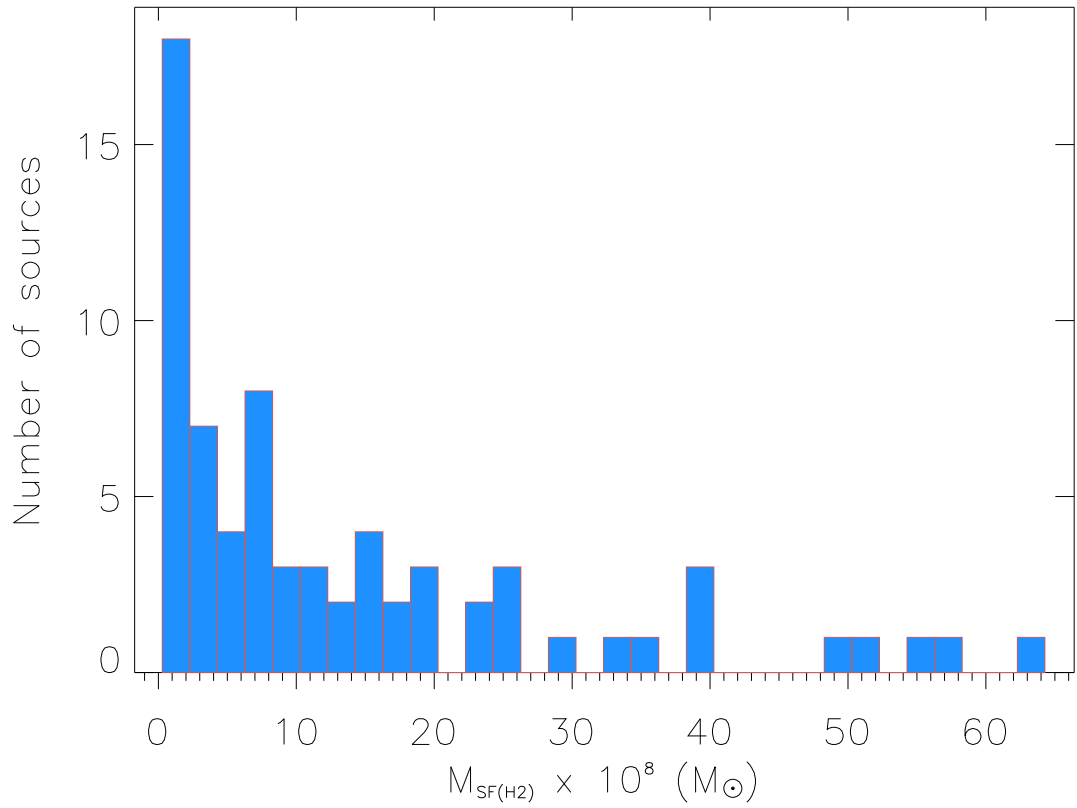}
\caption{The minimum molecular gas masses necessary for fueling
and Eddington-limited star formation in the LIRGs of our sample (see section 2.3). }
\end{figure}

\clearpage 

\begin{figure}
\epsscale{1.0}
\plotone{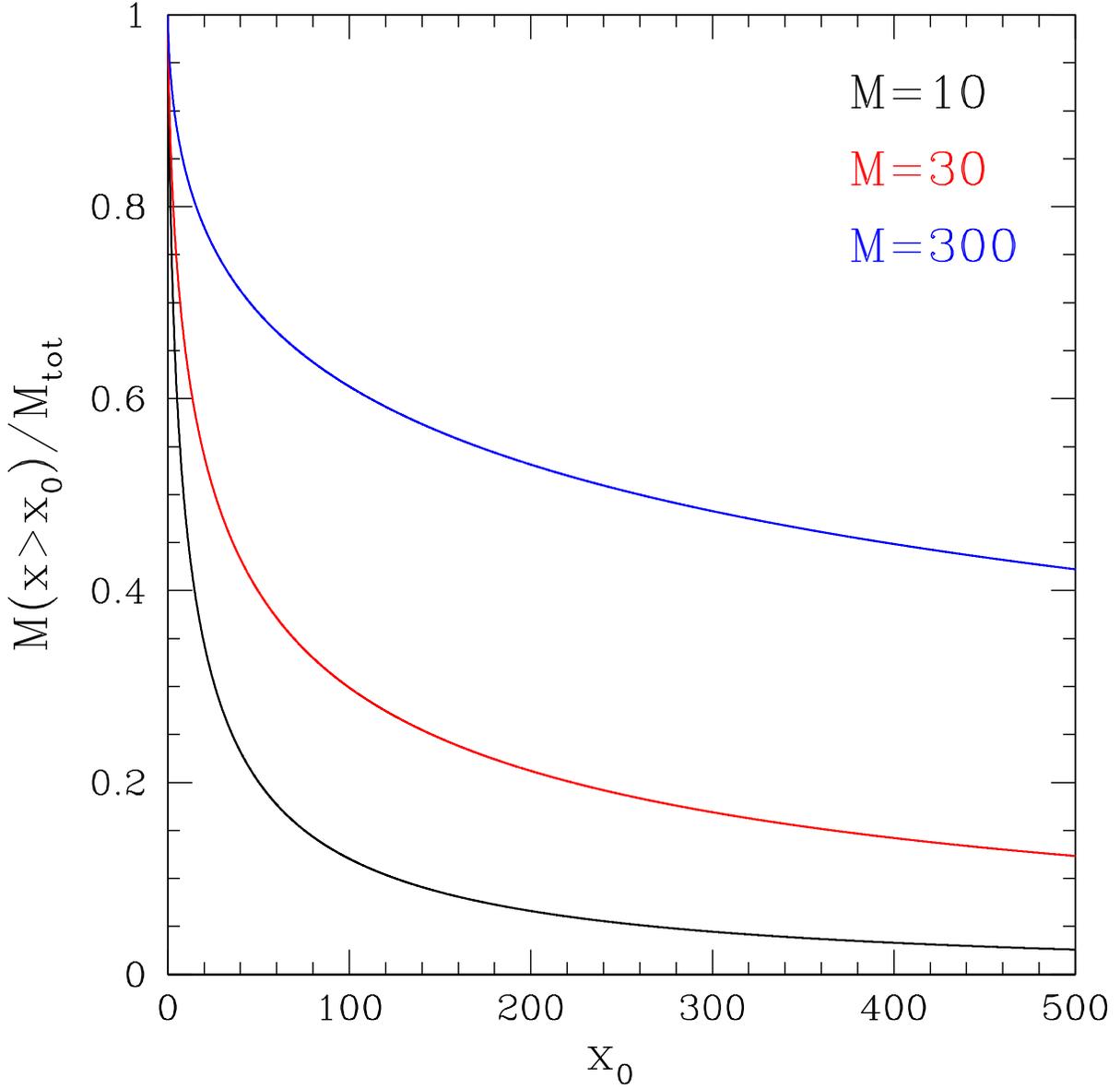}
\caption{The gas mass fraction  expected at overdensities $\rm x$$\geq
$$\rm x_{\circ}$  where x=$\rm n/\langle n \rangle$  (see Equation 13)
in supersonic  turbulent gas,  from spiral disks  (M=10) up  to ULIRGs
(M=300). }
\end{figure}

\clearpage

\begin{deluxetable}{lllccl}
\rotate
\tablecolumns{6}
\tablewidth{0pc}
\tablecaption{Molecular gas masses and $\rm X_{co}$ factors}
\tablehead{\colhead{Name} & \colhead{$\rm M_{tot}$($\rm X_{co}$)\tablenotemark{a}} & 
\colhead{$\rm M^{(2-ph)} _{tot}$($\rm X^{(2-ph)} _{co}$)\tablenotemark{b}} & 
\colhead{$\rm M_{SF}$\tablenotemark{c}} & 
\colhead{$\rm M_{h-ex}$\tablenotemark{d}} & \colhead{Remarks}\\
&\,\,\,\,\,($10^9$\,M$_{\odot}$)&($10^9$\,M$_{\odot}$) &($10^9$\,M$_{\odot}$)&($10^9$\,M$_{\odot}$)& }
\startdata
IRAS\,00057+4021 & 4.5 (1.1)  & 3.4-8.2(0.83-2.0) & 1.2 & 1.6 & degenerate $\rm X^{(l)}_{co}$ values\\
I\,Zw\,1         & 8.6 (1.5)  & 9.5 (1.65)        & 0.4 & 2.8 & Galactic  $\rm X^{(l)}_{co}$ adopted\\
NGC\,828         & 4.6-6.3 (0.8-1.1)& 4-14    (0.7-2.5) & 0.3 & 0.3  & degenerate $\rm X^{(l)} _{co}$  \\
IRAS\,02483+4302 & 1.6 (0.45) & 4.3-8.6 (1.2-2.4) & 1.4 & 3.3 & degenerate $\rm X^{(l)}_{co}$ values\\
IRAS\,03359+1523 & 22  (2.5)  &   \nodata         & 0.8 & \nodata & SF-quiescent ISM\\
VII\,Zw\,031     & 14.5 (1.25)&  39 (3.4)         & 2.0 & 3.1     &   Cold extended disk?  \\
IRAS\,05189--2524& 13.7 (3.5) &  9.4 (2.4)        & 3.6 & 3.6     & (h)-phase SLED assumed   \\
IRAS\,08030+5243 & 14.3-42.9 (1.5-4.5)& \nodata   & 2.4 & \nodata & degenerate 1-phase solutions\\
IRAS\,08572+3915 & 4.8-9.6 (3-6)&  \nodata        & 3.9 & \nodata & Highly excited CO J=6--5 line\\
Arp\,55          & 2.9-9.2 (0.25-0.8)& \nodata    & 0.9 & \nodata & degenerate 1-phase solutions\\
UGC\,05101       & 1.9-2.9 (0.4-0.6)& \nodata     & 1.6 & \nodata &   \\ 
NGC\,3310        & 0.03-0.15 (0.4-2.2)& \nodata   & 0.1 & \nodata & degenerate $\rm X_{co}$ values\\
IRAS\,10565+2448 & 3.8-4.8 (0.6-0.75) & 4.8 (0.75)& 2.6 & 2.6     & (h)-phase SLED assumed  \\
Arp\,299         & 1.2 (0.42)         & \nodata   & 1.9 & \nodata &  \\
IRAS\,12112+0305 & 15.2-53.5 (1.5-5.3)& 7.1-34.3 (0.7-3.4)&4.8&4  & Unconstrained $\rm X^{(l)} _{co}$\\
Mrk\,231         & 1.8-3.2 (0.25-0.45)& 21-49 (3-7)& 5.5 & 8-25   & CO 6-5, HCN lines used in 2-phase model\\
Arp\,193         & 1.6-3.5 (0.35-0.75)& 3.3-8.1 (0.7-1.8)& 0.9   &1.7-7.1&  CO 3--2, HCN lines used 
in 2-phase model\\
NGC\,5135        & 1.1-1.4 (0.35-0.45)&  \nodata  & 0.3 & \nodata &    \\
Mrk\,273         & 10.4 (2.0)         &  13  (2.5)& 3.3 &  6      & Cold extended disk?    \\
3C\,293          & 2.3-7.4 (0.5-1.6)  & 13.4-18 (2.9-3.9) & 0.05  & 1.1-3.5 & Galactic $\rm X^{(l)} _{co}$ adopted \\
IRAS\,14348--1447& 11-17   (0.65-1)   &  24-30  (1.4-1.7)& 5.6    & 5.6  & degenerate (l)-phase    \\
Zw\,049.057      &0.30-0.52 (0.35-0.6)& \nodata                   & 0.47  & \nodata &          \\
Arp\,220         & 1.85 (0.30)        &14.7-27.5 (2.4-4.5)& 4     & 12.6-25.2 & HCN lines used in 2-phase model  \\
NGC\,6240        & 2.5 (0.30)         &8.4-27.8  (1-3.3)  & 1.5   & 4.9-24.5  & HCN lines used  in 2-phase model \\
IRAS\,17208--0014& 5.2 (0.40)         &9.8-33   (0.75-2.5)& 6.2   & 3.5-6   & degenerate $\rm X^{(l)}_{co}$ values\\
$``$\, $``$      &  $``$              &34-77    (2.6-5.9) & $``$  & 27-70   &  HCN lines used in 2-phase model\tablenotemark{e}\\
IRAS\,22491--1808& 6.3-22 (0.70-2.4)  & \nodata           & 5.2   & \nodata &  degenerate $\rm X_{co}$ values\\
NGC\,7469        & 2.5       (0.72)   & 6.5      (1.88)   & 0.65  & 1.65    & Cold extended disk present\\
IRAS\,23365+3604 & 7.3-9.5 (1-1.3)    & 3.7-22   (0.5-3)  & 2.5   & 1.4     & Unconstrained  $\rm X^{(l)} _{co}$ \\
\enddata
\tablenotetext{a}{Total molecular gas mass  from Equation 3 and  the best 
one-phase LVG model parameters (the corresponding $\rm X_{co}$ value in $\rm X_l$ units),
with average values adopted in cases of significant LVG solution range degeneracy.}
\tablenotetext{b}{Total molecular gas mass from Equation 10 and a 2-phase fit
 (the corresponding $\rm X_{co}$ value in $\rm X_l$ units).}
\tablenotetext{c}{The minimum molecular gas mass necessary for an
Eddington-limited star formation rate (see 2.3).}
\tablenotetext{d}{The  gas mass of the (h)-phase in a 2-phase model when this is used.
 If a standard (h)-phase SLED and mass normalization is used (see  2.4) then
 $\rm M_{h-ex}$=$\rm M_{SF}$.}
\tablenotetext{e}{For LVG solutions with $\rm K_{vir}$(HCN)$\sim $1-6  (see B.24).}
\end{deluxetable}

\end{document}